\documentclass[aps,pra,
superscriptaddress,twocolumn,10pt]{revtex4-2}
\usepackage{color,ifthen,amsthm,amsmath,amsxtra,amsfonts,dsfont,graphicx,bm,tikz,scalerel,wasysym,bbm,graphicx,amsthm, braket,physics,empheq}
\usepackage[colorlinks=true,linkcolor=blue, citecolor=blue, urlcolor=blue, bookmarks]{hyperref}
\usepackage[inline]{enumitem}

\newcommand{\be}{\begin{equation}}
\newcommand{\ee}{\end{equation}}
\newcommand{\bea}{\begin{eqnarray}}
\newcommand{\eea}{\end{eqnarray}}
\newcommand{\bse}{\begin{subequations}}
\newcommand{\ese}{\end{subequations}}

\theoremstyle{plain}

\theoremstyle{plain}

\theoremstyle{plain}

\usepackage{color,ifthen,amsthm,amsmath,amsxtra,amsfonts,dsfont,graphicx,bm,tikz,scalerel,wasysym, physics, bbm,  graphicx}
\usepackage[colorlinks=true,linkcolor=blue, citecolor=blue, urlcolor=blue, bookmarks]{hyperref}

\usepackage{amsthm}

\begin{document}

\title{Transport and entanglement across integrable impurities from Generalized Hydrodynamics}

\author{Colin Rylands}
\affiliation{SISSA and INFN Sezione di Trieste, via Bonomea 265, 34136 Trieste, Italy}

\author{Pasquale Calabrese}
\affiliation{SISSA and INFN Sezione di Trieste, via Bonomea 265, 34136 Trieste, Italy}
\affiliation{International Centre for Theoretical Physics (ICTP), Strada Costiera 11, 34151 Trieste, Italy}

\begin{abstract}
 Quantum impurity models (QIMs) are ubiquitous throughout physics. As simplified toy models they provide crucial insights for understanding more complicated strongly correlated systems, while in their own right are accurate descriptions of many experimental platforms. In equilibrium, their physics is well understood and have proven a testing ground for many powerful theoretical tools, both numerical and analytical, in use today. Their non-equilibrium physics is much less studied and understood. However, the recent advancements in non equilibrium integrable  quantum systems through the development of generalized hydrodynamics (GHD) coupled with the fact that many archetypal QIMs are in fact integrable presents an enticing opportunity to enhance our understanding of these systems.  We take a step towards this  by expanding the framework of GHD to incorporate integrable interacting QIMs.  We present a set of Bethe-Boltzmann type equations which incorporate the effects of impurity scattering and discuss the new aspects which include entropy production. These impurity GHD equations are then used to study a bipartioning quench wherein a relevant backscattering impurity is included at the location of the bipartition.  The density and current profiles are studied as a function of the impurity strength and expressions for the entanglement entropy  and full counting statistics are derived. 

\end{abstract}
\maketitle
\textit{Introduction.---} The role of impurities in quantum physics has long been known to be central in our understanding of physical phenomena. Classical scattering from impurities in metals gives a finite lifetime to quasi particles, taming unphysical divergences in conductivities.  Interaction between electrons and magnetic impurities leads to the formation of clouds of many correlated particles resulting in strongly correlated compounds~\cite{hewson1993kondo}. Pronounced effects in emission and absorption spectra are explained via the orthogonality catastrophe, the thermodynamic vanishing of overlaps between ground states of pure and impure systems~\cite{mahan2000many}.  Quantum impurity models (QIMs) are simplified versions of the above scenarios, consisting of a single impurity with few degrees of freedom interacting with a much larger environment, which capture the essential physics. They provide an ideal platform to study strongly correlated phenomena such as dynamical scale generation or asymptotic freedom and as such have become a proving ground for many of the most widely used non perturbative techniques, both analytical and numerical of modern theoretical physics. They are also of interest in their own right as quantum impurity systems are now routinely engineered in the laboratory~\cite{goldhaber1998kondo,makarovski2007evolution,makarovski2007su4,mebrahtu2012quantum,
iftikhar2015two,pouse2023exotic}. 

\begin{figure}
 \includegraphics[trim=0 500 700 00 ,clip, width=1\columnwidth]{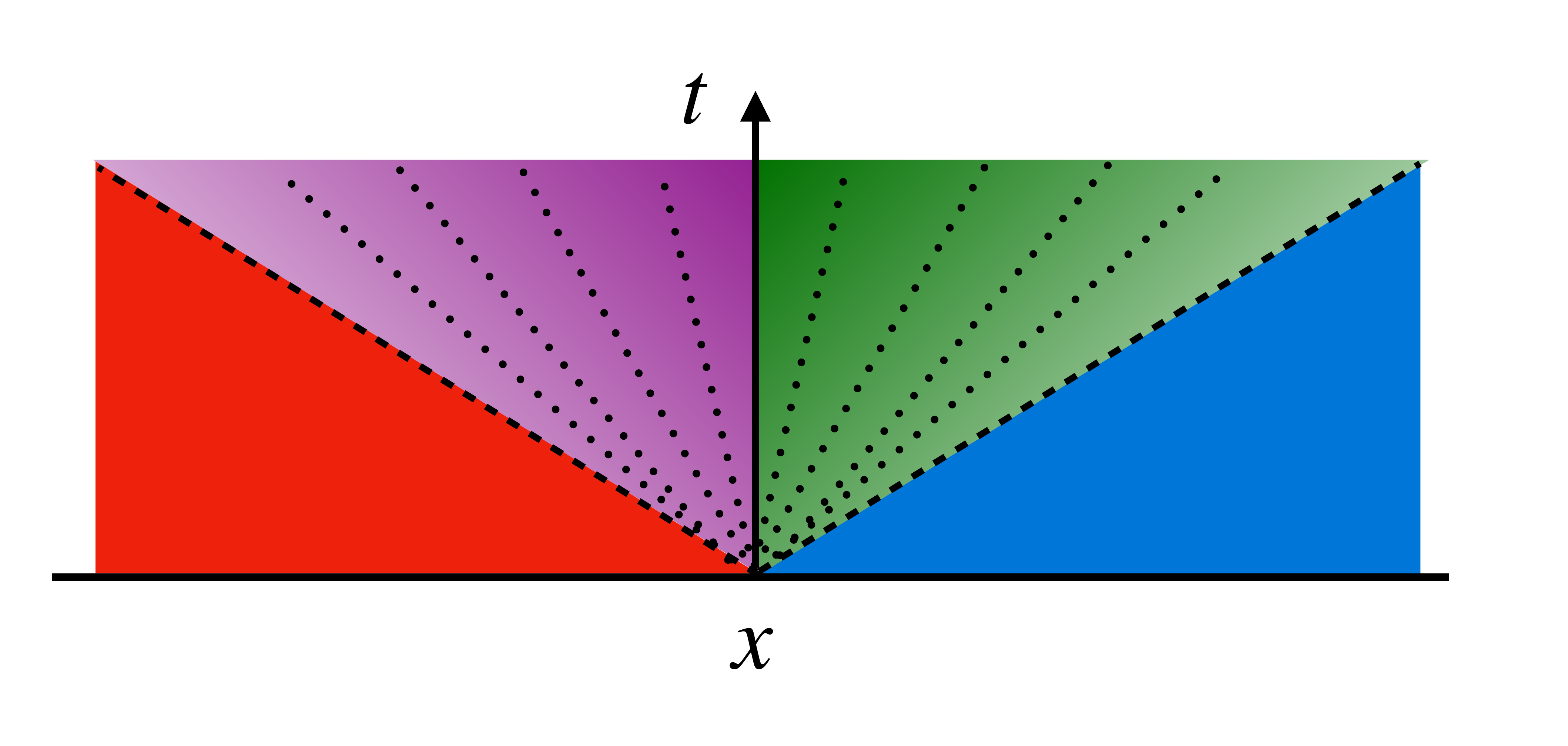} 
\caption{\label{Fig} The bipartioning protocol: The quantum system is prepared in two distinct states e.g.  all rapidities being filled up to a certain level, either side of a quantum impurity at the origin.  The system in then allowed to evolve leading to the formation of a light cone around the partitioning point.  The resulting local stationary state along rays of constant $x/t$ (dotted lines) differ from the pure case due to the reflection, transmission and scattering of particles from the impurity.  For the particular example of the Kane-Fisher model considered in the text there is no dependence on  the particular ray inside the light cone other than it being to the left or right of the impurity,  see ~\eqref{Rho}.}
\end{figure}

 Equilibrium properties of QIMs are quite well understood in large part due to the fact that many of the most important examples are integrable~\cite{andrei1983solution, tsvelik1983exact}. These include the well known Kondo~\cite{andrei1980diagonalization,vigman1980exact,wiegmann1981exact},  Anderson~\cite{kawakami1981exact,wiegmann1983exact} and Kane-Fisher models~\cite{fendley1994exact,rylands2016quantum} among others~\cite{tsvelik1985exact,andrei1984solution,
bedurftig1996integrable,bedfurtig1997exact,
 andrei1984heisenberg, lee1988integrable,jerez1998solution,jerez1995fermi,rylands2017quantum,rylands2018quantum,pasnoori2020kondo,pasnoori2022rise}. However experimental advances over recent decades have instigated a shift away from studies of equilibrium phenomena to those of far from equilibrium quantum physics and in particular of integrable interacting models~\cite{bastianello2022introduction,rylands2020nonequilibrium}.  At the same time key technical breakthroughs such as the quench action method~\cite{CauxEssler, Caux} and generalized hydrodynamics~\cite{CastroDoyonYoshimura,bertini2016transport} have facilitated our understanding of these systems. 
 These powerful analytical tools allow one to study the long time behaviour and emergent properties of non-equilibrium systems and have been applied to a plethora of scenarios, such as homogeneous~\cite{PiroliPozsgayVernier, DeNardis,Brockmann, wouters2014quenching, pozsgay2014correlations, MestyanPozsgayTakacsWerner, bertini2016quantum, MestyanBertiniPiroliCalabrese,BertiniSchurichtEssler,BertiniTartagliaCalabrese,PiroliVernierCalabresePozsgay1, piroli2016multiparticle, piroli2016quantum, alba2016the, piroli2016exact, denardis2015relaxation, mestyan2017exact,rylands2022integrable,rylands2023solution} and inhomogeneous~\cite{bulchandani2017solvable,bulchandani2018bethe,bastianello2018generalized,bastianello2019generalized,koch2021adiabatic,mestyan2019spin,collura2018analytic,bertini2019transport,
  scopa2022generalized,scopa2021real,alba2021generalized,piroli2017transport,gopalakrishnan2018hydrodynmaics,schemmer2019generalized,malvania2021generalized,
  zadnik2021folded,ruggiero2020quantum,collura2020domain,ruggiero2022quantum,scopa2021exact,scopa2022exact,scopa2023one,rylands2020many}  quantum quenches.  Despite their huge success and widespread use however these techniques have yet to be applied to interacting quantum impurity systems.  
In this Letter we take a step toward rectifying this and enlarge the framework of GHD to incorporate integrable quantum impurities. This is done by modifying the standard Bethe-Boltzmann type equations of GHD through the inclusion of an exactly determined collision integral describing the integrable scattering with the impurity~\eqref{GHD}.  We make some nontrivial checks on the resulting impurity GHD equations and comment on its new features.  Afterward we give an exact solution to these equations for a bipartioning quench in which an interacting backscattering impurity is located at the bipartition.  From this we calculate the density and current profiles as well as their full counting statistics and the half system entanglement entropy.

\textit{Integrable impurity models.---}
 Integrable quantum systems generically possess an extensive number of local conserved charges which heavily impact upon the static and dynamic properties of the system. Their spectra consist of a set of stable quasiparticle species,  parameterized by a species index $j$ and a rapidity $\lambda$ and whose  properties are described through  kinematic data which includes their value under the conserved charges such as  their  energy $\epsilon_j(\lambda)$ and momentum $p_j(\lambda)$,  in addition to the two-particle scattering kernel $T_{jk}(\lambda)$.  The state of a system is specified by the types of quasiparticles present and in the thermodynamic limit, this can be done through their distribution in rapidity space,  denoted by $\rho_j(\lambda)$.  It is also convenient to introduce  $\rho_j^h(\lambda)$ the distribution of unoccupied quasiparticles as well as the occupation function $\vartheta_j(\lambda)=\rho_j(\lambda)/\rho^t_j(\lambda),~\rho_j^t(\lambda)=\rho_j(\lambda)+\rho^h_j(\lambda)$.  These quantities are not unrelated and obey the Thermodynamic Bethe Ansatz equations, a set of coupled integral equations arising from the quantization conditions $\rho^t_j(\lambda)=|p'_j(\lambda)|/2\pi-\sum_k\int {\rm d}\mu T_{jk}(\lambda-\mu)\rho_k(\mu)$,  with $'$ denoting $d/d\lambda$.  Moreover in the presence of many excitations the quasiparticle properties become dressed due to the interactions we e.g.  $\epsilon_j(\lambda)\to [\epsilon_j(\lambda)]^{\rm Dr}$.  This dressing of physical quantities can be defined through a second dressing action $[\epsilon'_j(\lambda)]^{\rm dr}=\epsilon'_j(\lambda)-\sum_{k}\int d\mu T_{jk}(\lambda-\mu)\vartheta_k(\mu)[\epsilon'_j(\mu)]^{\rm dr}$ with the  physical dressed quantities then given by integrating the resulting function, $[\epsilon_j(\lambda)]^{\rm Dr}=\int^\lambda{\rm d}{ \mu} [\epsilon'_j(\mu)]^{\rm dr}$.  
 
 For an integrable QIM additional information is required to describe the interaction between the bulk quasiparticles and the impurity.  This can be encoded via two different bases: (i) the \textit{scattering basis} or (ii) the \textit{diagonal basis}.  In (i) the bulk quasiparticles are as described above and their interaction with the impurity  is given by the impurity $S$-matrix which is in general non diagonal, giving rise to scattering between different quasiparticle species.  We characterize this through its off diagonal components, $R_{ij}(\lambda)$ which are the bare reflection coefficients for a particle $i$ to scatter to a particle $j$.  Evidently this basis does not diagonalize the Hamiltonian but is useful when considering transport properties and has been utilized in earlier studies of non-equilibrium QIMs~\cite{fendley1995exact,fendley1995exact2,fendley1995exact3,mehta2006nonequilibrium,mehta2007nonequilibrium,konik2001transport,konik2002transport}.  In (ii) the bulk quasiparticle basis is rotated so as to diagonalize the impurity $S$-matrix and therefore also the Hamiltonian.  Their interaction  with the impurity is characterized by the impurity phase shifts $\varphi_j(\lambda)$ which the quasiparticles acquire when passing through the impurity~\cite{Note2}.   This basis is the natural one for describing equilibrium properties of the system.  The two bases are straightforwardly related as are their scattering data,  in particular we shall emphasize this by writing the reflection coefficients as $R_{ij}[\varphi(\lambda)]$.  In what follows we ignore the shift of the Bethe equations arising from the impurity as it does not contribute to leading order in the system size. The effect of the impurity is considered solely to cause scattering between quasiparticle species.  
 
~\footnotetext[1]{We use the same index for quasiparticle species in both bases with the correct type of quasiparticle being understood from context. } 
 
\textit{Generalized hydrodynamics with impurities.---}
GHD provides the long wavelength description of integrable models in an inhomogeneous far from equilibrium setting.   It is obtained by considering the transport of the conserved charges through the system via their continuity equations  which are then cast in terms of the quasiparticle distributions.    The result is a set  of coupled Euler equations for the quasiparticle rapidity distributions which become functions of space and time; $\rho_j(\lambda,x,t)$. They can be viewed semiclassically as describing the ballistic propagation of quasiparticles through the system.   A crucial step in this procedure is to express the conserved currents in terms of $\rho_j(\lambda,x,t)$,  which was at first conjectured on general grounds and later proven microscopically~\cite{borsi2020current}.   In the spirit of~\cite{CastroDoyonYoshimura,bertini2016transport,bulchandani2018bethe} we incorporate the presence of an integrable impurity at the origin through the inclusion of a collision integral term in the GHD equations,
\begin{eqnarray}\label{GHD}
\partial_t\rho_j(\lambda,x,t)+\partial_x [v_j(\lambda,x,t)\rho_j(\lambda,x,t)]=\delta(x)\mathcal{I}_j(\lambda,t).
\end{eqnarray}
Here the left hand side is the standard GHD equation with $v_j(\lambda,x,t)=[\epsilon'_j(\lambda)]^{\rm dr}/[p'_j(\lambda)]^{\rm dr}$ being the dressed quasiparticle velocity.  The right hand side is new and is given by $\mathcal{I}_j=\sum_{k\neq j} \mathcal{I}_{jk}$ with
\begin{equation}\label{CollisionIntegral}
\mathcal{I}_{jk}= |R_{kj}([\varphi]^{\rm Dr})|^2\rho_k\left[1-\vartheta_j\right]-|R_{jk}([\varphi]^{\rm Dr})|^2\rho_j\left[1-\vartheta_k\right].
\end{equation}
The first term here represents the scattering of species $k$ into $j$ while the second is for the reverse process.  The factors $1-\vartheta_i(\lambda)$ appear as we have assumed that the quasiparticles obey Pauli exclusion however other statistics can also be incorporated through appropriate replacements~\cite{doyon2018exact}. The terms $|R_{jk}([\varphi]^{\rm Dr})|^2$ are the reflection amplitudes of the \textit{dressed} quasiparticles and when combined with the other factors present give the total rate of scattering into and out of the species $j$ due to the species $k$.  As a consequence of the integrability of the impurity this preserves the total number of quasiparticles in the system but breaks some of the conservation laws,  specifically those for which $j$ and $k$ have different charges.  It is important to note that while the GHD equations are for those in the scattering basis it is the phase shift from the diagonal basis that undergoes dressing rather than the reflection coefficients themselves.  This is done locally at the impurity site,  i.e.  using the occupation functions at $x=0$.  
Equations \eqref{GHD} and \eqref{CollisionIntegral} constitute the main result of our work.   Through them one can study the nonequilibrium dynamics resulting from inhomogeneous quenches such as the bipartioning protocol (see below and Fig.~\ref{Fig}) in the presence of integrable interacting impurities.  
They represent a very natural extension of the formalism,  indeed such a protocol was the setting for the first appearance of the GHD equations,  albeit only a purely reflecting defect was considered~\cite{bertini2016determination}.  It should be noted however that the scattering between quasiparticle species caused by the impurity leads to many nontrivial effects which are absent when the defect is purely reflecting (or equivalently purely transmitting).  These include, the generation of a strong coupling scale, a feature of interacting QIMs, entropy production and nontrivial charge and current fluctuations. 

\textit{Checks.---} We can perform some simple analytic checks of our result.  First we consider a noninteracting model of one species type,  with energy $\epsilon(\lambda)$ and momentum $p(\lambda)$ such that $\text{sgn}[\epsilon'(\lambda)/p'(\lambda)]=\text{sgn}[\lambda]$,  coupled to a non interacting impurity at the origin. The impurity allows for both transmission and reflection i.e.  a flip of the sign of the momentum of an incident particle $p(\lambda)\to -p(\lambda)$ with  reflection coefficient $R(\lambda)$.  We take the system to be initially decoupled from the impurity and prepared in its ground state with different Fermi levels, $\Lambda_{L,R}$  to the left and right of the origin, see Fig~\ref{Fig}. The impurity is then suddenly turned on and the system allowed to evolve. This situation has been studied several times in both lattice and continuum systems~\cite{bertini2017lightcone,ljubotina2019non,capizzi2023domain,gouraud2022stationary,gouraud2022quench}
(see also~\cite{bastianello2018nonequilibrium,bastianello2018superluminal} for a moving defect). To reproduce those results we treat the quasiparticles with $\lambda>0$ and $\lambda<0$ as different species with the impurity causing scattering between the two.  As interactions are absent, $T_{jk}=0$, no dressing occurs and one can straightforwardly solve~\eqref{GHD} finding that the total current through the impurity is $J(t)=\int_{\Lambda_L}^{\Lambda_R} {\rm d}\lambda\,|p'(\lambda)|\, [1-|R(\lambda)|^2]/2\pi$, the usual Landauer-Buttiker result previously obtained.

As a more nontrivial check let us consider an interacting QIM with two quasiparticle species and an impurity which mixes the two.  For the bare impurity $S$-matrix we take the generic form \begin{eqnarray}
S(\lambda)=e^{i\alpha(\lambda)}\begin{pmatrix}
\cos{\chi(\lambda)}& i\sin{\chi(\lambda)}\\
i\sin\chi(\lambda)& \cos{\chi(\lambda)}
\end{pmatrix}\label{Smatrix}
\end{eqnarray}
such that the diagonal basis consists of symmetric and anti symmetric combination of the quasiparticles with phase shifts $\alpha(\lambda)\pm\chi(\lambda)$.  Suppose now we take the ground state of the system at finite density so that rapidities $\lambda>\Lambda$ are unoccuppied.  To this we add a single scattering quasiparticle at $\lambda=\lambda^p>\Lambda$.  According to~\eqref{CollisionIntegral} this particle is then scattered by the impurity from one species to the other at a rate $\mathcal{I}=|\sin{[\chi^{\rm Dr}}(\lambda^p)]|^2$.  This is the inverse lifetime of the quasiparticle and as $\lambda^p\to\Lambda$ can be related to the zero temperature resistivity of the system which has been calculated in QIMs such as the Kondo model.  Specializing to that case we find agreement with the known exact result calculated using the $T$-matrix formalism~\cite{andrei1982calculation, Note2}.

\footnotetext[2]{See supplementary material that contains (i) additional information on the resistance of QIMs using Bethe ansatz (ii) the derivation of \eqref{entropyrate} (iii) details concerning the bipartite quench in the Kane-Fisher model.}

\textit{Entropy production.---} In the absence of the impurity the GHD equations preserve entropy, however  diffusive corrections to this have been calculated which allow for the transfer of entropy between scales of the system~\cite{denardis2018hydrodynamic,denardis2019diffusion}.  The collision integral $\mathcal{I}_j$ plays a similar role here and results in the production of entropy even from a zero temperature state.  To see this we note that we can derive a GHD equation also for the occupation functions albeit with a modified impurity term,
\begin{eqnarray}
\partial_t\vartheta_j(\lambda,x,t)+v_j(\lambda,x,t)\partial_x \vartheta_j(\lambda,x,t)=\delta(x)\mathcal{I}^\vartheta_j(\lambda,x,t)
\end{eqnarray}
where $[\mathcal{I}^\vartheta_j(\lambda,t)]^{\rm dr}\rho^t_j(\lambda,0,t)=\mathcal{I}_j(\lambda,t)$.  Likewise an Euler type equation can be derived for $\rho^h_j$ also.  Combining these along with the definition of the Yang-Yang entropy~\cite{takahashi1972one} we find that the total entropy production rate in the system is~\cite{Note2}
\begin{eqnarray}\label{entropyrate}
\partial_t \mathcal{S}(t)&=&\sum_j\int\mathrm{d}\lambda s_j(\lambda,0,t)\mathcal{I}_j(\lambda,t)
\end{eqnarray}
where $s_j(\lambda,x ,t)=\log\left[\rho^h_j(\lambda,x,t)/\rho_j(\lambda,x,t)\right]+\sum _k\int \mathrm{d}\mu T_{jk}(\lambda-\mu) \log[1-\vartheta_k(\mu,x,t)]$.  The establishment of a non-equilibrium steady state therefore leads to a linear in time increase in the total entropy of the system.

\textit{Kane-Fisher model.---} We now look to implement this framework in a specific QIM, the Kane-Fisher model describing a backscattering impurity in a Luttinger liquid.  The Hamiltonian is given by
\begin{eqnarray}\nonumber
H=\int\mathrm{d} x \,\psi^\dag(x)[-i \sigma^z\partial_x]\psi(x)+U\delta(x)\psi^\dag(x)\sigma^x\psi(x)\\+\frac{g}{2}\left([\psi^\dag(x)\psi(x)]^2-[\psi^\dag(x)\sigma^z\psi(x)]^2\right).
\end{eqnarray}
Here $\psi=(\psi_r,\psi_l)^T$ are two component Weyl fermions with linear dispersion and positive $(r)$ or negative $(l)$ momentum.  They interact with each other via a four fermion interaction of strength $g$ and with an impurity allowing both transmission and reflection at the origin of strength $U$.  In the absence of the impurity, the model has two $U(1)$ charges the total charge $N=\int{\rm d}x \,\psi^\dag\psi$,  and chiral charge  $J=\int {\rm d}x\,\psi^\dag\sigma^z\psi$, however when $U\neq 0$ the latter is no longer conserved and a current is generated by the impurity.  The impurity is RG relevant for $g>0$ and leads to a dynamically generated scale, $T_U$ and vanishing conductance at zero temperature in equilibrium~\cite{kane1992transmission}.  The model is integrable and can be solved either through bosonization and mapping onto the boundary sine-Gordon model~\cite{fendley1994exact,fendley1995exact} or directly in fermionic form using coordinate Bethe ansatz~\cite{rylands2016quantum}.  

We study a bipartioning quench, where the system is prepared in the ground state of the $U=0$ system at different chemical potentials to the left and right of the origin and then allowed to evolve according $H$ at non zero $U$. This models the sudden coupling of two disjoint quantum wires through a quantum point contact.  In this context the system contains two quasiparticle species labeled $\pm$. Their energy and momenta are $\epsilon_\pm(\lambda)=\pm p(\lambda)=e^{\lambda}$ while $T_{\pm\pm}(\lambda)=T_{\pm\mp}(\lambda)\equiv T(\lambda)$,
\begin{eqnarray}
T(\lambda)=\int \frac{{\rm d}\omega }{2\pi}e^{-i\omega \lambda} \frac{\sinh(\frac{\pi}{2}(\gamma-1)\omega)}{2\sinh{(\frac{\pi}{2}\gamma\omega)}\cosh{(\frac{\pi}{2}\omega)}},
\end{eqnarray}
where $\gamma^{-1}\approx 1+2g/\pi$, which we take to be an integer.  The quasiparticles scatter from the impurity with an $S$-matrix of the form~\eqref{Smatrix} with $\chi(\lambda)=\pi/2-\arctan{e^{(\lambda-\lambda_U)/\gamma}}$ where we have introduced $\lambda_U\approx (1+1/\gamma)\log{U}$ which sets the impurity scale,  $T_U\sim e^{\lambda_U}$ and $\alpha(\lambda)$ given in the supplement.

Initially the system is taken to be populated only to the left of the impurity according to $\vartheta_\pm(\lambda,x,0)=\Theta(-x)\Theta(\Lambda-\lambda)$ with $\Theta(x)$ the Heaviside function and $E_F=e^\Lambda$ being the Fermi level for the quasiparticles.  The corresponding rapidity distribution, $\rho_0(\lambda)$ can then be determined analytically through the Wiener-Hopf method~\cite{Note2}.   From this initial condition the solution to \eqref{GHD} is given by
\begin{eqnarray}\nonumber
\rho_-(\lambda, x, t)&=&\Theta(-x)\left[\Theta(-x-t)+\Theta(t+x)\mathcal{R}(\lambda)\right]
\rho_0(\lambda)\\\label{RhoMain}
\rho_+(\lambda,x,t)&=&\left[\Theta(-x)+\Theta(x)\Theta(t-x)\mathcal{T}(\lambda)\right]\rho_0(\lambda)\end{eqnarray}
where we introduced $\mathcal{R}(\lambda)=|R([\chi(\lambda)]^{\rm Dr})|^2$ the dressed reflection amplitude $\mathcal{T}(\lambda)=1-\mathcal{R}(\lambda)$ the transmission amplitude.  A similar form holds also for the occupation functions with the replacement $\rho_0(\lambda)\to \Theta(\Lambda-\lambda)$ and $\mathcal{R}(\lambda)\to\mathcal{R}^\vartheta(\lambda
)$ such that $[\mathcal{R}^\vartheta(\lambda)]^{\rm dr}=\mathcal{R}(\lambda)\Theta(\Lambda-\lambda)$ and $\mathcal{T}^\vartheta(\lambda)=1-\mathcal{R}^\vartheta(\lambda)$.    The dressed phase shift  $\chi^{\rm Dr}$ can be determined analytically providing a full analytic solution to the problem~\cite{Note2}.   From this one finds that for $\Lambda\ll \lambda_U$ the dressing has negligible effect, in essence the natural scale of the system as set by the impurity, $T_U$ is much large than the one set by the quench, $E_F$ and system remains close to equilibrium.  In the opposite limit however the dressed phase shift can be approximated by 
\begin{equation}\label{eq:chifourier}
\chi^{\rm Dr}(\lambda)\simeq\frac{\pi}{2}+\int_{-\infty}^\infty \frac{{\rm d}\omega}{2  i\omega}\frac{\tanh{\left(\frac{\pi}{2}\gamma\omega\right)}\cosh{(\frac{\pi}{2}\omega)}}{\sinh{\left(\frac{\pi}{2}(1+\gamma)\omega\right)}}e^{-i\omega(\lambda-\lambda_U)}.
\end{equation}
An important feature here is that for $\lambda_U\ll\lambda$ the phase shift is constant $\chi^{\rm Dr}(\lambda)\simeq \frac{\pi}{2}\frac{1-\gamma}{1+\gamma}$.  

\begin{figure}
 \includegraphics[trim=0 200 0 220 ,clip, width=1\columnwidth]{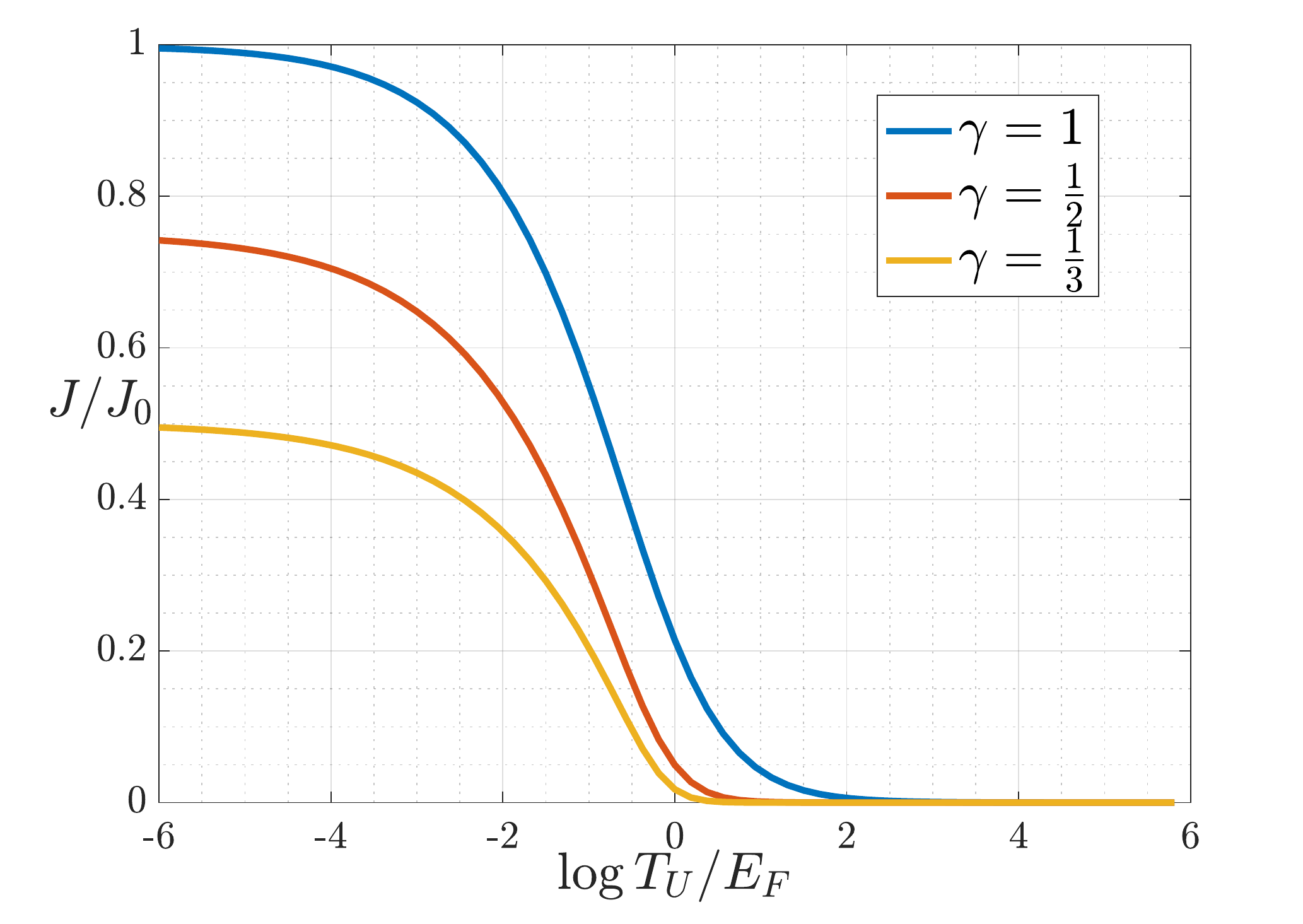} 
\caption{\label{Fig2} The current within the light cone about $x=0$,  rescaled by its value without the impurity $J(x,t)/J_0(x,t)$~\eqref{KFcurrent} as a function of $\log{T_U/E_F}$. The different curves correspond to  interaction strength $\gamma=1,\frac{1}{2},\frac{1}{3}$.  The asymptotic values for large $E_F$ are given by $\cos^2(\frac{\pi}{2}\frac{1-\gamma}{1+\gamma})$ as explained in the text. }
\end{figure}

 The current and charge density are straightforwardly obtained from~\eqref{RhoMain}.  The former is nonzero only within the light cone surrounding the impurity and takes a Landauer-Buttiker form,
\begin{eqnarray}\label{KFcurrent}
J(x,t)=\Theta(t-|x|)\int {\rm d}\lambda \mathcal{T}(\lambda)\rho_0(\lambda).
\end{eqnarray} 
Using our asymptotic result for the phase shift we see that in the far from equilibrium regime $J(x,t)\simeq \cos^2(\frac{\pi}{2}\frac{1-\gamma}{1+\gamma})J_0(x,t)$ where $J_0(x,t)$ is the current at  $U=0$.  In Fig.~\ref{Fig2} we plot the current for different values $\gamma$ as a function of $T_U/E_F$ normalized by $J_0$.  
The density also only deviates from its initial value within the light cone. Within this region we have 
\begin{eqnarray}\label{KFdensity}
N(x,t)=\Theta(\pm x)\int{\rm d}\lambda [1\mp \mathcal{R}(\lambda)]\rho_0(\lambda)
\end{eqnarray}
The density therefore exhibits a finite jump across the impurity, i.e. a potential difference of $2\int{\rm d}\lambda  \mathcal{R}(\lambda)\rho_0(\lambda)$.   In addition, using the Friedel sum rule  and $\alpha^{\rm Dr}(\lambda)$ we may compute the impurity induced charge deficit at $x=0$  obtaining to leading order in $E_F/T_U$,  $\delta N_\text{imp}=-\frac{2}{1+\gamma}$~\cite{Note2}. Thus despite being a theory of marcoscopic length scales we can use GHD to infer microscopic properties.

We may go beyond the expectation values of the current and density and calculate their fluctuations as well.  For this we introduce the time integrated current  $\mathcal{J}(t)=\int^t_0 {\rm d}\tau J(0^+,\tau)$ as well as its generating function
\begin{eqnarray}
G(t,\beta)=\text{Tr}[\varrho(t)\, e^{\beta \mathcal{J}(t)}]
\end{eqnarray}
where $\varrho$ is the density matrix of the system.  Then upon using the continuity equation $\partial_tN(x,t)+\partial_x J(x,t)=0$ this will give the fluctuations of the charge across the impurity.  This quantity obeys a large deviation principle and has been studied recently in the context of both integrable and nonintegrable models~\cite{krajnik2022exact,krajnik2022universal,gopalakrishnan2022theory,mccullough2023full,
myers2020transport, bertini2022full,bertini2023wip}.  Using  recent results for the full counting statistics of integrable models~\cite{myers2020transport, bertini2022full,bertini2023wip} we find
\begin{align}\nonumber
&\log{G(t,\beta)}=t\!\int\!\frac{{\rm d}\lambda}{2\pi}e^\lambda\log{\left[\mathcal{R}^\vartheta\!(\lambda)+\mathcal{T}^\vartheta\!(\lambda)e^{y(\lambda,\beta)}\right]}\\
&y(\lambda,\beta)=\beta-\int \!{\rm d}\mu \,T(\lambda-\mu)\log{\left[\mathcal{R}^\vartheta\!(\mu)+\mathcal{T}^\vartheta\!(\mu)e^{y(\mu,\beta)}\right]}.
\end{align}
The function $y(\lambda,\beta)$ is related to the effective charge carried by the quasiparticles,  indeed differentiating the second line above we have that $\partial_\beta y(\lambda,\beta)=q^{\rm dr}$ the dressed quasiparticle charge.  These results can be contrasted with those obtained for the related boundary sine-Gordon model~\cite{fendley1995exact,fendley1995exact2,fendley1995exact3}.  In these works the impurity scattering is also treated using a Boltzmann equation however the dressing of the scattering amplitudes does not appear. This difference is a result of the physical setup of the problem.  Therein the current was studied directly in the nonequilibrium steady state which emerges from an adiabatic turning on of the impurity and the potential difference~\cite{bazhanov1997integrable, bazhanov1999on}. In that circumstance one can envisage starting in a dilute limit where bare quasiparticles scatter one by one off the impurity,  without any dressing and then slowly increasing the density.  In the quench problem considered here this dilute limit cannot be used and we must instead turn to GHD which is a theory of dressed quasiparticles as opposed to bare ones.  Nevertheless when $\Lambda\ll \lambda_U$ one obtains the expressions for the current of~\cite{fendley1995exact,fendley1995exact2,fendley1995exact3}.  

Lastly we examine the entanglement entropy between the two halves of the system, using $S_{R}=-\text{Tr}\varrho_R(t)\log\varrho_R(t)$ where $\varrho_R$ is the reduced density matrix of the right half of the system which has been studied previously in noninteracting systems~\cite{fraenkel2022extensive,capizzi2023domain,gouraud2022stationary, bertini2018entanglement}.
We do so here by using the quasiparticle picture~\citep{CalabreseCardy,AlbaCalabrese1,AlbaCalabrese2,
alba2018entanglement}
 which essentially counts the number of pairs of quasipartlces which are entangled and are shared between the left and right halves of the system.  Since pairs are shared within the lightcone this will be linear in time and by equating the entanglement entropy with the thermodynamic entropy we obtain
 \begin{equation}
 S_R=-t\int \!\frac{{\rm d}\lambda}{2\pi}e^\lambda\left[\mathcal{R}^\vartheta(\lambda)\log{\mathcal{R}^\vartheta(\lambda)}+\mathcal{T}^\vartheta(\lambda)\log{\mathcal{T}^\vartheta(\lambda)}\right].
 \end{equation}
This reduces to the known expression in noninteracting limit~\cite{fraenkel2022extensive} and vanishes when the impurity either purely transmits or reflects. 

\textit{Discussion \& Conclusions.---} In this Letter we have expanded the framework of GHD to include interacting quantum impurity models through the addition of an impurity collision integral which can be determined exactly from integrability.  After performing some non trivial analytic checks on this expression, we presented an exact solution of the impurity GHD equations for a bipartite quench with an interacting impurity.  Using this we then derived several results on the current, entanglement entropy and full counting statistics of the model.  In addition we also showed how the approach can be used to determine microscopic properties of the impurity like the charge deficit. 

%{{\color{blue}Our impurity GHD equations were determined along the lines of the original GHD papers~\cite{CastroDoyonYoshimura,bertini2016transport} using a semiclassical picture of quasiparticle scattering, however, a more rigorous approach may be possible  by adapting exact results for conserved currents in integrable models~\cite{borsi2020current} and incorporating modifications of the conservation laws due to the impurity~\cite{bertini2016determination, fagotti2017charges}.}}

While ostensibly a theory of integrable dynamics, GHD facilitates the inclusion of certain mild integrability breaking terms such as external potentials ~\cite{doyon2017note}, inhomogenous interactions~\cite{bastianello2019generalized}, atom losses~\cite{bouchoule2020effect} or extended nonintegrable defects~\cite{delvecchio2022transport} through the use of collision integrals~\cite{bastianello2021hydrodynamics}.   These naturally limit the applicability of the theory to be shorter than the quasiparticle lifetime, which however may still be quite large.  A similar approach can be adopted here through the inclusion of near-integrable impurities  with approximate reflection coefficients determined via Fermi's golden rule and compared to numerics~\cite{groha2017spinon, collura2013entanglement}.  Alternatively we can consider including multiple widely separated impurities and determine transport through a system with finite but small impurity concentration.  

We are grateful to Fabian Essler for enlightening discussions  and Bruno Bertini for valuable comments on the manuscript. This work has been supported by the ERC under Consolidator grant number 771536 NEMO.

\bibliographystyle{apsrev4-2}
\bibliography{ImpGHD.bib}  

%apsrev4-2.bst 2019-01-14 (MD) hand-edited version of apsrev4-1.bst
%Control: key (0)
%Control: author (72) initials jnrlst
%Control: editor formatted (1) identically to author
%Control: production of article title (-1) disabled
%Control: page (0) single
%Control: year (1) truncated
%Control: production of eprint (0) enabled
\begin{thebibliography}{123}%
\makeatletter
\providecommand \@ifxundefined [1]{%
 \@ifx{#1\undefined}
}%
\providecommand \@ifnum [1]{%
 \ifnum #1\expandafter \@firstoftwo
 \else \expandafter \@secondoftwo
 \fi
}%
\providecommand \@ifx [1]{%
 \ifx #1\expandafter \@firstoftwo
 \else \expandafter \@secondoftwo
 \fi
}%
\providecommand \natexlab [1]{#1}%
\providecommand \enquote  [1]{``#1''}%
\providecommand \bibnamefont  [1]{#1}%
\providecommand \bibfnamefont [1]{#1}%
\providecommand \citenamefont [1]{#1}%
\providecommand \href@noop [0]{\@secondoftwo}%
\providecommand \href [0]{\begingroup \@sanitize@url \@href}%
\providecommand \@href[1]{\@@startlink{#1}\@@href}%
\providecommand \@@href[1]{\endgroup#1\@@endlink}%
\providecommand \@sanitize@url [0]{\catcode `\\12\catcode `\$12\catcode
  `\&12\catcode `\#12\catcode `\^12\catcode `\_12\catcode `\%12\relax}%
\providecommand \@@startlink[1]{}%
\providecommand \@@endlink[0]{}%
\providecommand \url  [0]{\begingroup\@sanitize@url \@url }%
\providecommand \@url [1]{\endgroup\@href {#1}{\urlprefix }}%
\providecommand \urlprefix  [0]{URL }%
\providecommand \Eprint [0]{\href }%
\providecommand \doibase [0]{https://doi.org/}%
\providecommand \selectlanguage [0]{\@gobble}%
\providecommand \bibinfo  [0]{\@secondoftwo}%
\providecommand \bibfield  [0]{\@secondoftwo}%
\providecommand \translation [1]{[#1]}%
\providecommand \BibitemOpen [0]{}%
\providecommand \bibitemStop [0]{}%
\providecommand \bibitemNoStop [0]{.\EOS\space}%
\providecommand \EOS [0]{\spacefactor3000\relax}%
\providecommand \BibitemShut  [1]{\csname bibitem#1\endcsname}%
\let\auto@bib@innerbib\@empty
%</preamble>
\bibitem [{\citenamefont {Hewson}(1993)}]{hewson1993kondo}%
  \BibitemOpen
  \bibfield  {author} {\bibinfo {author} {\bibfnamefont {A.~C.}\ \bibnamefont
  {Hewson}},\ }\href {https://doi.org/10.1017/CBO9780511470752} {\emph
  {\bibinfo {title} {The Kondo Problem to Heavy Fermions}}},\ Cambridge Studies
  in Magnetism\ (\bibinfo  {publisher} {Cambridge University Press},\ \bibinfo
  {year} {1993})\BibitemShut {NoStop}%
\bibitem [{\citenamefont {Mahan}(2000)}]{mahan2000many}%
  \BibitemOpen
  \bibfield  {author} {\bibinfo {author} {\bibfnamefont {G.~D.}\ \bibnamefont
  {Mahan}},\ }\href {https://doi.org/https://doi.org/10.1007/978-1-4757-5714-9}
  {\emph {\bibinfo {title} {Many Particle Physics, Third Edition}}}\ (\bibinfo
  {publisher} {Plenum},\ \bibinfo {year} {2000})\BibitemShut {NoStop}%
\bibitem [{\citenamefont {{Goldhaber-Gordon}}\ \emph
  {et~al.}(1998)\citenamefont {{Goldhaber-Gordon}}, \citenamefont
  {{Shtrikman}}, \citenamefont {{Mahalu}}, \citenamefont {{Abusch-Magder}},
  \citenamefont {{Meirav}},\ and\ \citenamefont
  {{Kastner}}}]{goldhaber1998kondo}%
  \BibitemOpen
  \bibfield  {author} {\bibinfo {author} {\bibfnamefont {D.}~\bibnamefont
  {{Goldhaber-Gordon}}}, \bibinfo {author} {\bibfnamefont {H.}~\bibnamefont
  {{Shtrikman}}}, \bibinfo {author} {\bibfnamefont {D.}~\bibnamefont
  {{Mahalu}}}, \bibinfo {author} {\bibfnamefont {D.}~\bibnamefont
  {{Abusch-Magder}}}, \bibinfo {author} {\bibfnamefont {U.}~\bibnamefont
  {{Meirav}}},\ and\ \bibinfo {author} {\bibfnamefont {M.~A.}\ \bibnamefont
  {{Kastner}}},\ }\href {https://doi.org/10.1038/34373} {\bibfield  {journal}
  {\bibinfo  {journal} {\nat}\ }\textbf {\bibinfo {volume} {391}},\ \bibinfo
  {pages} {156} (\bibinfo {year} {1998})}\BibitemShut {NoStop}%
\bibitem [{\citenamefont {Makarovski}\ \emph
  {et~al.}(2007{\natexlab{a}})\citenamefont {Makarovski}, \citenamefont {Liu},\
  and\ \citenamefont {Finkelstein}}]{makarovski2007evolution}%
  \BibitemOpen
  \bibfield  {author} {\bibinfo {author} {\bibfnamefont {A.}~\bibnamefont
  {Makarovski}}, \bibinfo {author} {\bibfnamefont {J.}~\bibnamefont {Liu}},\
  and\ \bibinfo {author} {\bibfnamefont {G.}~\bibnamefont {Finkelstein}},\
  }\href {https://doi.org/10.1103/PhysRevLett.99.066801} {\bibfield  {journal}
  {\bibinfo  {journal} {Phys. Rev. Lett.}\ }\textbf {\bibinfo {volume} {99}},\
  \bibinfo {pages} {066801} (\bibinfo {year} {2007}{\natexlab{a}})}\BibitemShut
  {NoStop}%
\bibitem [{\citenamefont {Makarovski}\ \emph
  {et~al.}(2007{\natexlab{b}})\citenamefont {Makarovski}, \citenamefont
  {Zhukov}, \citenamefont {Liu},\ and\ \citenamefont
  {Finkelstein}}]{makarovski2007su4}%
  \BibitemOpen
  \bibfield  {author} {\bibinfo {author} {\bibfnamefont {A.}~\bibnamefont
  {Makarovski}}, \bibinfo {author} {\bibfnamefont {A.}~\bibnamefont {Zhukov}},
  \bibinfo {author} {\bibfnamefont {J.}~\bibnamefont {Liu}},\ and\ \bibinfo
  {author} {\bibfnamefont {G.}~\bibnamefont {Finkelstein}},\ }\href
  {https://doi.org/10.1103/PhysRevB.75.241407} {\bibfield  {journal} {\bibinfo
  {journal} {Phys. Rev. B}\ }\textbf {\bibinfo {volume} {75}},\ \bibinfo
  {pages} {241407} (\bibinfo {year} {2007}{\natexlab{b}})}\BibitemShut
  {NoStop}%
\bibitem [{\citenamefont {Mebrahtu}\ \emph {et~al.}(2012)\citenamefont
  {Mebrahtu}, \citenamefont {Borzenets}, \citenamefont {Liu}, \citenamefont
  {Zheng}, \citenamefont {Bomze}, \citenamefont {Smirnov}, \citenamefont
  {Baranger},\ and\ \citenamefont {Finkelstein}}]{mebrahtu2012quantum}%
  \BibitemOpen
  \bibfield  {author} {\bibinfo {author} {\bibfnamefont {H.~T.}\ \bibnamefont
  {Mebrahtu}}, \bibinfo {author} {\bibfnamefont {I.~V.}\ \bibnamefont
  {Borzenets}}, \bibinfo {author} {\bibfnamefont {D.~E.}\ \bibnamefont {Liu}},
  \bibinfo {author} {\bibfnamefont {H.}~\bibnamefont {Zheng}}, \bibinfo
  {author} {\bibfnamefont {Y.~V.}\ \bibnamefont {Bomze}}, \bibinfo {author}
  {\bibfnamefont {A.~I.}\ \bibnamefont {Smirnov}}, \bibinfo {author}
  {\bibfnamefont {H.~U.}\ \bibnamefont {Baranger}},\ and\ \bibinfo {author}
  {\bibfnamefont {G.}~\bibnamefont {Finkelstein}},\ }\href
  {https://doi.org/10.1038/nature11265} {\bibfield  {journal} {\bibinfo
  {journal} {Nature}\ }\textbf {\bibinfo {volume} {488}},\ \bibinfo {pages}
  {61} (\bibinfo {year} {2012})}\BibitemShut {NoStop}%
\bibitem [{\citenamefont {Iftikhar}\ \emph {et~al.}(2015)\citenamefont
  {Iftikhar}, \citenamefont {Jezouin}, \citenamefont {Anthore}, \citenamefont
  {Gennser}, \citenamefont {Parmentier}, \citenamefont {Cavanna},\ and\
  \citenamefont {Pierre}}]{iftikhar2015two}%
  \BibitemOpen
  \bibfield  {author} {\bibinfo {author} {\bibfnamefont {Z.}~\bibnamefont
  {Iftikhar}}, \bibinfo {author} {\bibfnamefont {S.}~\bibnamefont {Jezouin}},
  \bibinfo {author} {\bibfnamefont {A.}~\bibnamefont {Anthore}}, \bibinfo
  {author} {\bibfnamefont {U.}~\bibnamefont {Gennser}}, \bibinfo {author}
  {\bibfnamefont {F.~D.}\ \bibnamefont {Parmentier}}, \bibinfo {author}
  {\bibfnamefont {A.}~\bibnamefont {Cavanna}},\ and\ \bibinfo {author}
  {\bibfnamefont {F.}~\bibnamefont {Pierre}},\ }\href
  {https://doi.org/10.1038/nature15384} {\bibfield  {journal} {\bibinfo
  {journal} {Nature}\ }\textbf {\bibinfo {volume} {526}},\ \bibinfo {pages}
  {233} (\bibinfo {year} {2015})}\BibitemShut {NoStop}%
\bibitem [{\citenamefont {Pouse}\ \emph {et~al.}(2023)\citenamefont {Pouse},
  \citenamefont {Peeters}, \citenamefont {Hsueh}, \citenamefont {Gennser},
  \citenamefont {Cavanna}, \citenamefont {Kastner}, \citenamefont {Mitchell},\
  and\ \citenamefont {Goldhaber-Gordon}}]{pouse2023exotic}%
  \BibitemOpen
  \bibfield  {author} {\bibinfo {author} {\bibfnamefont {W.}~\bibnamefont
  {Pouse}}, \bibinfo {author} {\bibfnamefont {L.}~\bibnamefont {Peeters}},
  \bibinfo {author} {\bibfnamefont {C.~L.}\ \bibnamefont {Hsueh}}, \bibinfo
  {author} {\bibfnamefont {U.}~\bibnamefont {Gennser}}, \bibinfo {author}
  {\bibfnamefont {A.}~\bibnamefont {Cavanna}}, \bibinfo {author} {\bibfnamefont
  {M.~A.}\ \bibnamefont {Kastner}}, \bibinfo {author} {\bibfnamefont {A.~K.}\
  \bibnamefont {Mitchell}},\ and\ \bibinfo {author} {\bibfnamefont
  {D.}~\bibnamefont {Goldhaber-Gordon}},\ }\bibfield  {journal} {\bibinfo
  {journal} {Nature Physics}\ }\href
  {https://doi.org/10.1038/s41567-022-01905-4} {10.1038/s41567-022-01905-4}
  (\bibinfo {year} {2023})\BibitemShut {NoStop}%
\bibitem [{\citenamefont {Andrei}\ \emph {et~al.}(1983)\citenamefont {Andrei},
  \citenamefont {Furuya},\ and\ \citenamefont
  {Lowenstein}}]{andrei1983solution}%
  \BibitemOpen
  \bibfield  {author} {\bibinfo {author} {\bibfnamefont {N.}~\bibnamefont
  {Andrei}}, \bibinfo {author} {\bibfnamefont {K.}~\bibnamefont {Furuya}},\
  and\ \bibinfo {author} {\bibfnamefont {J.~H.}\ \bibnamefont {Lowenstein}},\
  }\href {https://doi.org/10.1103/RevModPhys.55.331} {\bibfield  {journal}
  {\bibinfo  {journal} {Rev. Mod. Phys.}\ }\textbf {\bibinfo {volume} {55}},\
  \bibinfo {pages} {331} (\bibinfo {year} {1983})}\BibitemShut {NoStop}%
\bibitem [{\citenamefont {Tsvelick}\ and\ \citenamefont
  {Wiegmann}(1983)}]{tsvelik1983exact}%
  \BibitemOpen
  \bibfield  {author} {\bibinfo {author} {\bibfnamefont {A.}~\bibnamefont
  {Tsvelick}}\ and\ \bibinfo {author} {\bibfnamefont {P.}~\bibnamefont
  {Wiegmann}},\ }\href {https://doi.org/10.1080/00018738300101581} {\bibfield
  {journal} {\bibinfo  {journal} {Advances in Physics}\ }\textbf {\bibinfo
  {volume} {32}},\ \bibinfo {pages} {453} (\bibinfo {year} {1983})}\BibitemShut
  {NoStop}%
\bibitem [{\citenamefont {Andrei}(1980)}]{andrei1980diagonalization}%
  \BibitemOpen
  \bibfield  {author} {\bibinfo {author} {\bibfnamefont {N.}~\bibnamefont
  {Andrei}},\ }\href {https://doi.org/10.1103/PhysRevLett.45.379} {\bibfield
  {journal} {\bibinfo  {journal} {Phys. Rev. Lett.}\ }\textbf {\bibinfo
  {volume} {45}},\ \bibinfo {pages} {379} (\bibinfo {year} {1980})}\BibitemShut
  {NoStop}%
\bibitem [{\citenamefont {{Vigman}}(1980)}]{vigman1980exact}%
  \BibitemOpen
  \bibfield  {author} {\bibinfo {author} {\bibfnamefont {P.~B.}\ \bibnamefont
  {{Vigman}}},\ }\href@noop {} {\bibfield  {journal} {\bibinfo  {journal}
  {Soviet Journal of Experimental and Theoretical Physics Letters}\ }\textbf
  {\bibinfo {volume} {31}},\ \bibinfo {pages} {364} (\bibinfo {year}
  {1980})}\BibitemShut {NoStop}%
\bibitem [{\citenamefont {Wiegmann}(1981)}]{wiegmann1981exact}%
  \BibitemOpen
  \bibfield  {author} {\bibinfo {author} {\bibfnamefont {P.~B.}\ \bibnamefont
  {Wiegmann}},\ }\href {https://doi.org/10.1088/0022-3719/14/10/014} {\bibfield
   {journal} {\bibinfo  {journal} {Journal of Physics C: Solid State Physics}\
  }\textbf {\bibinfo {volume} {14}},\ \bibinfo {pages} {1463} (\bibinfo {year}
  {1981})}\BibitemShut {NoStop}%
\bibitem [{\citenamefont {Kawakami}\ and\ \citenamefont
  {Okiji}(1981)}]{kawakami1981exact}%
  \BibitemOpen
  \bibfield  {author} {\bibinfo {author} {\bibfnamefont {N.}~\bibnamefont
  {Kawakami}}\ and\ \bibinfo {author} {\bibfnamefont {A.}~\bibnamefont
  {Okiji}},\ }\href
  {https://doi.org/https://doi.org/10.1016/0375-9601(81)90663-0} {\bibfield
  {journal} {\bibinfo  {journal} {Physics Letters A}\ }\textbf {\bibinfo
  {volume} {86}},\ \bibinfo {pages} {483} (\bibinfo {year} {1981})}\BibitemShut
  {NoStop}%
\bibitem [{\citenamefont {Wiegmann}\ and\ \citenamefont
  {Tsvelick}(1983)}]{wiegmann1983exact}%
  \BibitemOpen
  \bibfield  {author} {\bibinfo {author} {\bibfnamefont {P.~B.}\ \bibnamefont
  {Wiegmann}}\ and\ \bibinfo {author} {\bibfnamefont {A.~M.}\ \bibnamefont
  {Tsvelick}},\ }\href {https://doi.org/10.1088/0022-3719/16/12/017} {\bibfield
   {journal} {\bibinfo  {journal} {Journal of Physics C: Solid State Physics}\
  }\textbf {\bibinfo {volume} {16}},\ \bibinfo {pages} {2281} (\bibinfo {year}
  {1983})}\BibitemShut {NoStop}%
\bibitem [{\citenamefont {Fendley}\ \emph {et~al.}(1994)\citenamefont
  {Fendley}, \citenamefont {Saleur},\ and\ \citenamefont
  {Warner}}]{fendley1994exact}%
  \BibitemOpen
  \bibfield  {author} {\bibinfo {author} {\bibfnamefont {P.}~\bibnamefont
  {Fendley}}, \bibinfo {author} {\bibfnamefont {H.}~\bibnamefont {Saleur}},\
  and\ \bibinfo {author} {\bibfnamefont {N.}~\bibnamefont {Warner}},\ }\href
  {https://doi.org/10.1016/0550-3213(94)90160-0} {\bibfield  {journal}
  {\bibinfo  {journal} {Nuclear Physics B}\ }\textbf {\bibinfo {volume}
  {430}},\ \bibinfo {pages} {577} (\bibinfo {year} {1994})}\BibitemShut
  {NoStop}%
\bibitem [{\citenamefont {Rylands}\ and\ \citenamefont
  {Andrei}(2016)}]{rylands2016quantum}%
  \BibitemOpen
  \bibfield  {author} {\bibinfo {author} {\bibfnamefont {C.}~\bibnamefont
  {Rylands}}\ and\ \bibinfo {author} {\bibfnamefont {N.}~\bibnamefont
  {Andrei}},\ }\href {https://doi.org/10.1103/PhysRevB.94.115142} {\bibfield
  {journal} {\bibinfo  {journal} {Phys. Rev. B}\ }\textbf {\bibinfo {volume}
  {94}},\ \bibinfo {pages} {115142} (\bibinfo {year} {2016})}\BibitemShut
  {NoStop}%
\bibitem [{\citenamefont {{Tsvelick}}\ and\ \citenamefont
  {{Wiegmann}}(1985)}]{tsvelik1985exact}%
  \BibitemOpen
  \bibfield  {author} {\bibinfo {author} {\bibfnamefont {A.~M.}\ \bibnamefont
  {{Tsvelick}}}\ and\ \bibinfo {author} {\bibfnamefont {P.~B.}\ \bibnamefont
  {{Wiegmann}}},\ }\href {https://doi.org/10.1007/BF01017853} {\bibfield
  {journal} {\bibinfo  {journal} {Journal of Statistical Physics}\ }\textbf
  {\bibinfo {volume} {38}},\ \bibinfo {pages} {125} (\bibinfo {year}
  {1985})}\BibitemShut {NoStop}%
\bibitem [{\citenamefont {Andrei}\ and\ \citenamefont
  {Destri}(1984)}]{andrei1984solution}%
  \BibitemOpen
  \bibfield  {author} {\bibinfo {author} {\bibfnamefont {N.}~\bibnamefont
  {Andrei}}\ and\ \bibinfo {author} {\bibfnamefont {C.}~\bibnamefont
  {Destri}},\ }\href {https://doi.org/10.1103/PhysRevLett.52.364} {\bibfield
  {journal} {\bibinfo  {journal} {Phys. Rev. Lett.}\ }\textbf {\bibinfo
  {volume} {52}},\ \bibinfo {pages} {364} (\bibinfo {year} {1984})}\BibitemShut
  {NoStop}%
\bibitem [{\citenamefont {Bed\"urftig}\ \emph {et~al.}(1996)\citenamefont
  {Bed\"urftig}, \citenamefont {E\ss{}ler},\ and\ \citenamefont
  {Frahm}}]{bedurftig1996integrable}%
  \BibitemOpen
  \bibfield  {author} {\bibinfo {author} {\bibfnamefont {G.}~\bibnamefont
  {Bed\"urftig}}, \bibinfo {author} {\bibfnamefont {F.~H.~L.}\ \bibnamefont
  {E\ss{}ler}},\ and\ \bibinfo {author} {\bibfnamefont {H.}~\bibnamefont
  {Frahm}},\ }\href {https://doi.org/10.1103/PhysRevLett.77.5098} {\bibfield
  {journal} {\bibinfo  {journal} {Phys. Rev. Lett.}\ }\textbf {\bibinfo
  {volume} {77}},\ \bibinfo {pages} {5098} (\bibinfo {year}
  {1996})}\BibitemShut {NoStop}%
\bibitem [{\citenamefont {{Bed{\"u}rftig}}\ \emph {et~al.}(1997)\citenamefont
  {{Bed{\"u}rftig}}, \citenamefont {{E{\ss}ler}},\ and\ \citenamefont
  {{Frahm}}}]{bedfurtig1997exact}%
  \BibitemOpen
  \bibfield  {author} {\bibinfo {author} {\bibfnamefont {G.}~\bibnamefont
  {{Bed{\"u}rftig}}}, \bibinfo {author} {\bibfnamefont {F.~H.~L.}\ \bibnamefont
  {{E{\ss}ler}}},\ and\ \bibinfo {author} {\bibfnamefont {H.}~\bibnamefont
  {{Frahm}}},\ }\href {https://doi.org/10.1016/S0550-3213(97)00059-X}
  {\bibfield  {journal} {\bibinfo  {journal} {Nuclear Physics B}\ }\textbf
  {\bibinfo {volume} {489}},\ \bibinfo {pages} {697} (\bibinfo {year}
  {1997})}\BibitemShut {NoStop}%
\bibitem [{\citenamefont {{Andrei}}\ and\ \citenamefont
  {{Johannesson}}(1984)}]{andrei1984heisenberg}%
  \BibitemOpen
  \bibfield  {author} {\bibinfo {author} {\bibfnamefont {N.}~\bibnamefont
  {{Andrei}}}\ and\ \bibinfo {author} {\bibfnamefont {H.}~\bibnamefont
  {{Johannesson}}},\ }\href {https://doi.org/10.1016/0375-9601(84)90675-3}
  {\bibfield  {journal} {\bibinfo  {journal} {Physics Letters A}\ }\textbf
  {\bibinfo {volume} {100}},\ \bibinfo {pages} {108} (\bibinfo {year}
  {1984})}\BibitemShut {NoStop}%
\bibitem [{\citenamefont {Lee}\ and\ \citenamefont
  {Schlottmann}(1988)}]{lee1988integrable}%
  \BibitemOpen
  \bibfield  {author} {\bibinfo {author} {\bibfnamefont {K.-J.-B.}\
  \bibnamefont {Lee}}\ and\ \bibinfo {author} {\bibfnamefont {P.}~\bibnamefont
  {Schlottmann}},\ }\href {https://doi.org/10.1103/PhysRevB.37.379} {\bibfield
  {journal} {\bibinfo  {journal} {Phys. Rev. B}\ }\textbf {\bibinfo {volume}
  {37}},\ \bibinfo {pages} {379} (\bibinfo {year} {1988})}\BibitemShut
  {NoStop}%
\bibitem [{\citenamefont {Jerez}\ \emph {et~al.}(1998)\citenamefont {Jerez},
  \citenamefont {Andrei},\ and\ \citenamefont {Zar\'and}}]{jerez1998solution}%
  \BibitemOpen
  \bibfield  {author} {\bibinfo {author} {\bibfnamefont {A.}~\bibnamefont
  {Jerez}}, \bibinfo {author} {\bibfnamefont {N.}~\bibnamefont {Andrei}},\ and\
  \bibinfo {author} {\bibfnamefont {G.}~\bibnamefont {Zar\'and}},\ }\href
  {https://doi.org/10.1103/PhysRevB.58.3814} {\bibfield  {journal} {\bibinfo
  {journal} {Phys. Rev. B}\ }\textbf {\bibinfo {volume} {58}},\ \bibinfo
  {pages} {3814} (\bibinfo {year} {1998})}\BibitemShut {NoStop}%
\bibitem [{\citenamefont {Andrei}\ and\ \citenamefont
  {Jerez}(1995)}]{jerez1995fermi}%
  \BibitemOpen
  \bibfield  {author} {\bibinfo {author} {\bibfnamefont {N.}~\bibnamefont
  {Andrei}}\ and\ \bibinfo {author} {\bibfnamefont {A.}~\bibnamefont {Jerez}},\
  }\href {https://doi.org/10.1103/PhysRevLett.74.4507} {\bibfield  {journal}
  {\bibinfo  {journal} {Phys. Rev. Lett.}\ }\textbf {\bibinfo {volume} {74}},\
  \bibinfo {pages} {4507} (\bibinfo {year} {1995})}\BibitemShut {NoStop}%
\bibitem [{\citenamefont {Rylands}\ and\ \citenamefont
  {Andrei}(2017)}]{rylands2017quantum}%
  \BibitemOpen
  \bibfield  {author} {\bibinfo {author} {\bibfnamefont {C.}~\bibnamefont
  {Rylands}}\ and\ \bibinfo {author} {\bibfnamefont {N.}~\bibnamefont
  {Andrei}},\ }\href {https://doi.org/10.1103/PhysRevB.96.115424} {\bibfield
  {journal} {\bibinfo  {journal} {Phys. Rev. B}\ }\textbf {\bibinfo {volume}
  {96}},\ \bibinfo {pages} {115424} (\bibinfo {year} {2017})}\BibitemShut
  {NoStop}%
\bibitem [{\citenamefont {Rylands}\ and\ \citenamefont
  {Andrei}(2018)}]{rylands2018quantum}%
  \BibitemOpen
  \bibfield  {author} {\bibinfo {author} {\bibfnamefont {C.}~\bibnamefont
  {Rylands}}\ and\ \bibinfo {author} {\bibfnamefont {N.}~\bibnamefont
  {Andrei}},\ }\href {https://doi.org/10.1103/PhysRevB.97.155426} {\bibfield
  {journal} {\bibinfo  {journal} {Phys. Rev. B}\ }\textbf {\bibinfo {volume}
  {97}},\ \bibinfo {pages} {155426} (\bibinfo {year} {2018})}\BibitemShut
  {NoStop}%
\bibitem [{\citenamefont {Pasnoori}\ \emph {et~al.}(2020)\citenamefont
  {Pasnoori}, \citenamefont {Rylands},\ and\ \citenamefont
  {Andrei}}]{pasnoori2020kondo}%
  \BibitemOpen
  \bibfield  {author} {\bibinfo {author} {\bibfnamefont {P.~R.}\ \bibnamefont
  {Pasnoori}}, \bibinfo {author} {\bibfnamefont {C.}~\bibnamefont {Rylands}},\
  and\ \bibinfo {author} {\bibfnamefont {N.}~\bibnamefont {Andrei}},\ }\href
  {https://doi.org/10.1103/PhysRevResearch.2.013006} {\bibfield  {journal}
  {\bibinfo  {journal} {Phys. Rev. Res.}\ }\textbf {\bibinfo {volume} {2}},\
  \bibinfo {pages} {013006} (\bibinfo {year} {2020})}\BibitemShut {NoStop}%
\bibitem [{\citenamefont {Pasnoori}\ \emph {et~al.}(2022)\citenamefont
  {Pasnoori}, \citenamefont {Andrei}, \citenamefont {Rylands},\ and\
  \citenamefont {Azaria}}]{pasnoori2022rise}%
  \BibitemOpen
  \bibfield  {author} {\bibinfo {author} {\bibfnamefont {P.~R.}\ \bibnamefont
  {Pasnoori}}, \bibinfo {author} {\bibfnamefont {N.}~\bibnamefont {Andrei}},
  \bibinfo {author} {\bibfnamefont {C.}~\bibnamefont {Rylands}},\ and\ \bibinfo
  {author} {\bibfnamefont {P.}~\bibnamefont {Azaria}},\ }\href
  {https://doi.org/10.1103/PhysRevB.105.174517} {\bibfield  {journal} {\bibinfo
   {journal} {Phys. Rev. B}\ }\textbf {\bibinfo {volume} {105}},\ \bibinfo
  {pages} {174517} (\bibinfo {year} {2022})}\BibitemShut {NoStop}%
\bibitem [{\citenamefont {Bastianello}\ \emph {et~al.}(2022)\citenamefont
  {Bastianello}, \citenamefont {Bertini}, \citenamefont {Doyon},\ and\
  \citenamefont {Vasseur}}]{bastianello2022introduction}%
  \BibitemOpen
  \bibfield  {author} {\bibinfo {author} {\bibfnamefont {A.}~\bibnamefont
  {Bastianello}}, \bibinfo {author} {\bibfnamefont {B.}~\bibnamefont
  {Bertini}}, \bibinfo {author} {\bibfnamefont {B.}~\bibnamefont {Doyon}},\
  and\ \bibinfo {author} {\bibfnamefont {R.}~\bibnamefont {Vasseur}},\ }\href
  {https://doi.org/10.1088/1742-5468/ac3e6a} {\bibfield  {journal} {\bibinfo
  {journal} {J. Stat. Mech. Theory Exp.}\ }\textbf {\bibinfo {volume} {2022}},\
  \bibinfo {pages} {014001} (\bibinfo {year} {2022})}\BibitemShut {NoStop}%
\bibitem [{\citenamefont {Rylands}\ and\ \citenamefont
  {Andrei}(2020)}]{rylands2020nonequilibrium}%
  \BibitemOpen
  \bibfield  {author} {\bibinfo {author} {\bibfnamefont {C.}~\bibnamefont
  {Rylands}}\ and\ \bibinfo {author} {\bibfnamefont {N.}~\bibnamefont
  {Andrei}},\ }\href {https://doi.org/10.1146/annurev-conmatphys-031119-050630}
  {\bibfield  {journal} {\bibinfo  {journal} {Annual Review of Condensed Matter
  Physics}\ }\textbf {\bibinfo {volume} {11}},\ \bibinfo {pages} {147}
  (\bibinfo {year} {2020})}\BibitemShut {NoStop}%
\bibitem [{\citenamefont {Caux}\ and\ \citenamefont
  {Essler}(2013)}]{CauxEssler}%
  \BibitemOpen
  \bibfield  {author} {\bibinfo {author} {\bibfnamefont {J.-S.}\ \bibnamefont
  {Caux}}\ and\ \bibinfo {author} {\bibfnamefont {F.~H.~L.}\ \bibnamefont
  {Essler}},\ }\href {https://doi.org/10.1103/PhysRevLett.110.257203}
  {\bibfield  {journal} {\bibinfo  {journal} {Phys. Rev. Lett.}\ }\textbf
  {\bibinfo {volume} {110}},\ \bibinfo {pages} {257203} (\bibinfo {year}
  {2013})}\BibitemShut {NoStop}%
\bibitem [{\citenamefont {Caux}(2016)}]{Caux}%
  \BibitemOpen
  \bibfield  {author} {\bibinfo {author} {\bibfnamefont {J.-S.}\ \bibnamefont
  {Caux}},\ }\href {https://doi.org/10.1088/1742-5468/2016/06/064006}
  {\bibfield  {journal} {\bibinfo  {journal} {J. Stat. Mech. Theory Exp.}\
  }\textbf {\bibinfo {volume} {2016}},\ \bibinfo {pages} {064006} (\bibinfo
  {year} {2016})}\BibitemShut {NoStop}%
\bibitem [{\citenamefont {Castro-Alvaredo}\ \emph {et~al.}(2016)\citenamefont
  {Castro-Alvaredo}, \citenamefont {Doyon},\ and\ \citenamefont
  {Yoshimura}}]{CastroDoyonYoshimura}%
  \BibitemOpen
  \bibfield  {author} {\bibinfo {author} {\bibfnamefont {O.~A.}\ \bibnamefont
  {Castro-Alvaredo}}, \bibinfo {author} {\bibfnamefont {B.}~\bibnamefont
  {Doyon}},\ and\ \bibinfo {author} {\bibfnamefont {T.}~\bibnamefont
  {Yoshimura}},\ }\href {https://doi.org/10.1103/PhysRevX.6.041065} {\bibfield
  {journal} {\bibinfo  {journal} {Phys. Rev. X}\ }\textbf {\bibinfo {volume}
  {6}},\ \bibinfo {pages} {041065} (\bibinfo {year} {2016})}\BibitemShut
  {NoStop}%
\bibitem [{\citenamefont {Bertini}\ \emph
  {et~al.}(2016{\natexlab{a}})\citenamefont {Bertini}, \citenamefont {Collura},
  \citenamefont {De~Nardis},\ and\ \citenamefont
  {Fagotti}}]{bertini2016transport}%
  \BibitemOpen
  \bibfield  {author} {\bibinfo {author} {\bibfnamefont {B.}~\bibnamefont
  {Bertini}}, \bibinfo {author} {\bibfnamefont {M.}~\bibnamefont {Collura}},
  \bibinfo {author} {\bibfnamefont {J.}~\bibnamefont {De~Nardis}},\ and\
  \bibinfo {author} {\bibfnamefont {M.}~\bibnamefont {Fagotti}},\ }\href
  {https://doi.org/10.1103/PhysRevLett.117.207201} {\bibfield  {journal}
  {\bibinfo  {journal} {Phys. Rev. Lett.}\ }\textbf {\bibinfo {volume} {117}},\
  \bibinfo {pages} {207201} (\bibinfo {year} {2016}{\natexlab{a}})}\BibitemShut
  {NoStop}%
\bibitem [{\citenamefont {{Piroli}}\ \emph {et~al.}(2017)\citenamefont
  {{Piroli}}, \citenamefont {{Pozsgay}},\ and\ \citenamefont
  {{Vernier}}}]{PiroliPozsgayVernier}%
  \BibitemOpen
  \bibfield  {author} {\bibinfo {author} {\bibfnamefont {L.}~\bibnamefont
  {{Piroli}}}, \bibinfo {author} {\bibfnamefont {B.}~\bibnamefont
  {{Pozsgay}}},\ and\ \bibinfo {author} {\bibfnamefont {E.}~\bibnamefont
  {{Vernier}}},\ }\href {https://doi.org/10.1016/j.nuclphysb.2017.10.012}
  {\bibfield  {journal} {\bibinfo  {journal} {Nucl. Phys. B.}\ }\textbf
  {\bibinfo {volume} {925}},\ \bibinfo {pages} {362} (\bibinfo {year}
  {2017})}\BibitemShut {NoStop}%
\bibitem [{\citenamefont {De~Nardis}\ \emph {et~al.}(2014)\citenamefont
  {De~Nardis}, \citenamefont {Wouters}, \citenamefont {Brockmann},\ and\
  \citenamefont {Caux}}]{DeNardis}%
  \BibitemOpen
  \bibfield  {author} {\bibinfo {author} {\bibfnamefont {J.}~\bibnamefont
  {De~Nardis}}, \bibinfo {author} {\bibfnamefont {B.}~\bibnamefont {Wouters}},
  \bibinfo {author} {\bibfnamefont {M.}~\bibnamefont {Brockmann}},\ and\
  \bibinfo {author} {\bibfnamefont {J.-S.}\ \bibnamefont {Caux}},\ }\href
  {https://doi.org/10.1103/PhysRevA.89.033601} {\bibfield  {journal} {\bibinfo
  {journal} {Phys. Rev. A}\ }\textbf {\bibinfo {volume} {89}},\ \bibinfo
  {pages} {033601} (\bibinfo {year} {2014})}\BibitemShut {NoStop}%
\bibitem [{\citenamefont {Brockmann}\ \emph {et~al.}(2014)\citenamefont
  {Brockmann}, \citenamefont {Wouters}, \citenamefont {Fioretto}, \citenamefont
  {Nardis}, \citenamefont {Vlijm},\ and\ \citenamefont {Caux}}]{Brockmann}%
  \BibitemOpen
  \bibfield  {author} {\bibinfo {author} {\bibfnamefont {M.}~\bibnamefont
  {Brockmann}}, \bibinfo {author} {\bibfnamefont {B.}~\bibnamefont {Wouters}},
  \bibinfo {author} {\bibfnamefont {D.}~\bibnamefont {Fioretto}}, \bibinfo
  {author} {\bibfnamefont {J.~D.}\ \bibnamefont {Nardis}}, \bibinfo {author}
  {\bibfnamefont {R.}~\bibnamefont {Vlijm}},\ and\ \bibinfo {author}
  {\bibfnamefont {J.-S.}\ \bibnamefont {Caux}},\ }\href
  {https://doi.org/10.1088/1742-5468/2014/12/p12009} {\bibfield  {journal}
  {\bibinfo  {journal} {J. Stat. Mech. Theory Exp.}\ }\textbf {\bibinfo
  {volume} {2014}},\ \bibinfo {pages} {P12009} (\bibinfo {year}
  {2014})}\BibitemShut {NoStop}%
\bibitem [{\citenamefont {Wouters}\ \emph {et~al.}(2014)\citenamefont
  {Wouters}, \citenamefont {De~Nardis}, \citenamefont {Brockmann},
  \citenamefont {Fioretto}, \citenamefont {Rigol},\ and\ \citenamefont
  {Caux}}]{wouters2014quenching}%
  \BibitemOpen
  \bibfield  {author} {\bibinfo {author} {\bibfnamefont {B.}~\bibnamefont
  {Wouters}}, \bibinfo {author} {\bibfnamefont {J.}~\bibnamefont {De~Nardis}},
  \bibinfo {author} {\bibfnamefont {M.}~\bibnamefont {Brockmann}}, \bibinfo
  {author} {\bibfnamefont {D.}~\bibnamefont {Fioretto}}, \bibinfo {author}
  {\bibfnamefont {M.}~\bibnamefont {Rigol}},\ and\ \bibinfo {author}
  {\bibfnamefont {J.-S.}\ \bibnamefont {Caux}},\ }\href
  {https://doi.org/10.1103/PhysRevLett.113.117202} {\bibfield  {journal}
  {\bibinfo  {journal} {Phys. Rev. Lett.}\ }\textbf {\bibinfo {volume} {113}},\
  \bibinfo {pages} {117202} (\bibinfo {year} {2014})}\BibitemShut {NoStop}%
\bibitem [{\citenamefont {Pozsgay}\ \emph {et~al.}(2014)\citenamefont
  {Pozsgay}, \citenamefont {Mesty\'an}, \citenamefont {Werner}, \citenamefont
  {Kormos}, \citenamefont {Zar\'and},\ and\ \citenamefont
  {Tak\'acs}}]{pozsgay2014correlations}%
  \BibitemOpen
  \bibfield  {author} {\bibinfo {author} {\bibfnamefont {B.}~\bibnamefont
  {Pozsgay}}, \bibinfo {author} {\bibfnamefont {M.}~\bibnamefont {Mesty\'an}},
  \bibinfo {author} {\bibfnamefont {M.~A.}\ \bibnamefont {Werner}}, \bibinfo
  {author} {\bibfnamefont {M.}~\bibnamefont {Kormos}}, \bibinfo {author}
  {\bibfnamefont {G.}~\bibnamefont {Zar\'and}},\ and\ \bibinfo {author}
  {\bibfnamefont {G.}~\bibnamefont {Tak\'acs}},\ }\href
  {https://doi.org/10.1103/PhysRevLett.113.117203} {\bibfield  {journal}
  {\bibinfo  {journal} {Phys. Rev. Lett.}\ }\textbf {\bibinfo {volume} {113}},\
  \bibinfo {pages} {117203} (\bibinfo {year} {2014})}\BibitemShut {NoStop}%
\bibitem [{\citenamefont {Mesty{\'{a}}n}\ \emph {et~al.}(2015)\citenamefont
  {Mesty{\'{a}}n}, \citenamefont {Pozsgay}, \citenamefont {Tak{\'{a}}cs},\ and\
  \citenamefont {Werner}}]{MestyanPozsgayTakacsWerner}%
  \BibitemOpen
  \bibfield  {author} {\bibinfo {author} {\bibfnamefont {M.}~\bibnamefont
  {Mesty{\'{a}}n}}, \bibinfo {author} {\bibfnamefont {B.}~\bibnamefont
  {Pozsgay}}, \bibinfo {author} {\bibfnamefont {G.}~\bibnamefont
  {Tak{\'{a}}cs}},\ and\ \bibinfo {author} {\bibfnamefont {M.~A.}\ \bibnamefont
  {Werner}},\ }\href {https://doi.org/10.1088/1742-5468/2015/04/p04001}
  {\bibfield  {journal} {\bibinfo  {journal} {J. Stat. Mech. Theory Exp.}\
  }\textbf {\bibinfo {volume} {2015}},\ \bibinfo {pages} {P04001} (\bibinfo
  {year} {2015})}\BibitemShut {NoStop}%
\bibitem [{\citenamefont {Bertini}\ \emph
  {et~al.}(2016{\natexlab{b}})\citenamefont {Bertini}, \citenamefont {Piroli},\
  and\ \citenamefont {Calabrese}}]{bertini2016quantum}%
  \BibitemOpen
  \bibfield  {author} {\bibinfo {author} {\bibfnamefont {B.}~\bibnamefont
  {Bertini}}, \bibinfo {author} {\bibfnamefont {L.}~\bibnamefont {Piroli}},\
  and\ \bibinfo {author} {\bibfnamefont {P.}~\bibnamefont {Calabrese}},\ }\href
  {https://doi.org/10.1088/1742-5468/2016/06/063102} {\bibfield  {journal}
  {\bibinfo  {journal} {J. Stat. Mech. Theory Exp.}\ }\textbf {\bibinfo
  {volume} {2016}},\ \bibinfo {pages} {063102} (\bibinfo {year}
  {2016}{\natexlab{b}})}\BibitemShut {NoStop}%
\bibitem [{\citenamefont {Mesty\'an}\ \emph
  {et~al.}(2019{\natexlab{a}})\citenamefont {Mesty\'an}, \citenamefont
  {Bertini}, \citenamefont {Piroli},\ and\ \citenamefont
  {Calabrese}}]{MestyanBertiniPiroliCalabrese}%
  \BibitemOpen
  \bibfield  {author} {\bibinfo {author} {\bibfnamefont {M.}~\bibnamefont
  {Mesty\'an}}, \bibinfo {author} {\bibfnamefont {B.}~\bibnamefont {Bertini}},
  \bibinfo {author} {\bibfnamefont {L.}~\bibnamefont {Piroli}},\ and\ \bibinfo
  {author} {\bibfnamefont {P.}~\bibnamefont {Calabrese}},\ }\href
  {https://doi.org/10.1103/PhysRevB.99.014305} {\bibfield  {journal} {\bibinfo
  {journal} {Phys. Rev. B}\ }\textbf {\bibinfo {volume} {99}},\ \bibinfo
  {pages} {014305} (\bibinfo {year} {2019}{\natexlab{a}})}\BibitemShut
  {NoStop}%
\bibitem [{\citenamefont {Bertini}\ \emph {et~al.}(2014)\citenamefont
  {Bertini}, \citenamefont {Schuricht},\ and\ \citenamefont
  {Essler}}]{BertiniSchurichtEssler}%
  \BibitemOpen
  \bibfield  {author} {\bibinfo {author} {\bibfnamefont {B.}~\bibnamefont
  {Bertini}}, \bibinfo {author} {\bibfnamefont {D.}~\bibnamefont {Schuricht}},\
  and\ \bibinfo {author} {\bibfnamefont {F.~H.~L.}\ \bibnamefont {Essler}},\
  }\href {https://doi.org/10.1088/1742-5468/2014/10/p10035} {\bibfield
  {journal} {\bibinfo  {journal} {J. Stat. Mech. Theory Exp.}\ }\textbf
  {\bibinfo {volume} {2014}},\ \bibinfo {pages} {P10035} (\bibinfo {year}
  {2014})}\BibitemShut {NoStop}%
\bibitem [{\citenamefont {Bertini}\ \emph {et~al.}(2017)\citenamefont
  {Bertini}, \citenamefont {Tartaglia},\ and\ \citenamefont
  {Calabrese}}]{BertiniTartagliaCalabrese}%
  \BibitemOpen
  \bibfield  {author} {\bibinfo {author} {\bibfnamefont {B.}~\bibnamefont
  {Bertini}}, \bibinfo {author} {\bibfnamefont {E.}~\bibnamefont {Tartaglia}},\
  and\ \bibinfo {author} {\bibfnamefont {P.}~\bibnamefont {Calabrese}},\ }\href
  {https://doi.org/10.1088/1742-5468/aa8c2c} {\bibfield  {journal} {\bibinfo
  {journal} {J. Stat. Mech. Theory Exp.}\ }\textbf {\bibinfo {volume} {2017}},\
  \bibinfo {pages} {103107} (\bibinfo {year} {2017})}\BibitemShut {NoStop}%
\bibitem [{\citenamefont {{Piroli}}\ \emph {et~al.}(2019)\citenamefont
  {{Piroli}}, \citenamefont {{Vernier}}, \citenamefont {{Calabrese}},\ and\
  \citenamefont {{Pozsgay}}}]{PiroliVernierCalabresePozsgay1}%
  \BibitemOpen
  \bibfield  {author} {\bibinfo {author} {\bibfnamefont {L.}~\bibnamefont
  {{Piroli}}}, \bibinfo {author} {\bibfnamefont {E.}~\bibnamefont {{Vernier}}},
  \bibinfo {author} {\bibfnamefont {P.}~\bibnamefont {{Calabrese}}},\ and\
  \bibinfo {author} {\bibfnamefont {B.}~\bibnamefont {{Pozsgay}}},\ }\href
  {https://doi.org/10.1088/1742-5468/ab1c51} {\bibfield  {journal} {\bibinfo
  {journal} {J. Stat. Mech. Theory Exp.}\ }\textbf {\bibinfo {volume} {6}},\
  \bibinfo {pages} {063103} (\bibinfo {year} {2019})}\BibitemShut {NoStop}%
\bibitem [{\citenamefont {Piroli}\ \emph
  {et~al.}(2016{\natexlab{a}})\citenamefont {Piroli}, \citenamefont
  {Calabrese},\ and\ \citenamefont {Essler}}]{piroli2016multiparticle}%
  \BibitemOpen
  \bibfield  {author} {\bibinfo {author} {\bibfnamefont {L.}~\bibnamefont
  {Piroli}}, \bibinfo {author} {\bibfnamefont {P.}~\bibnamefont {Calabrese}},\
  and\ \bibinfo {author} {\bibfnamefont {F.~H.~L.}\ \bibnamefont {Essler}},\
  }\href {https://doi.org/10.1103/PhysRevLett.116.070408} {\bibfield  {journal}
  {\bibinfo  {journal} {Phys. Rev. Lett.}\ }\textbf {\bibinfo {volume} {116}},\
  \bibinfo {pages} {070408} (\bibinfo {year} {2016}{\natexlab{a}})}\BibitemShut
  {NoStop}%
\bibitem [{\citenamefont {Piroli}\ \emph
  {et~al.}(2016{\natexlab{b}})\citenamefont {Piroli}, \citenamefont
  {Calabrese},\ and\ \citenamefont {Essler}}]{piroli2016quantum}%
  \BibitemOpen
  \bibfield  {author} {\bibinfo {author} {\bibfnamefont {L.}~\bibnamefont
  {Piroli}}, \bibinfo {author} {\bibfnamefont {P.}~\bibnamefont {Calabrese}},\
  and\ \bibinfo {author} {\bibfnamefont {F.~H.~L.}\ \bibnamefont {Essler}},\
  }\href {https://doi.org/10.21468/SciPostPhys.1.1.001} {\bibfield  {journal}
  {\bibinfo  {journal} {SciPost Phys.}\ }\textbf {\bibinfo {volume} {1}},\
  \bibinfo {pages} {001} (\bibinfo {year} {2016}{\natexlab{b}})}\BibitemShut
  {NoStop}%
\bibitem [{\citenamefont {Alba}\ and\ \citenamefont
  {Calabrese}(2016)}]{alba2016the}%
  \BibitemOpen
  \bibfield  {author} {\bibinfo {author} {\bibfnamefont {V.}~\bibnamefont
  {Alba}}\ and\ \bibinfo {author} {\bibfnamefont {P.}~\bibnamefont
  {Calabrese}},\ }\href {https://doi.org/10.1088/1742-5468/2016/04/043105}
  {\bibfield  {journal} {\bibinfo  {journal} {J. Stat. Mech. Theory Exp.}\
  }\textbf {\bibinfo {volume} {2016}},\ \bibinfo {pages} {043105} (\bibinfo
  {year} {2016})}\BibitemShut {NoStop}%
\bibitem [{\citenamefont {Piroli}\ \emph
  {et~al.}(2016{\natexlab{c}})\citenamefont {Piroli}, \citenamefont {Vernier},\
  and\ \citenamefont {Calabrese}}]{piroli2016exact}%
  \BibitemOpen
  \bibfield  {author} {\bibinfo {author} {\bibfnamefont {L.}~\bibnamefont
  {Piroli}}, \bibinfo {author} {\bibfnamefont {E.}~\bibnamefont {Vernier}},\
  and\ \bibinfo {author} {\bibfnamefont {P.}~\bibnamefont {Calabrese}},\ }\href
  {https://doi.org/10.1103/PhysRevB.94.054313} {\bibfield  {journal} {\bibinfo
  {journal} {Phys. Rev. B}\ }\textbf {\bibinfo {volume} {94}},\ \bibinfo
  {pages} {054313} (\bibinfo {year} {2016}{\natexlab{c}})}\BibitemShut
  {NoStop}%
\bibitem [{\citenamefont {Nardis}\ \emph {et~al.}(2015)\citenamefont {Nardis},
  \citenamefont {Piroli},\ and\ \citenamefont {Caux}}]{denardis2015relaxation}%
  \BibitemOpen
  \bibfield  {author} {\bibinfo {author} {\bibfnamefont {J.~D.}\ \bibnamefont
  {Nardis}}, \bibinfo {author} {\bibfnamefont {L.}~\bibnamefont {Piroli}},\
  and\ \bibinfo {author} {\bibfnamefont {J.-S.}\ \bibnamefont {Caux}},\ }\href
  {https://doi.org/10.1088/1751-8113/48/43/43ft01} {\bibfield  {journal}
  {\bibinfo  {journal} {J. Phys. A Math. Theor.}\ }\textbf {\bibinfo {volume}
  {48}},\ \bibinfo {pages} {43FT01} (\bibinfo {year} {2015})}\BibitemShut
  {NoStop}%
\bibitem [{\citenamefont {Mesty{\'{a}}n}\ \emph {et~al.}(2017)\citenamefont
  {Mesty{\'{a}}n}, \citenamefont {Bertini}, \citenamefont {Piroli},\ and\
  \citenamefont {Calabrese}}]{mestyan2017exact}%
  \BibitemOpen
  \bibfield  {author} {\bibinfo {author} {\bibfnamefont {M.}~\bibnamefont
  {Mesty{\'{a}}n}}, \bibinfo {author} {\bibfnamefont {B.}~\bibnamefont
  {Bertini}}, \bibinfo {author} {\bibfnamefont {L.}~\bibnamefont {Piroli}},\
  and\ \bibinfo {author} {\bibfnamefont {P.}~\bibnamefont {Calabrese}},\ }\href
  {https://doi.org/10.1088/1742-5468/aa7df0} {\bibfield  {journal} {\bibinfo
  {journal} {J. Stat. Mech. Theory Exp.}\ }\textbf {\bibinfo {volume} {2017}},\
  \bibinfo {pages} {083103} (\bibinfo {year} {2017})}\BibitemShut {NoStop}%
\bibitem [{\citenamefont {Rylands}\ \emph {et~al.}(2022)\citenamefont
  {Rylands}, \citenamefont {Bertini},\ and\ \citenamefont
  {Calabrese}}]{rylands2022integrable}%
  \BibitemOpen
  \bibfield  {author} {\bibinfo {author} {\bibfnamefont {C.}~\bibnamefont
  {Rylands}}, \bibinfo {author} {\bibfnamefont {B.}~\bibnamefont {Bertini}},\
  and\ \bibinfo {author} {\bibfnamefont {P.}~\bibnamefont {Calabrese}},\ }\href
  {https://doi.org/10.1088/1742-5468/ac98be} {\bibfield  {journal} {\bibinfo
  {journal} {J. Stat. Mech. Theory Exp.}\ }\textbf {\bibinfo {volume} {2022}},\
  \bibinfo {pages} {103103} (\bibinfo {year} {2022})}\BibitemShut {NoStop}%
\bibitem [{\citenamefont {Rylands}\ \emph {et~al.}(2023)\citenamefont
  {Rylands}, \citenamefont {Calabrese},\ and\ \citenamefont
  {Bertini}}]{rylands2023solution}%
  \BibitemOpen
  \bibfield  {author} {\bibinfo {author} {\bibfnamefont {C.}~\bibnamefont
  {Rylands}}, \bibinfo {author} {\bibfnamefont {P.}~\bibnamefont {Calabrese}},\
  and\ \bibinfo {author} {\bibfnamefont {B.}~\bibnamefont {Bertini}},\ }\href
  {https://doi.org/10.1103/PhysRevLett.130.023001} {\bibfield  {journal}
  {\bibinfo  {journal} {Phys. Rev. Lett.}\ }\textbf {\bibinfo {volume} {130}},\
  \bibinfo {pages} {023001} (\bibinfo {year} {2023})}\BibitemShut {NoStop}%
\bibitem [{\citenamefont {Bulchandani}\ \emph {et~al.}(2017)\citenamefont
  {Bulchandani}, \citenamefont {Vasseur}, \citenamefont {Karrasch},\ and\
  \citenamefont {Moore}}]{bulchandani2017solvable}%
  \BibitemOpen
  \bibfield  {author} {\bibinfo {author} {\bibfnamefont {V.~B.}\ \bibnamefont
  {Bulchandani}}, \bibinfo {author} {\bibfnamefont {R.}~\bibnamefont
  {Vasseur}}, \bibinfo {author} {\bibfnamefont {C.}~\bibnamefont {Karrasch}},\
  and\ \bibinfo {author} {\bibfnamefont {J.~E.}\ \bibnamefont {Moore}},\ }\href
  {https://doi.org/10.1103/PhysRevLett.119.220604} {\bibfield  {journal}
  {\bibinfo  {journal} {Phys. Rev. Lett.}\ }\textbf {\bibinfo {volume} {119}},\
  \bibinfo {pages} {220604} (\bibinfo {year} {2017})}\BibitemShut {NoStop}%
\bibitem [{\citenamefont {Bulchandani}\ \emph {et~al.}(2018)\citenamefont
  {Bulchandani}, \citenamefont {Vasseur}, \citenamefont {Karrasch},\ and\
  \citenamefont {Moore}}]{bulchandani2018bethe}%
  \BibitemOpen
  \bibfield  {author} {\bibinfo {author} {\bibfnamefont {V.~B.}\ \bibnamefont
  {Bulchandani}}, \bibinfo {author} {\bibfnamefont {R.}~\bibnamefont
  {Vasseur}}, \bibinfo {author} {\bibfnamefont {C.}~\bibnamefont {Karrasch}},\
  and\ \bibinfo {author} {\bibfnamefont {J.~E.}\ \bibnamefont {Moore}},\ }\href
  {https://doi.org/10.1103/PhysRevB.97.045407} {\bibfield  {journal} {\bibinfo
  {journal} {Phys. Rev. B}\ }\textbf {\bibinfo {volume} {97}},\ \bibinfo
  {pages} {045407} (\bibinfo {year} {2018})}\BibitemShut {NoStop}%
\bibitem [{\citenamefont {Bastianello}\ \emph {et~al.}(2018)\citenamefont
  {Bastianello}, \citenamefont {Doyon}, \citenamefont {Watts},\ and\
  \citenamefont {Yoshimura}}]{bastianello2018generalized}%
  \BibitemOpen
  \bibfield  {author} {\bibinfo {author} {\bibfnamefont {A.}~\bibnamefont
  {Bastianello}}, \bibinfo {author} {\bibfnamefont {B.}~\bibnamefont {Doyon}},
  \bibinfo {author} {\bibfnamefont {G.}~\bibnamefont {Watts}},\ and\ \bibinfo
  {author} {\bibfnamefont {T.}~\bibnamefont {Yoshimura}},\ }\href
  {https://doi.org/10.21468/SciPostPhys.4.6.045} {\bibfield  {journal}
  {\bibinfo  {journal} {SciPost Phys.}\ }\textbf {\bibinfo {volume} {4}},\
  \bibinfo {pages} {045} (\bibinfo {year} {2018})}\BibitemShut {NoStop}%
\bibitem [{\citenamefont {Bastianello}\ \emph {et~al.}(2019)\citenamefont
  {Bastianello}, \citenamefont {Alba},\ and\ \citenamefont
  {Caux}}]{bastianello2019generalized}%
  \BibitemOpen
  \bibfield  {author} {\bibinfo {author} {\bibfnamefont {A.}~\bibnamefont
  {Bastianello}}, \bibinfo {author} {\bibfnamefont {V.}~\bibnamefont {Alba}},\
  and\ \bibinfo {author} {\bibfnamefont {J.-S.}\ \bibnamefont {Caux}},\ }\href
  {https://doi.org/10.1103/PhysRevLett.123.130602} {\bibfield  {journal}
  {\bibinfo  {journal} {Phys. Rev. Lett.}\ }\textbf {\bibinfo {volume} {123}},\
  \bibinfo {pages} {130602} (\bibinfo {year} {2019})}\BibitemShut {NoStop}%
\bibitem [{\citenamefont {Koch}\ \emph {et~al.}(2021)\citenamefont {Koch},
  \citenamefont {Bastianello},\ and\ \citenamefont {Caux}}]{koch2021adiabatic}%
  \BibitemOpen
  \bibfield  {author} {\bibinfo {author} {\bibfnamefont {R.}~\bibnamefont
  {Koch}}, \bibinfo {author} {\bibfnamefont {A.}~\bibnamefont {Bastianello}},\
  and\ \bibinfo {author} {\bibfnamefont {J.-S.}\ \bibnamefont {Caux}},\ }\href
  {https://doi.org/10.1103/PhysRevB.103.165121} {\bibfield  {journal} {\bibinfo
   {journal} {Phys. Rev. B}\ }\textbf {\bibinfo {volume} {103}},\ \bibinfo
  {pages} {165121} (\bibinfo {year} {2021})}\BibitemShut {NoStop}%
\bibitem [{\citenamefont {Mesty\'an}\ \emph
  {et~al.}(2019{\natexlab{b}})\citenamefont {Mesty\'an}, \citenamefont
  {Bertini}, \citenamefont {Piroli},\ and\ \citenamefont
  {Calabrese}}]{mestyan2019spin}%
  \BibitemOpen
  \bibfield  {author} {\bibinfo {author} {\bibfnamefont {M.}~\bibnamefont
  {Mesty\'an}}, \bibinfo {author} {\bibfnamefont {B.}~\bibnamefont {Bertini}},
  \bibinfo {author} {\bibfnamefont {L.}~\bibnamefont {Piroli}},\ and\ \bibinfo
  {author} {\bibfnamefont {P.}~\bibnamefont {Calabrese}},\ }\href
  {https://doi.org/10.1103/PhysRevB.99.014305} {\bibfield  {journal} {\bibinfo
  {journal} {Phys. Rev. B}\ }\textbf {\bibinfo {volume} {99}},\ \bibinfo
  {pages} {014305} (\bibinfo {year} {2019}{\natexlab{b}})}\BibitemShut
  {NoStop}%
\bibitem [{\citenamefont {Collura}\ \emph {et~al.}(2018)\citenamefont
  {Collura}, \citenamefont {De~Luca},\ and\ \citenamefont
  {Viti}}]{collura2018analytic}%
  \BibitemOpen
  \bibfield  {author} {\bibinfo {author} {\bibfnamefont {M.}~\bibnamefont
  {Collura}}, \bibinfo {author} {\bibfnamefont {A.}~\bibnamefont {De~Luca}},\
  and\ \bibinfo {author} {\bibfnamefont {J.}~\bibnamefont {Viti}},\ }\href
  {https://doi.org/10.1103/PhysRevB.97.081111} {\bibfield  {journal} {\bibinfo
  {journal} {Phys. Rev. B}\ }\textbf {\bibinfo {volume} {97}},\ \bibinfo
  {pages} {081111} (\bibinfo {year} {2018})}\BibitemShut {NoStop}%
\bibitem [{\citenamefont {Bertini}\ \emph {et~al.}(2019)\citenamefont
  {Bertini}, \citenamefont {Piroli},\ and\ \citenamefont
  {Kormos}}]{bertini2019transport}%
  \BibitemOpen
  \bibfield  {author} {\bibinfo {author} {\bibfnamefont {B.}~\bibnamefont
  {Bertini}}, \bibinfo {author} {\bibfnamefont {L.}~\bibnamefont {Piroli}},\
  and\ \bibinfo {author} {\bibfnamefont {M.}~\bibnamefont {Kormos}},\ }\href
  {https://doi.org/10.1103/PhysRevB.100.035108} {\bibfield  {journal} {\bibinfo
   {journal} {Phys. Rev. B}\ }\textbf {\bibinfo {volume} {100}},\ \bibinfo
  {pages} {035108} (\bibinfo {year} {2019})}\BibitemShut {NoStop}%
\bibitem [{\citenamefont {Scopa}\ \emph
  {et~al.}(2022{\natexlab{a}})\citenamefont {Scopa}, \citenamefont
  {Calabrese},\ and\ \citenamefont {Piroli}}]{scopa2022generalized}%
  \BibitemOpen
  \bibfield  {author} {\bibinfo {author} {\bibfnamefont {S.}~\bibnamefont
  {Scopa}}, \bibinfo {author} {\bibfnamefont {P.}~\bibnamefont {Calabrese}},\
  and\ \bibinfo {author} {\bibfnamefont {L.}~\bibnamefont {Piroli}},\ }\href
  {https://doi.org/10.1103/PhysRevB.106.134314} {\bibfield  {journal} {\bibinfo
   {journal} {Phys. Rev. B}\ }\textbf {\bibinfo {volume} {106}},\ \bibinfo
  {pages} {134314} (\bibinfo {year} {2022}{\natexlab{a}})}\BibitemShut
  {NoStop}%
\bibitem [{\citenamefont {Scopa}\ \emph
  {et~al.}(2021{\natexlab{a}})\citenamefont {Scopa}, \citenamefont
  {Calabrese},\ and\ \citenamefont {Piroli}}]{scopa2021real}%
  \BibitemOpen
  \bibfield  {author} {\bibinfo {author} {\bibfnamefont {S.}~\bibnamefont
  {Scopa}}, \bibinfo {author} {\bibfnamefont {P.}~\bibnamefont {Calabrese}},\
  and\ \bibinfo {author} {\bibfnamefont {L.}~\bibnamefont {Piroli}},\ }\href
  {https://doi.org/10.1103/PhysRevB.104.115423} {\bibfield  {journal} {\bibinfo
   {journal} {Phys. Rev. B}\ }\textbf {\bibinfo {volume} {104}},\ \bibinfo
  {pages} {115423} (\bibinfo {year} {2021}{\natexlab{a}})}\BibitemShut
  {NoStop}%
\bibitem [{\citenamefont {Alba}\ \emph {et~al.}(2021)\citenamefont {Alba},
  \citenamefont {Bertini}, \citenamefont {Fagotti}, \citenamefont {Piroli},\
  and\ \citenamefont {Ruggiero}}]{alba2021generalized}%
  \BibitemOpen
  \bibfield  {author} {\bibinfo {author} {\bibfnamefont {V.}~\bibnamefont
  {Alba}}, \bibinfo {author} {\bibfnamefont {B.}~\bibnamefont {Bertini}},
  \bibinfo {author} {\bibfnamefont {M.}~\bibnamefont {Fagotti}}, \bibinfo
  {author} {\bibfnamefont {L.}~\bibnamefont {Piroli}},\ and\ \bibinfo {author}
  {\bibfnamefont {P.}~\bibnamefont {Ruggiero}},\ }\href
  {https://doi.org/10.1088/1742-5468/ac257d} {\bibfield  {journal} {\bibinfo
  {journal} {J. Stat. Mech. Theory Exp.}\ }\textbf {\bibinfo {volume} {2021}},\
  \bibinfo {pages} {114004} (\bibinfo {year} {2021})}\BibitemShut {NoStop}%
\bibitem [{\citenamefont {Piroli}\ \emph {et~al.}(2017)\citenamefont {Piroli},
  \citenamefont {De~Nardis}, \citenamefont {Collura}, \citenamefont {Bertini},\
  and\ \citenamefont {Fagotti}}]{piroli2017transport}%
  \BibitemOpen
  \bibfield  {author} {\bibinfo {author} {\bibfnamefont {L.}~\bibnamefont
  {Piroli}}, \bibinfo {author} {\bibfnamefont {J.}~\bibnamefont {De~Nardis}},
  \bibinfo {author} {\bibfnamefont {M.}~\bibnamefont {Collura}}, \bibinfo
  {author} {\bibfnamefont {B.}~\bibnamefont {Bertini}},\ and\ \bibinfo {author}
  {\bibfnamefont {M.}~\bibnamefont {Fagotti}},\ }\href
  {https://doi.org/10.1103/PhysRevB.96.115124} {\bibfield  {journal} {\bibinfo
  {journal} {Phys. Rev. B}\ }\textbf {\bibinfo {volume} {96}},\ \bibinfo
  {pages} {115124} (\bibinfo {year} {2017})}\BibitemShut {NoStop}%
\bibitem [{\citenamefont {Gopalakrishnan}\ \emph {et~al.}(2018)\citenamefont
  {Gopalakrishnan}, \citenamefont {Huse}, \citenamefont {Khemani},\ and\
  \citenamefont {Vasseur}}]{gopalakrishnan2018hydrodynmaics}%
  \BibitemOpen
  \bibfield  {author} {\bibinfo {author} {\bibfnamefont {S.}~\bibnamefont
  {Gopalakrishnan}}, \bibinfo {author} {\bibfnamefont {D.~A.}\ \bibnamefont
  {Huse}}, \bibinfo {author} {\bibfnamefont {V.}~\bibnamefont {Khemani}},\ and\
  \bibinfo {author} {\bibfnamefont {R.}~\bibnamefont {Vasseur}},\ }\href
  {https://doi.org/10.1103/PhysRevB.98.220303} {\bibfield  {journal} {\bibinfo
  {journal} {Phys. Rev. B}\ }\textbf {\bibinfo {volume} {98}},\ \bibinfo
  {pages} {220303} (\bibinfo {year} {2018})}\BibitemShut {NoStop}%
\bibitem [{\citenamefont {Schemmer}\ \emph {et~al.}(2019)\citenamefont
  {Schemmer}, \citenamefont {Bouchoule}, \citenamefont {Doyon},\ and\
  \citenamefont {Dubail}}]{schemmer2019generalized}%
  \BibitemOpen
  \bibfield  {author} {\bibinfo {author} {\bibfnamefont {M.}~\bibnamefont
  {Schemmer}}, \bibinfo {author} {\bibfnamefont {I.}~\bibnamefont {Bouchoule}},
  \bibinfo {author} {\bibfnamefont {B.}~\bibnamefont {Doyon}},\ and\ \bibinfo
  {author} {\bibfnamefont {J.}~\bibnamefont {Dubail}},\ }\href
  {https://doi.org/10.1103/PhysRevLett.122.090601} {\bibfield  {journal}
  {\bibinfo  {journal} {Phys. Rev. Lett.}\ }\textbf {\bibinfo {volume} {122}},\
  \bibinfo {pages} {090601} (\bibinfo {year} {2019})}\BibitemShut {NoStop}%
\bibitem [{\citenamefont {{Malvania}}\ \emph {et~al.}(2021)\citenamefont
  {{Malvania}}, \citenamefont {{Zhang}}, \citenamefont {{Le}}, \citenamefont
  {{Dubail}}, \citenamefont {{Rigol}},\ and\ \citenamefont
  {{Weiss}}}]{malvania2021generalized}%
  \BibitemOpen
  \bibfield  {author} {\bibinfo {author} {\bibfnamefont {N.}~\bibnamefont
  {{Malvania}}}, \bibinfo {author} {\bibfnamefont {Y.}~\bibnamefont {{Zhang}}},
  \bibinfo {author} {\bibfnamefont {Y.}~\bibnamefont {{Le}}}, \bibinfo {author}
  {\bibfnamefont {J.}~\bibnamefont {{Dubail}}}, \bibinfo {author}
  {\bibfnamefont {M.}~\bibnamefont {{Rigol}}},\ and\ \bibinfo {author}
  {\bibfnamefont {D.~S.}\ \bibnamefont {{Weiss}}},\ }\href
  {https://doi.org/10.1126/science.abf0147} {\bibfield  {journal} {\bibinfo
  {journal} {Science}\ }\textbf {\bibinfo {volume} {373}},\ \bibinfo {pages}
  {1129} (\bibinfo {year} {2021})}\BibitemShut {NoStop}%
\bibitem [{\citenamefont {Zadnik}\ \emph {et~al.}(2021)\citenamefont {Zadnik},
  \citenamefont {Bidzhiev},\ and\ \citenamefont {Fagotti}}]{zadnik2021folded}%
  \BibitemOpen
  \bibfield  {author} {\bibinfo {author} {\bibfnamefont {L.}~\bibnamefont
  {Zadnik}}, \bibinfo {author} {\bibfnamefont {K.}~\bibnamefont {Bidzhiev}},\
  and\ \bibinfo {author} {\bibfnamefont {M.}~\bibnamefont {Fagotti}},\ }\href
  {https://doi.org/10.21468/SciPostPhys.10.5.099} {\bibfield  {journal}
  {\bibinfo  {journal} {SciPost Phys.}\ }\textbf {\bibinfo {volume} {10}},\
  \bibinfo {pages} {099} (\bibinfo {year} {2021})}\BibitemShut {NoStop}%
\bibitem [{\citenamefont {Ruggiero}\ \emph {et~al.}(2020)\citenamefont
  {Ruggiero}, \citenamefont {Calabrese}, \citenamefont {Doyon},\ and\
  \citenamefont {Dubail}}]{ruggiero2020quantum}%
  \BibitemOpen
  \bibfield  {author} {\bibinfo {author} {\bibfnamefont {P.}~\bibnamefont
  {Ruggiero}}, \bibinfo {author} {\bibfnamefont {P.}~\bibnamefont {Calabrese}},
  \bibinfo {author} {\bibfnamefont {B.}~\bibnamefont {Doyon}},\ and\ \bibinfo
  {author} {\bibfnamefont {J.}~\bibnamefont {Dubail}},\ }\href
  {https://doi.org/10.1103/PhysRevLett.124.140603} {\bibfield  {journal}
  {\bibinfo  {journal} {Phys. Rev. Lett.}\ }\textbf {\bibinfo {volume} {124}},\
  \bibinfo {pages} {140603} (\bibinfo {year} {2020})}\BibitemShut {NoStop}%
\bibitem [{\citenamefont {Collura}\ \emph {et~al.}(2020)\citenamefont
  {Collura}, \citenamefont {De~Luca}, \citenamefont {Calabrese},\ and\
  \citenamefont {Dubail}}]{collura2020domain}%
  \BibitemOpen
  \bibfield  {author} {\bibinfo {author} {\bibfnamefont {M.}~\bibnamefont
  {Collura}}, \bibinfo {author} {\bibfnamefont {A.}~\bibnamefont {De~Luca}},
  \bibinfo {author} {\bibfnamefont {P.}~\bibnamefont {Calabrese}},\ and\
  \bibinfo {author} {\bibfnamefont {J.}~\bibnamefont {Dubail}},\ }\href
  {https://doi.org/10.1103/PhysRevB.102.180409} {\bibfield  {journal} {\bibinfo
   {journal} {Phys. Rev. B}\ }\textbf {\bibinfo {volume} {102}},\ \bibinfo
  {pages} {180409} (\bibinfo {year} {2020})}\BibitemShut {NoStop}%
\bibitem [{\citenamefont {Ruggiero}\ \emph {et~al.}(2021)\citenamefont
  {Ruggiero}, \citenamefont {Calabrese}, \citenamefont {Doyon},\ and\
  \citenamefont {Dubail}}]{ruggiero2022quantum}%
  \BibitemOpen
  \bibfield  {author} {\bibinfo {author} {\bibfnamefont {P.}~\bibnamefont
  {Ruggiero}}, \bibinfo {author} {\bibfnamefont {P.}~\bibnamefont {Calabrese}},
  \bibinfo {author} {\bibfnamefont {B.}~\bibnamefont {Doyon}},\ and\ \bibinfo
  {author} {\bibfnamefont {J.}~\bibnamefont {Dubail}},\ }\href
  {https://doi.org/10.1088/1751-8121/ac3d68} {\bibfield  {journal} {\bibinfo
  {journal} {Journal of Physics A: Mathematical and Theoretical}\ }\textbf
  {\bibinfo {volume} {55}},\ \bibinfo {pages} {024003} (\bibinfo {year}
  {2021})}\BibitemShut {NoStop}%
\bibitem [{\citenamefont {Scopa}\ \emph
  {et~al.}(2021{\natexlab{b}})\citenamefont {Scopa}, \citenamefont
  {Krajenbrink}, \citenamefont {Calabrese},\ and\ \citenamefont
  {Dubail}}]{scopa2021exact}%
  \BibitemOpen
  \bibfield  {author} {\bibinfo {author} {\bibfnamefont {S.}~\bibnamefont
  {Scopa}}, \bibinfo {author} {\bibfnamefont {A.}~\bibnamefont {Krajenbrink}},
  \bibinfo {author} {\bibfnamefont {P.}~\bibnamefont {Calabrese}},\ and\
  \bibinfo {author} {\bibfnamefont {J.}~\bibnamefont {Dubail}},\ }\href
  {https://doi.org/10.1088/1751-8121/ac20ee} {\bibfield  {journal} {\bibinfo
  {journal} {Journal of Physics A: Mathematical and Theoretical}\ }\textbf
  {\bibinfo {volume} {54}},\ \bibinfo {pages} {404002} (\bibinfo {year}
  {2021}{\natexlab{b}})}\BibitemShut {NoStop}%
\bibitem [{\citenamefont {Scopa}\ \emph
  {et~al.}(2022{\natexlab{b}})\citenamefont {Scopa}, \citenamefont
  {Calabrese},\ and\ \citenamefont {Dubail}}]{scopa2022exact}%
  \BibitemOpen
  \bibfield  {author} {\bibinfo {author} {\bibfnamefont {S.}~\bibnamefont
  {Scopa}}, \bibinfo {author} {\bibfnamefont {P.}~\bibnamefont {Calabrese}},\
  and\ \bibinfo {author} {\bibfnamefont {J.}~\bibnamefont {Dubail}},\ }\href
  {https://doi.org/10.21468/SciPostPhys.12.6.207} {\bibfield  {journal}
  {\bibinfo  {journal} {SciPost Phys.}\ }\textbf {\bibinfo {volume} {12}},\
  \bibinfo {pages} {207} (\bibinfo {year} {2022}{\natexlab{b}})}\BibitemShut
  {NoStop}%
\bibitem [{\citenamefont {{Scopa}}\ \emph {et~al.}(2023)\citenamefont
  {{Scopa}}, \citenamefont {{Ruggiero}}, \citenamefont {{Calabrese}},\ and\
  \citenamefont {{Dubail}}}]{scopa2023one}%
  \BibitemOpen
  \bibfield  {author} {\bibinfo {author} {\bibfnamefont {S.}~\bibnamefont
  {{Scopa}}}, \bibinfo {author} {\bibfnamefont {P.}~\bibnamefont {{Ruggiero}}},
  \bibinfo {author} {\bibfnamefont {P.}~\bibnamefont {{Calabrese}}},\ and\
  \bibinfo {author} {\bibfnamefont {J.}~\bibnamefont {{Dubail}}}\ }\href
  {https://doi.org/10.48550/arXiv.2301.04094} {10.48550/arXiv.2301.04094}
  (\bibinfo {year} {2023})\BibitemShut {NoStop}%
\bibitem [{\citenamefont {Rylands}\ \emph {et~al.}(2020)\citenamefont
  {Rylands}, \citenamefont {Rozenbaum}, \citenamefont {Galitski},\ and\
  \citenamefont {Konik}}]{rylands2020many}%
  \BibitemOpen
  \bibfield  {author} {\bibinfo {author} {\bibfnamefont {C.}~\bibnamefont
  {Rylands}}, \bibinfo {author} {\bibfnamefont {E.~B.}\ \bibnamefont
  {Rozenbaum}}, \bibinfo {author} {\bibfnamefont {V.}~\bibnamefont
  {Galitski}},\ and\ \bibinfo {author} {\bibfnamefont {R.}~\bibnamefont
  {Konik}},\ }\href {https://doi.org/10.1103/PhysRevLett.124.155302} {\bibfield
   {journal} {\bibinfo  {journal} {Phys. Rev. Lett.}\ }\textbf {\bibinfo
  {volume} {124}},\ \bibinfo {pages} {155302} (\bibinfo {year}
  {2020})}\BibitemShut {NoStop}%
\bibitem [{\citenamefont {Fendley}\ \emph
  {et~al.}(1995{\natexlab{a}})\citenamefont {Fendley}, \citenamefont {Ludwig},\
  and\ \citenamefont {Saleur}}]{fendley1995exact}%
  \BibitemOpen
  \bibfield  {author} {\bibinfo {author} {\bibfnamefont {P.}~\bibnamefont
  {Fendley}}, \bibinfo {author} {\bibfnamefont {A.~W.~W.}\ \bibnamefont
  {Ludwig}},\ and\ \bibinfo {author} {\bibfnamefont {H.}~\bibnamefont
  {Saleur}},\ }\href {https://doi.org/10.1103/PhysRevLett.74.3005} {\bibfield
  {journal} {\bibinfo  {journal} {Phys. Rev. Lett.}\ }\textbf {\bibinfo
  {volume} {74}},\ \bibinfo {pages} {3005} (\bibinfo {year}
  {1995}{\natexlab{a}})}\BibitemShut {NoStop}%
\bibitem [{\citenamefont {Fendley}\ \emph
  {et~al.}(1995{\natexlab{b}})\citenamefont {Fendley}, \citenamefont {Ludwig},\
  and\ \citenamefont {Saleur}}]{fendley1995exact2}%
  \BibitemOpen
  \bibfield  {author} {\bibinfo {author} {\bibfnamefont {P.}~\bibnamefont
  {Fendley}}, \bibinfo {author} {\bibfnamefont {A.~W.~W.}\ \bibnamefont
  {Ludwig}},\ and\ \bibinfo {author} {\bibfnamefont {H.}~\bibnamefont
  {Saleur}},\ }\href {https://doi.org/10.1103/PhysRevB.52.8934} {\bibfield
  {journal} {\bibinfo  {journal} {Phys. Rev. B}\ }\textbf {\bibinfo {volume}
  {52}},\ \bibinfo {pages} {8934} (\bibinfo {year}
  {1995}{\natexlab{b}})}\BibitemShut {NoStop}%
\bibitem [{\citenamefont {Fendley}\ \emph
  {et~al.}(1995{\natexlab{c}})\citenamefont {Fendley}, \citenamefont {Ludwig},\
  and\ \citenamefont {Saleur}}]{fendley1995exact3}%
  \BibitemOpen
  \bibfield  {author} {\bibinfo {author} {\bibfnamefont {P.}~\bibnamefont
  {Fendley}}, \bibinfo {author} {\bibfnamefont {A.~W.~W.}\ \bibnamefont
  {Ludwig}},\ and\ \bibinfo {author} {\bibfnamefont {H.}~\bibnamefont
  {Saleur}},\ }\href {https://doi.org/10.1103/PhysRevLett.75.2196} {\bibfield
  {journal} {\bibinfo  {journal} {Phys. Rev. Lett.}\ }\textbf {\bibinfo
  {volume} {75}},\ \bibinfo {pages} {2196} (\bibinfo {year}
  {1995}{\natexlab{c}})}\BibitemShut {NoStop}%
\bibitem [{\citenamefont {Mehta}\ and\ \citenamefont
  {Andrei}(2006)}]{mehta2006nonequilibrium}%
  \BibitemOpen
  \bibfield  {author} {\bibinfo {author} {\bibfnamefont {P.}~\bibnamefont
  {Mehta}}\ and\ \bibinfo {author} {\bibfnamefont {N.}~\bibnamefont {Andrei}},\
  }\href {https://doi.org/10.1103/PhysRevLett.96.216802} {\bibfield  {journal}
  {\bibinfo  {journal} {Phys. Rev. Lett.}\ }\textbf {\bibinfo {volume} {96}},\
  \bibinfo {pages} {216802} (\bibinfo {year} {2006})}\BibitemShut {NoStop}%
\bibitem [{\citenamefont {Mehta}\ and\ \citenamefont
  {Andrei}(2008)}]{mehta2007nonequilibrium}%
  \BibitemOpen
  \bibfield  {author} {\bibinfo {author} {\bibfnamefont {P.}~\bibnamefont
  {Mehta}}\ and\ \bibinfo {author} {\bibfnamefont {N.}~\bibnamefont {Andrei}},\
  }\href {https://doi.org/10.1103/PhysRevLett.100.086804} {\bibfield  {journal}
  {\bibinfo  {journal} {Phys. Rev. Lett.}\ }\textbf {\bibinfo {volume} {100}},\
  \bibinfo {pages} {086804} (\bibinfo {year} {2008})}\BibitemShut {NoStop}%
\bibitem [{\citenamefont {Konik}\ \emph {et~al.}(2001)\citenamefont {Konik},
  \citenamefont {Saleur},\ and\ \citenamefont {Ludwig}}]{konik2001transport}%
  \BibitemOpen
  \bibfield  {author} {\bibinfo {author} {\bibfnamefont {R.~M.}\ \bibnamefont
  {Konik}}, \bibinfo {author} {\bibfnamefont {H.}~\bibnamefont {Saleur}},\ and\
  \bibinfo {author} {\bibfnamefont {A.~W.~W.}\ \bibnamefont {Ludwig}},\ }\href
  {https://doi.org/10.1103/PhysRevLett.87.236801} {\bibfield  {journal}
  {\bibinfo  {journal} {Phys. Rev. Lett.}\ }\textbf {\bibinfo {volume} {87}},\
  \bibinfo {pages} {236801} (\bibinfo {year} {2001})}\BibitemShut {NoStop}%
\bibitem [{\citenamefont {Konik}\ \emph {et~al.}(2002)\citenamefont {Konik},
  \citenamefont {Saleur},\ and\ \citenamefont {Ludwig}}]{konik2002transport}%
  \BibitemOpen
  \bibfield  {author} {\bibinfo {author} {\bibfnamefont {R.~M.}\ \bibnamefont
  {Konik}}, \bibinfo {author} {\bibfnamefont {H.}~\bibnamefont {Saleur}},\ and\
  \bibinfo {author} {\bibfnamefont {A.}~\bibnamefont {Ludwig}},\ }\href
  {https://doi.org/10.1103/PhysRevB.66.125304} {\bibfield  {journal} {\bibinfo
  {journal} {Phys. Rev. B}\ }\textbf {\bibinfo {volume} {66}},\ \bibinfo
  {pages} {125304} (\bibinfo {year} {2002})}\BibitemShut {NoStop}%
\bibitem [{Note2()}]{Note2}%
  \BibitemOpen
  \bibinfo {note} {See supplementary material that contains (i) additional
  information on the resistance of QIMs using Bethe ansatz (ii) the derivation
  of \protect \textup {\hbox {\mathsurround \z@ \protect \normalfont
  (\ignorespaces \ref {entropyrate}\unskip \@@italiccorr )}} (iii) details
  concerning the bipartite quench in the Kane-Fisher model.}\BibitemShut
  {Stop}%
\bibitem [{\citenamefont {Borsi}\ \emph {et~al.}(2020)\citenamefont {Borsi},
  \citenamefont {Pozsgay},\ and\ \citenamefont
  {Pristy\'ak}}]{borsi2020current}%
  \BibitemOpen
  \bibfield  {author} {\bibinfo {author} {\bibfnamefont {M.}~\bibnamefont
  {Borsi}}, \bibinfo {author} {\bibfnamefont {B.}~\bibnamefont {Pozsgay}},\
  and\ \bibinfo {author} {\bibfnamefont {L.}~\bibnamefont {Pristy\'ak}},\
  }\href {https://doi.org/10.1103/PhysRevX.10.011054} {\bibfield  {journal}
  {\bibinfo  {journal} {Phys. Rev. X}\ }\textbf {\bibinfo {volume} {10}},\
  \bibinfo {pages} {011054} (\bibinfo {year} {2020})}\BibitemShut {NoStop}%
\bibitem [{\citenamefont {Doyon}(2018)}]{doyon2018exact}%
  \BibitemOpen
  \bibfield  {author} {\bibinfo {author} {\bibfnamefont {B.}~\bibnamefont
  {Doyon}},\ }\href {https://doi.org/10.21468/SciPostPhys.5.5.054} {\bibfield
  {journal} {\bibinfo  {journal} {SciPost Phys.}\ }\textbf {\bibinfo {volume}
  {5}},\ \bibinfo {pages} {054} (\bibinfo {year} {2018})}\BibitemShut {NoStop}%
\bibitem [{\citenamefont {Bertini}\ and\ \citenamefont
  {Fagotti}(2016)}]{bertini2016determination}%
  \BibitemOpen
  \bibfield  {author} {\bibinfo {author} {\bibfnamefont {B.}~\bibnamefont
  {Bertini}}\ and\ \bibinfo {author} {\bibfnamefont {M.}~\bibnamefont
  {Fagotti}},\ }\href {https://doi.org/10.1103/PhysRevLett.117.130402}
  {\bibfield  {journal} {\bibinfo  {journal} {Phys. Rev. Lett.}\ }\textbf
  {\bibinfo {volume} {117}},\ \bibinfo {pages} {130402} (\bibinfo {year}
  {2016})}\BibitemShut {NoStop}%
\bibitem [{\citenamefont {Bertini}(2017)}]{bertini2017lightcone}%
  \BibitemOpen
  \bibfield  {author} {\bibinfo {author} {\bibfnamefont {B.}~\bibnamefont
  {Bertini}},\ }\href {https://doi.org/10.1103/PhysRevB.95.075153} {\bibfield
  {journal} {\bibinfo  {journal} {Phys. Rev. B}\ }\textbf {\bibinfo {volume}
  {95}},\ \bibinfo {pages} {075153} (\bibinfo {year} {2017})}\BibitemShut
  {NoStop}%
\bibitem [{\citenamefont {Ljubotina}\ \emph {et~al.}(2019)\citenamefont
  {Ljubotina}, \citenamefont {Sotiriadis},\ and\ \citenamefont
  {Prosen}}]{ljubotina2019non}%
  \BibitemOpen
  \bibfield  {author} {\bibinfo {author} {\bibfnamefont {M.}~\bibnamefont
  {Ljubotina}}, \bibinfo {author} {\bibfnamefont {S.}~\bibnamefont
  {Sotiriadis}},\ and\ \bibinfo {author} {\bibfnamefont {T.}~\bibnamefont
  {Prosen}},\ }\href {https://doi.org/10.21468/SciPostPhys.6.1.004} {\bibfield
  {journal} {\bibinfo  {journal} {SciPost Phys.}\ }\textbf {\bibinfo {volume}
  {6}},\ \bibinfo {pages} {004} (\bibinfo {year} {2019})}\BibitemShut {NoStop}%
\bibitem [{\citenamefont {Capizzi}\ \emph {et~al.}(2023)\citenamefont
  {Capizzi}, \citenamefont {Scopa}, \citenamefont {Rottoli},\ and\
  \citenamefont {Calabrese}}]{capizzi2023domain}%
  \BibitemOpen
  \bibfield  {author} {\bibinfo {author} {\bibfnamefont {L.}~\bibnamefont
  {Capizzi}}, \bibinfo {author} {\bibfnamefont {S.}~\bibnamefont {Scopa}},
  \bibinfo {author} {\bibfnamefont {F.}~\bibnamefont {Rottoli}},\ and\ \bibinfo
  {author} {\bibfnamefont {P.}~\bibnamefont {Calabrese}},\ }\href
  {https://doi.org/10.1209/0295-5075/acb50a} {\bibfield  {journal} {\bibinfo
  {journal} {Europhysics Letters}\ }\textbf {\bibinfo {volume} {141}},\
  \bibinfo {pages} {31002} (\bibinfo {year} {2023})}\BibitemShut {NoStop}%
\bibitem [{\citenamefont {Gouraud}\ \emph {et~al.}()\citenamefont {Gouraud},
  \citenamefont {Doussal},\ and\ \citenamefont
  {Schehr}}]{gouraud2022stationary}%
  \BibitemOpen
  \bibfield  {author} {\bibinfo {author} {\bibfnamefont {G.}~\bibnamefont
  {Gouraud}}, \bibinfo {author} {\bibfnamefont {P.~L.}\ \bibnamefont
  {Doussal}},\ and\ \bibinfo {author} {\bibfnamefont {G.}~\bibnamefont
  {Schehr}}\ }\href {https://doi.org/10.48550/arxiv.2211.15447}
  {10.48550/arxiv.2211.15447}\BibitemShut {NoStop}%
\bibitem [{\citenamefont {Gouraud}\ \emph {et~al.}(2022)\citenamefont
  {Gouraud}, \citenamefont {Doussal},\ and\ \citenamefont
  {Schehr}}]{gouraud2022quench}%
  \BibitemOpen
  \bibfield  {author} {\bibinfo {author} {\bibfnamefont {G.}~\bibnamefont
  {Gouraud}}, \bibinfo {author} {\bibfnamefont {P.~L.}\ \bibnamefont
  {Doussal}},\ and\ \bibinfo {author} {\bibfnamefont {G.}~\bibnamefont
  {Schehr}},\ }\href {https://doi.org/10.1088/1751-8121/ac83fb} {\bibfield
  {journal} {\bibinfo  {journal} {Journal of Physics A: Mathematical and
  Theoretical}\ }\textbf {\bibinfo {volume} {55}},\ \bibinfo {pages} {395001}
  (\bibinfo {year} {2022})}\BibitemShut {NoStop}%
\bibitem [{\citenamefont {Bastianello}\ and\ \citenamefont
  {De~Luca}(2018{\natexlab{a}})}]{bastianello2018nonequilibrium}%
  \BibitemOpen
  \bibfield  {author} {\bibinfo {author} {\bibfnamefont {A.}~\bibnamefont
  {Bastianello}}\ and\ \bibinfo {author} {\bibfnamefont {A.}~\bibnamefont
  {De~Luca}},\ }\href {https://doi.org/10.1103/PhysRevLett.120.060602}
  {\bibfield  {journal} {\bibinfo  {journal} {Phys. Rev. Lett.}\ }\textbf
  {\bibinfo {volume} {120}},\ \bibinfo {pages} {060602} (\bibinfo {year}
  {2018}{\natexlab{a}})}\BibitemShut {NoStop}%
\bibitem [{\citenamefont {Bastianello}\ and\ \citenamefont
  {De~Luca}(2018{\natexlab{b}})}]{bastianello2018superluminal}%
  \BibitemOpen
  \bibfield  {author} {\bibinfo {author} {\bibfnamefont {A.}~\bibnamefont
  {Bastianello}}\ and\ \bibinfo {author} {\bibfnamefont {A.}~\bibnamefont
  {De~Luca}},\ }\href {https://doi.org/10.1103/PhysRevB.98.064304} {\bibfield
  {journal} {\bibinfo  {journal} {Phys. Rev. B}\ }\textbf {\bibinfo {volume}
  {98}},\ \bibinfo {pages} {064304} (\bibinfo {year}
  {2018}{\natexlab{b}})}\BibitemShut {NoStop}%
\bibitem [{\citenamefont {Andrei}(1982)}]{andrei1982calculation}%
  \BibitemOpen
  \bibfield  {author} {\bibinfo {author} {\bibfnamefont {N.}~\bibnamefont
  {Andrei}},\ }\href
  {https://doi.org/https://doi.org/10.1016/0375-9601(82)90702-2} {\bibfield
  {journal} {\bibinfo  {journal} {Physics Letters A}\ }\textbf {\bibinfo
  {volume} {87}},\ \bibinfo {pages} {299} (\bibinfo {year} {1982})}\BibitemShut
  {NoStop}%
\bibitem [{\citenamefont {De~Nardis}\ \emph {et~al.}(2018)\citenamefont
  {De~Nardis}, \citenamefont {Bernard},\ and\ \citenamefont
  {Doyon}}]{denardis2018hydrodynamic}%
  \BibitemOpen
  \bibfield  {author} {\bibinfo {author} {\bibfnamefont {J.}~\bibnamefont
  {De~Nardis}}, \bibinfo {author} {\bibfnamefont {D.}~\bibnamefont {Bernard}},\
  and\ \bibinfo {author} {\bibfnamefont {B.}~\bibnamefont {Doyon}},\ }\href
  {https://doi.org/10.1103/PhysRevLett.121.160603} {\bibfield  {journal}
  {\bibinfo  {journal} {Phys. Rev. Lett.}\ }\textbf {\bibinfo {volume} {121}},\
  \bibinfo {pages} {160603} (\bibinfo {year} {2018})}\BibitemShut {NoStop}%
\bibitem [{\citenamefont {Nardis}\ \emph {et~al.}(2019)\citenamefont {Nardis},
  \citenamefont {Bernard},\ and\ \citenamefont
  {Doyon}}]{denardis2019diffusion}%
  \BibitemOpen
  \bibfield  {author} {\bibinfo {author} {\bibfnamefont {J.~D.}\ \bibnamefont
  {Nardis}}, \bibinfo {author} {\bibfnamefont {D.}~\bibnamefont {Bernard}},\
  and\ \bibinfo {author} {\bibfnamefont {B.}~\bibnamefont {Doyon}},\ }\href
  {https://doi.org/10.21468/SciPostPhys.6.4.049} {\bibfield  {journal}
  {\bibinfo  {journal} {SciPost Phys.}\ }\textbf {\bibinfo {volume} {6}},\
  \bibinfo {pages} {049} (\bibinfo {year} {2019})}\BibitemShut {NoStop}%
\bibitem [{\citenamefont {Takahashi}(1972)}]{takahashi1972one}%
  \BibitemOpen
  \bibfield  {author} {\bibinfo {author} {\bibfnamefont {M.}~\bibnamefont
  {Takahashi}},\ }\href {https://doi.org/10.1143/PTP.47.69} {\bibfield
  {journal} {\bibinfo  {journal} {Prog. Theor. Phys.}\ }\textbf {\bibinfo
  {volume} {47}},\ \bibinfo {pages} {69} (\bibinfo {year} {1972})}\BibitemShut
  {NoStop}%
\bibitem [{\citenamefont {Kane}\ and\ \citenamefont
  {Fisher}(1992)}]{kane1992transmission}%
  \BibitemOpen
  \bibfield  {author} {\bibinfo {author} {\bibfnamefont {C.~L.}\ \bibnamefont
  {Kane}}\ and\ \bibinfo {author} {\bibfnamefont {M.~P.~A.}\ \bibnamefont
  {Fisher}},\ }\href {https://doi.org/10.1103/PhysRevB.46.15233} {\bibfield
  {journal} {\bibinfo  {journal} {Phys. Rev. B}\ }\textbf {\bibinfo {volume}
  {46}},\ \bibinfo {pages} {15233} (\bibinfo {year} {1992})}\BibitemShut
  {NoStop}%
\bibitem [{\citenamefont {Krajnik}\ \emph
  {et~al.}(2022{\natexlab{a}})\citenamefont {Krajnik}, \citenamefont {Schmidt},
  \citenamefont {Pasquier}, \citenamefont {Ilievski},\ and\ \citenamefont
  {Prosen}}]{krajnik2022exact}%
  \BibitemOpen
  \bibfield  {author} {\bibinfo {author} {\bibfnamefont {{\ifmmode
  \check{Z}\else \v{Z}\fi{}}.}~\bibnamefont {Krajnik}}, \bibinfo {author}
  {\bibfnamefont {J.}~\bibnamefont {Schmidt}}, \bibinfo {author} {\bibfnamefont
  {V.}~\bibnamefont {Pasquier}}, \bibinfo {author} {\bibfnamefont
  {E.}~\bibnamefont {Ilievski}},\ and\ \bibinfo {author} {\bibfnamefont
  {T.}~\bibnamefont {Prosen}},\ }\href
  {https://doi.org/10.1103/PhysRevLett.128.160601} {\bibfield  {journal}
  {\bibinfo  {journal} {Phys. Rev. Lett.}\ }\textbf {\bibinfo {volume} {128}},\
  \bibinfo {pages} {160601} (\bibinfo {year} {2022}{\natexlab{a}})}\BibitemShut
  {NoStop}%
\bibitem [{\citenamefont {Krajnik}\ \emph
  {et~al.}(2022{\natexlab{b}})\citenamefont {Krajnik}, \citenamefont {Schmidt},
  \citenamefont {Pasquier}, \citenamefont {Prosen},\ and\ \citenamefont
  {Ilievski}}]{krajnik2022universal}%
  \BibitemOpen
  \bibfield  {author} {\bibinfo {author} {\bibfnamefont {{\ifmmode
  \check{Z}\else \v{Z}\fi{}}.}~\bibnamefont {Krajnik}}, \bibinfo {author}
  {\bibfnamefont {J.}~\bibnamefont {Schmidt}}, \bibinfo {author} {\bibfnamefont
  {V.}~\bibnamefont {Pasquier}}, \bibinfo {author} {\bibfnamefont
  {T.}~\bibnamefont {Prosen}},\ and\ \bibinfo {author} {\bibfnamefont
  {E.}~\bibnamefont {Ilievski}}\ }\href
  {https://doi.org/10.48550/arxiv.2208.01463} {10.48550/arxiv.2208.01463}
  (\bibinfo {year} {2022}{\natexlab{b}})\BibitemShut {NoStop}%
\bibitem [{\citenamefont {Gopalakrishnan}\ \emph {et~al.}(2022)\citenamefont
  {Gopalakrishnan}, \citenamefont {Morningstar}, \citenamefont {Vasseur},\ and\
  \citenamefont {Khemani}}]{gopalakrishnan2022theory}%
  \BibitemOpen
  \bibfield  {author} {\bibinfo {author} {\bibfnamefont {S.}~\bibnamefont
  {Gopalakrishnan}}, \bibinfo {author} {\bibfnamefont {A.}~\bibnamefont
  {Morningstar}}, \bibinfo {author} {\bibfnamefont {R.}~\bibnamefont
  {Vasseur}},\ and\ \bibinfo {author} {\bibfnamefont {V.}~\bibnamefont
  {Khemani}}\ }\href {https://doi.org/10.48550/arxiv.2203.09526}
  {10.48550/arxiv.2203.09526} (\bibinfo {year} {2022})\BibitemShut {NoStop}%
\bibitem [{\citenamefont {McCulloch}\ \emph {et~al.}(2023)\citenamefont
  {McCulloch}, \citenamefont {De~Nardis}, \citenamefont {Gopalakrishnan},\ and\
  \citenamefont {Vasseur}}]{mccullough2023full}%
  \BibitemOpen
  \bibfield  {author} {\bibinfo {author} {\bibfnamefont {E.}~\bibnamefont
  {McCulloch}}, \bibinfo {author} {\bibfnamefont {J.}~\bibnamefont
  {De~Nardis}}, \bibinfo {author} {\bibfnamefont {S.}~\bibnamefont
  {Gopalakrishnan}},\ and\ \bibinfo {author} {\bibfnamefont {R.}~\bibnamefont
  {Vasseur}}\ }\href {https://doi.org/10.48550/arxiv.2302.01355}
  {10.48550/arxiv.2302.01355} (\bibinfo {year} {2023})\BibitemShut {NoStop}%
\bibitem [{\citenamefont {Myers}\ \emph {et~al.}(2020)\citenamefont {Myers},
  \citenamefont {Bhaseen}, \citenamefont {Harris},\ and\ \citenamefont
  {Doyon}}]{myers2020transport}%
  \BibitemOpen
  \bibfield  {author} {\bibinfo {author} {\bibfnamefont {J.}~\bibnamefont
  {Myers}}, \bibinfo {author} {\bibfnamefont {M.~J.}\ \bibnamefont {Bhaseen}},
  \bibinfo {author} {\bibfnamefont {R.~J.}\ \bibnamefont {Harris}},\ and\
  \bibinfo {author} {\bibfnamefont {B.}~\bibnamefont {Doyon}},\ }\href
  {https://doi.org/10.21468/SciPostPhys.8.1.007} {\bibfield  {journal}
  {\bibinfo  {journal} {SciPost Phys.}\ }\textbf {\bibinfo {volume} {8}},\
  \bibinfo {pages} {007} (\bibinfo {year} {2020})}\BibitemShut {NoStop}%
\bibitem [{\citenamefont {{Bertini}}\ \emph {et~al.}(2022)\citenamefont
  {{Bertini}}, \citenamefont {{Calabrese}}, \citenamefont {{Collura}},
  \citenamefont {{Klobas}},\ and\ \citenamefont {{Rylands}}}]{bertini2022full}%
  \BibitemOpen
  \bibfield  {author} {\bibinfo {author} {\bibfnamefont {B.}~\bibnamefont
  {{Bertini}}}, \bibinfo {author} {\bibfnamefont {P.}~\bibnamefont
  {{Calabrese}}}, \bibinfo {author} {\bibfnamefont {M.}~\bibnamefont
  {{Collura}}}, \bibinfo {author} {\bibfnamefont {K.}~\bibnamefont
  {{Klobas}}},\ and\ \bibinfo {author} {\bibfnamefont {C.}~\bibnamefont
  {{Rylands}}},\ }\Eprint {https://arxiv.org/abs/2212.06188} {arXiv:2212.06188}
   (\bibinfo {year} {2022})\BibitemShut {NoStop}%
\bibitem [{\citenamefont {{Bertini}}\ \emph {et~al.}(2023)\citenamefont
  {{Bertini}}, \citenamefont {{Calabrese}}, \citenamefont {{Collura}},
  \citenamefont {{Klobas}},\ and\ \citenamefont {{Rylands}}}]{bertini2023wip}%
  \BibitemOpen
  \bibfield  {author} {\bibinfo {author} {\bibfnamefont {B.}~\bibnamefont
  {{Bertini}}}, \bibinfo {author} {\bibfnamefont {P.}~\bibnamefont
  {{Calabrese}}}, \bibinfo {author} {\bibfnamefont {M.}~\bibnamefont
  {{Collura}}}, \bibinfo {author} {\bibfnamefont {K.}~\bibnamefont
  {{Klobas}}},\ and\ \bibinfo {author} {\bibfnamefont {C.}~\bibnamefont
  {{Rylands}}},\ }\href@noop {} {\bibinfo {title} {in preparation}} (\bibinfo
  {year} {2023})\BibitemShut {NoStop}%
\bibitem [{\citenamefont {Bazhanov}\ \emph {et~al.}(1997)\citenamefont
  {Bazhanov}, \citenamefont {Lukyanov},\ and\ \citenamefont
  {Zamolodchikov}}]{bazhanov1997integrable}%
  \BibitemOpen
  \bibfield  {author} {\bibinfo {author} {\bibfnamefont {V.~V.}\ \bibnamefont
  {Bazhanov}}, \bibinfo {author} {\bibfnamefont {S.~L.}\ \bibnamefont
  {Lukyanov}},\ and\ \bibinfo {author} {\bibfnamefont {A.~B.}\ \bibnamefont
  {Zamolodchikov}},\ }\href {https://doi.org/10.1007/s002200050240} {\bibfield
  {journal} {\bibinfo  {journal} {Communications in Mathematical Physics}\
  }\textbf {\bibinfo {volume} {190}},\ \bibinfo {pages} {247} (\bibinfo {year}
  {1997})}\BibitemShut {NoStop}%
\bibitem [{\citenamefont {Bazhanov}\ \emph {et~al.}(1999)\citenamefont
  {Bazhanov}, \citenamefont {Lukyanov},\ and\ \citenamefont
  {Zamolodchikov}}]{bazhanov1999on}%
  \BibitemOpen
  \bibfield  {author} {\bibinfo {author} {\bibfnamefont {V.~V.}\ \bibnamefont
  {Bazhanov}}, \bibinfo {author} {\bibfnamefont {S.~L.}\ \bibnamefont
  {Lukyanov}},\ and\ \bibinfo {author} {\bibfnamefont {A.~B.}\ \bibnamefont
  {Zamolodchikov}},\ }\href {https://doi.org/10.1016/s0550-3213(99)00198-4}
  {\bibfield  {journal} {\bibinfo  {journal} {Nuclear Physics B}\ }\textbf
  {\bibinfo {volume} {549}},\ \bibinfo {pages} {529} (\bibinfo {year}
  {1999})}\BibitemShut {NoStop}%
\bibitem [{\citenamefont {Fraenkel}\ and\ \citenamefont
  {Goldstein}(2022)}]{fraenkel2022extensive}%
  \BibitemOpen
  \bibfield  {author} {\bibinfo {author} {\bibfnamefont {S.}~\bibnamefont
  {Fraenkel}}\ and\ \bibinfo {author} {\bibfnamefont {M.}~\bibnamefont
  {Goldstein}}\ }\href {https://doi.org/10.48550/arxiv.2205.12991}
  {10.48550/arxiv.2205.12991} (\bibinfo {year} {2022})\BibitemShut {NoStop}%
\bibitem [{\citenamefont {Bertini}\ \emph {et~al.}(2018)\citenamefont
  {Bertini}, \citenamefont {Fagotti}, \citenamefont {Piroli},\ and\
  \citenamefont {Calabrese}}]{bertini2018entanglement}%
  \BibitemOpen
  \bibfield  {author} {\bibinfo {author} {\bibfnamefont {B.}~\bibnamefont
  {Bertini}}, \bibinfo {author} {\bibfnamefont {M.}~\bibnamefont {Fagotti}},
  \bibinfo {author} {\bibfnamefont {L.}~\bibnamefont {Piroli}},\ and\ \bibinfo
  {author} {\bibfnamefont {P.}~\bibnamefont {Calabrese}},\ }\href
  {https://doi.org/10.1088/1751-8121/aad82e} {\bibfield  {journal} {\bibinfo
  {journal} {Journal of Physics A: Mathematical and Theoretical}\ }\textbf
  {\bibinfo {volume} {51}},\ \bibinfo {pages} {39LT01} (\bibinfo {year}
  {2018})}\BibitemShut {NoStop}%
\bibitem [{\citenamefont {Calabrese}\ and\ \citenamefont
  {Cardy}(2005)}]{CalabreseCardy}%
  \BibitemOpen
  \bibfield  {author} {\bibinfo {author} {\bibfnamefont {P.}~\bibnamefont
  {Calabrese}}\ and\ \bibinfo {author} {\bibfnamefont {J.}~\bibnamefont
  {Cardy}},\ }\href {https://doi.org/10.1088/1742-5468/2005/04/p04010}
  {\bibfield  {journal} {\bibinfo  {journal} {J. Stat. Mech. Theory Exp.}\
  }\textbf {\bibinfo {volume} {2005}},\ \bibinfo {pages} {P04010} (\bibinfo
  {year} {2005})}\BibitemShut {NoStop}%
\bibitem [{\citenamefont {{Alba}}\ and\ \citenamefont
  {{Calabrese}}(2017)}]{AlbaCalabrese1}%
  \BibitemOpen
  \bibfield  {author} {\bibinfo {author} {\bibfnamefont {V.}~\bibnamefont
  {{Alba}}}\ and\ \bibinfo {author} {\bibfnamefont {P.}~\bibnamefont
  {{Calabrese}}},\ }\href {https://doi.org/10.1073/pnas.1703516114} {\bibfield
  {journal} {\bibinfo  {journal} {PNAS}\ }\textbf {\bibinfo {volume} {114}},\
  \bibinfo {pages} {7947} (\bibinfo {year} {2017})}\BibitemShut {NoStop}%
\bibitem [{\citenamefont {Alba}\ and\ \citenamefont
  {Calabrese}(2018)}]{AlbaCalabrese2}%
  \BibitemOpen
  \bibfield  {author} {\bibinfo {author} {\bibfnamefont {V.}~\bibnamefont
  {Alba}}\ and\ \bibinfo {author} {\bibfnamefont {P.}~\bibnamefont
  {Calabrese}},\ }\href {https://doi.org/10.21468/SciPostPhys.4.3.017}
  {\bibfield  {journal} {\bibinfo  {journal} {SciPost Phys.}\ }\textbf
  {\bibinfo {volume} {4}},\ \bibinfo {pages} {17} (\bibinfo {year}
  {2018})}\BibitemShut {NoStop}%
\bibitem [{\citenamefont {Alba}(2018)}]{alba2018entanglement}%
  \BibitemOpen
  \bibfield  {author} {\bibinfo {author} {\bibfnamefont {V.}~\bibnamefont
  {Alba}},\ }\href {https://doi.org/10.1103/PhysRevB.97.245135} {\bibfield
  {journal} {\bibinfo  {journal} {Phys. Rev. B}\ }\textbf {\bibinfo {volume}
  {97}},\ \bibinfo {pages} {245135} (\bibinfo {year} {2018})}\BibitemShut
  {NoStop}%
\bibitem [{\citenamefont {Doyon}\ and\ \citenamefont
  {Yoshimura}(2017)}]{doyon2017note}%
  \BibitemOpen
  \bibfield  {author} {\bibinfo {author} {\bibfnamefont {B.}~\bibnamefont
  {Doyon}}\ and\ \bibinfo {author} {\bibfnamefont {T.}~\bibnamefont
  {Yoshimura}},\ }\href {https://doi.org/10.21468/SciPostPhys.2.2.014}
  {\bibfield  {journal} {\bibinfo  {journal} {SciPost Phys.}\ }\textbf
  {\bibinfo {volume} {2}},\ \bibinfo {pages} {014} (\bibinfo {year}
  {2017})}\BibitemShut {NoStop}%
\bibitem [{\citenamefont {Bouchoule}\ \emph {et~al.}(2020)\citenamefont
  {Bouchoule}, \citenamefont {Doyon},\ and\ \citenamefont
  {Dubail}}]{bouchoule2020effect}%
  \BibitemOpen
  \bibfield  {author} {\bibinfo {author} {\bibfnamefont {I.}~\bibnamefont
  {Bouchoule}}, \bibinfo {author} {\bibfnamefont {B.}~\bibnamefont {Doyon}},\
  and\ \bibinfo {author} {\bibfnamefont {J.}~\bibnamefont {Dubail}},\ }\href
  {https://doi.org/10.21468/SciPostPhys.9.4.044} {\bibfield  {journal}
  {\bibinfo  {journal} {SciPost Phys.}\ }\textbf {\bibinfo {volume} {9}},\
  \bibinfo {pages} {044} (\bibinfo {year} {2020})}\BibitemShut {NoStop}%
\bibitem [{\citenamefont {Vecchio}\ \emph {et~al.}(2022)\citenamefont
  {Vecchio}, \citenamefont {Luca},\ and\ \citenamefont
  {Bastianello}}]{delvecchio2022transport}%
  \BibitemOpen
  \bibfield  {author} {\bibinfo {author} {\bibfnamefont {G.~D. V.~D.}\
  \bibnamefont {Vecchio}}, \bibinfo {author} {\bibfnamefont {A.~D.}\
  \bibnamefont {Luca}},\ and\ \bibinfo {author} {\bibfnamefont
  {A.}~\bibnamefont {Bastianello}},\ }\href
  {https://doi.org/10.21468/SciPostPhys.12.2.060} {\bibfield  {journal}
  {\bibinfo  {journal} {SciPost Phys.}\ }\textbf {\bibinfo {volume} {12}},\
  \bibinfo {pages} {060} (\bibinfo {year} {2022})}\BibitemShut {NoStop}%
\bibitem [{\citenamefont {Bastianello}\ \emph {et~al.}(2021)\citenamefont
  {Bastianello}, \citenamefont {Luca},\ and\ \citenamefont
  {Vasseur}}]{bastianello2021hydrodynamics}%
  \BibitemOpen
  \bibfield  {author} {\bibinfo {author} {\bibfnamefont {A.}~\bibnamefont
  {Bastianello}}, \bibinfo {author} {\bibfnamefont {A.~D.}\ \bibnamefont
  {Luca}},\ and\ \bibinfo {author} {\bibfnamefont {R.}~\bibnamefont
  {Vasseur}},\ }\href {https://doi.org/10.1088/1742-5468/ac26b2} {\bibfield
  {journal} {\bibinfo  {journal} {Journal of Statistical Mechanics: Theory and
  Experiment}\ }\textbf {\bibinfo {volume} {2021}},\ \bibinfo {pages} {114003}
  (\bibinfo {year} {2021})}\BibitemShut {NoStop}%
\bibitem [{\citenamefont {Groha}\ and\ \citenamefont
  {Essler}(2017)}]{groha2017spinon}%
  \BibitemOpen
  \bibfield  {author} {\bibinfo {author} {\bibfnamefont {S.}~\bibnamefont
  {Groha}}\ and\ \bibinfo {author} {\bibfnamefont {F.~H.~L.}\ \bibnamefont
  {Essler}},\ }\href {https://doi.org/10.1088/1751-8121/aa7d41} {\bibfield
  {journal} {\bibinfo  {journal} {Journal of Physics A: Mathematical and
  Theoretical}\ }\textbf {\bibinfo {volume} {50}},\ \bibinfo {pages} {334002}
  (\bibinfo {year} {2017})}\BibitemShut {NoStop}%
\bibitem [{\citenamefont {Collura}\ and\ \citenamefont
  {Calabrese}(2013)}]{collura2013entanglement}%
  \BibitemOpen
  \bibfield  {author} {\bibinfo {author} {\bibfnamefont {M.}~\bibnamefont
  {Collura}}\ and\ \bibinfo {author} {\bibfnamefont {P.}~\bibnamefont
  {Calabrese}},\ }\href {https://doi.org/10.1088/1751-8113/46/17/175001}
  {\bibfield  {journal} {\bibinfo  {journal} {Journal of Physics A:
  Mathematical and Theoretical}\ }\textbf {\bibinfo {volume} {46}},\ \bibinfo
  {pages} {175001} (\bibinfo {year} {2013})}\BibitemShut {NoStop}%
\bibitem [{\citenamefont {{Korepin}}\ \emph {et~al.}(1997)\citenamefont
  {{Korepin}}, \citenamefont {{Bogoliubov}},\ and\ \citenamefont
  {{Izergin}}}]{KorepinBogoliubovIzergin}%
  \BibitemOpen
  \bibfield  {author} {\bibinfo {author} {\bibfnamefont {V.~E.}\ \bibnamefont
  {{Korepin}}}, \bibinfo {author} {\bibfnamefont {N.~M.}\ \bibnamefont
  {{Bogoliubov}}},\ and\ \bibinfo {author} {\bibfnamefont {A.~G.}\ \bibnamefont
  {{Izergin}}},\ }\href@noop {} {\emph {\bibinfo {title} {{Quantum Inverse
  Scattering Method and Correlation Functions}}}}\ (\bibinfo {year}
  {1997})\BibitemShut {NoStop}%
\bibitem [{\citenamefont {{Takahashi}}(1999)}]{Takahashi}%
  \BibitemOpen
  \bibfield  {author} {\bibinfo {author} {\bibfnamefont {M.}~\bibnamefont
  {{Takahashi}}},\ }\href@noop {} {\emph {\bibinfo {title} {Thermodynamics of
  One-Dimensional Solvable Models, by Minoru Takahashi, Cambridge, UK:
  Cambridge University Press, 1999}}}\ (\bibinfo {year} {1999})\BibitemShut
  {NoStop}%
\end{thebibliography}%

\onecolumngrid
\newpage 
\newcounter{equationSM}
\newcounter{figureSM}
\newcounter{tableSM}
\stepcounter{equationSM}
\setcounter{equation}{0}
\setcounter{figure}{0}
\setcounter{table}{0}
\setcounter{section}{0}
\makeatletter
\renewcommand{\theequation}{\textsc{sm}-\arabic{equation}}
\renewcommand{\thefigure}{\textsc{sm}-\arabic{figure}}
\renewcommand{\thetable}{\textsc{sm}-\arabic{table}}

\begin{center}
  {\large{\bf Supplemental Material for\\
  ``Transport and entanglement across integrable impurities from Generalized Hydrodynamics''}}
\end{center}
Here we report some useful information complementing the main text. In particular
\begin{itemize}
  \item[-]  In Sec.~\ref{sec:resistivity} we discuss the calculation of the resistivity in  a specific integrable quantum impurity model, the Kondo model.
  \item[-] In Sec.\ref{sec:Derivations} we present the derivation of the GHD equations for the occupation functions and the rate of entropy production.
  \item[-]  In Sec.~\ref{sec:KFequations} we present some relevant details of the bipartite quench of the Kane-Fisher model. 
\end{itemize}

\section{Resistivity in integrable quantum impurity models }
\label{sec:resistivity}
In the main text we have compared our collision integral given in equation (2) to the previous calculations in QIMs of the inverse lifetime and resistivity.  In this section we elaborate on this using a specific example,  the two-lead Kondo model.  The Hamiltonian is given by 
\begin{eqnarray}
H=\sum_{j=1,2}\sum_{\xi=\uparrow,\downarrow}\int \mathrm{d}x\, \psi_{\xi,j}^\dag(x)(-i\partial_x)\psi_{\xi,j}(x)+J\delta(x)\sum_{\xi'=\uparrow,\downarrow}\sum_{l=1,2}\psi^{\dag}_{\xi,j}(x)\vec{\sigma}_{\xi\xi'}\cdot\vec{S}\,\psi_{\xi',l}(x).
\end{eqnarray}
The model describes two types/leads of spinful fermions $\psi_{\alpha,j}^\dag(x),\psi_{\alpha,j}(x)$ with $\alpha=\uparrow,\downarrow$ and lead index $j=1,2$ interacting with an immobile  impurity spin $\vec{S}$ through a Kondo interaction of strength $J$.  One can consider the system for arbitrary representations of spin but here we restrict to a spin $1/2$ impurity and fermions with  $\vec{\sigma}=(\sigma^x,\sigma^y,\sigma^z)$ being Pauli matrices.  The Hamiltonian is the standard one to model quantum dot experiments in the Kondo regime but is nevertheless  integrable~\cite{andrei1980diagonalization,wiegmann1981exact} and many of its equilibrium properties have been investigated in an analytic fashion through its exact solution.  The spectrum consists of two decoupled sectors, \textit{even} and \textit{odd} which consist of symmetric and anti-symmetric combinations of the two leads.  The latter of these decouples form the impurity while the former undergoes nontrivial scattering with the impurity.  The spectrum of the system is given by $E=\sum_{l=1}^{N_e} k_l+\sum_{l=1}^{N_o} q_l,$ where $N_{e,o}$ are the number of even/odd fermions and $k_l,q_l$ are single particle wavevectors in these sectors.  In a finite size $L$ they are quantized according to $e^{iq_lL}=1$ in the odd sector and
\begin{eqnarray}\label{Suppeq:KondoTBA}
e^{ik_lL}&=&\prod_{\alpha=1}^{M}\frac{\lambda_\alpha+i/2}{\lambda_\alpha-i/2}\\
\left[\frac{\lambda_\alpha+i/2}{\lambda_\alpha-i/2}\right]^{N^e}&=&\frac{\lambda_\alpha+c-i/2}{\lambda_\alpha+c+i/2}\prod_{\beta\neq \alpha}^M\frac{\lambda_\alpha-\lambda_\beta+i}{\lambda_\alpha-\lambda_\beta-i}
\end{eqnarray}
in the even sector.  Above $\lambda_\alpha$ are Bethe rapidities which describe the spin part of the even sector, $M$ is the number of down spin even fermions and $c\approx1/4J$ encodes the Kondo interaction strength. This latter quantity is related to the Kondo scale $T_K=De^{-\pi c}$ where $D=N_e/L$ is the density in the even lead and also serving as a momentum cutoff.  In equilibrium $T_K$ marks a crossover between strongly correlated physics at energies below this scale and single paricle physics above it.

By inserting \eqref{Suppeq:KondoTBA} into the energy we find that 
\begin{eqnarray}\label{Suppee:Energy}
E&=&\sum_l^{N_e} \frac{2\pi}{L}n_l^e+N^e\frac{2\pi}{L}\sum_{\alpha}^M\phi_1(\lambda_\alpha)+\sum_l^{N_o} \frac{2\pi}{L}n_l^o\\\label{Suppeq:Energy2}
&=&\sum_l^{N_e} \frac{2\pi}{L}n_l^e-\frac{2\pi}{L}\sum_{\alpha}^M\phi_1(\lambda_\alpha+c)+\sum_\alpha^M\frac{2\pi}{L}I_\alpha+\sum_l^{N_o} \frac{2\pi}{L}n_l^o
\end{eqnarray}
where we have introduced 
\begin{eqnarray}
\phi_n(\lambda)=\frac{1}{2\pi i}\log\left[\frac{\lambda+in /2}{\lambda-in/2}\right]
\end{eqnarray}
and $n^{e/o}_j$ are integers or half integer quantum numbers belonging to the charge degrees of freedom and $I_\alpha$ are the quantum numbers of the spin degrees of freedom.  The remaining term in  \eqref{Suppeq:Energy2} is the contribution of the impurity to the energy allowing us to identify $-\phi_1(\lambda+c)$ as the bare quasiparticle impurity phase shift.  As  discussed in the  main text this quantity will be dressed by the presence of many excitations due to the interactions in the model. 

In the thermodynamic limit the rapidities fill the real line and can be described by a set of distributions $\rho(\lambda), \rho^h(\lambda)$ for the occupied and unoccupied rapidities.  It is also useful to introduce the occupation function $\vartheta(\lambda)=\rho(\lambda)/\rho^t(\lambda)$,~$\rho^t(\lambda)=\rho(\lambda)+\rho^h(\lambda)$.  At finite temperature the Bethe eqautions allow for complex solutions known as strings however for simplicity we consider only real $\lambda$ with the extension to include these strings being a straightforward generalization accommodated using the thermodynamic Bethe ansatz (TBA).  These distributions satisfy an intergal equation
\begin{eqnarray}\label{Suppeq:GSBAE}
\rho^t(\lambda)=N^ea_1(\lambda)+a_1(\lambda)-\int_{-\infty}^\infty\! \mathrm{d}\mu \,a_2(\lambda-\mu)\vartheta(\mu)\rho^t(\mu)
\end{eqnarray}
where $a_j(\lambda)=\frac{d}{d\lambda}\phi_j(\lambda)$. In practice the specific state is obtained by specifying an occupation of modes $\vartheta(\lambda)$ and then solving for $\rho^t(\lambda)$ to obtain $\rho(\lambda)$, for instance in the ground state $\vartheta(\lambda)=1$ and this equation is solved by Fourier transform while in the presence of a magnetic field a number of holes are present in the ground state and  $\vartheta(\lambda)=\Theta(\lambda-B)$ for some parameter $B$ related to the field.

\subsection{Dressed phase shift}
We wish to determine the impurity phase shift of an excitation (in the diagonal basis) over an arbitrary state specified by $\vartheta$.  This consists of increasing $N^e\to N^e+1$ which in turn causes a hole in the distribution of $\lambda$'s~\cite{andrei1983solution}.  Let us denote this hole by $\lambda^h$ and the corresponding quantum number $I^h$.  This then induces a shift in the remaining $\lambda_\alpha\to\tilde{\lambda}_\alpha$ and accordingly the energy given by 
\begin{eqnarray}
\Delta E&=&-\frac{2\pi}{L}I^h-\sum_\alpha^m\left[\phi_1(\tilde{\lambda}_\alpha+c)-\phi_1({\lambda_\alpha}+c)\right]+\phi_1(\lambda^h+c)\\
&=&-\frac{2\pi}{L}I^h+\frac{1}{L}\varphi^{\rm Dr}(\lambda^h)
\end{eqnarray}
where we have identified in the last line the dressed impurity phase shift of the excitation $\varphi^{\rm Dr}(\lambda)$.   In the thermodynamic limit this is given by 
\begin{eqnarray}\label{Suppeq:dressedphaseshift}
\varphi^{\rm Dr}(\lambda^h)&=&-\phi_1(\lambda^h+c)-\sum_\alpha^m\left[\phi_1(\tilde{\lambda}_\alpha+c)-\phi_1({\lambda_\alpha}+c)\right]\\\label{Suppeq:dressedphaseshift2}
&=&-\phi_1(\lambda^h+c)-\int_{-\infty}^\infty \! \mathrm{d}\lambda\, \vartheta(\lambda)F(\lambda|\lambda^h)a_1(\lambda+c)
\end{eqnarray}
where we have introduced the shift function $F(\lambda|\lambda^h)$ which describes the shift in the rapidities $\lambda$ due to the presence of a hole at $\lambda^h$~\cite{KorepinBogoliubovIzergin}.  It satisfies the following integral equation
\begin{eqnarray}
F(\lambda|\lambda^h)=-\phi_2(\lambda-\lambda^h)-\int_{-\infty}^\infty\! \mathrm{d}{\mu}\, \vartheta(\mu)a_2(\lambda-\mu)F(\mu|\lambda^h).
\end{eqnarray}
As discussed in the main text it is useful to define another dressing procedure denoted by $[f]^{\rm dr}$.  For the derivative of the bare impurity phase shift this is
\begin{eqnarray}
[\varphi'(\lambda)]^{\rm{dr}}=-a_1(\lambda+c)-\int \mathrm{d}\mu \,\vartheta(\mu)a_2(\lambda-\mu)[\varphi'(\mu)]^{\rm{dr}}
\end{eqnarray}
Comparing this with~\eqref{Suppeq:dressedphaseshift2} we find that 
\begin{eqnarray}
\varphi^{\rm Dr}(\lambda)=\int^\lambda \!\!\rm{d}\mu \,[\varphi'(\mu)]^{\rm{dr}}.
\end{eqnarray}
In the limit where $\vartheta(\lambda)$ describes the zero field ground state, $ \vartheta\to 1$ the two notions of dressing coincide and we have $\varphi^{\rm{Dr}}=\varphi^{\rm{dr}}$.

\subsection{Resistivity}
The above discussion centered around the diagonal basis of states in the model which can either be in the even or odd sectors with the latter scattering trivially off the impurity while the former acquiring a phase shift $\varphi$ which is dressed to $\varphi^{\rm Dr}$.  To compute transport properties, however, like the inverse lifetime or the resisitivity we must consider the final state of excitations which are created in a certain lead and allowed to scatter of the impurity.   Since the excitations in the even/odd basis are related to the lead basis by taking symmetric and anti-symmetric combinations we can determine that the dressed impurity S-matrix in the scattering basis is 
\begin{eqnarray}
S=\frac{1}{2}\begin{pmatrix}
e^{i\varphi^{\rm Dr}(\lambda)}+1& e^{i\varphi^{\rm Dr}(\lambda)}-1\\
e^{i\varphi^{\rm Dr}(\lambda)}-1& e^{i\varphi^{\rm Dr}(\lambda)}+1
\end{pmatrix}
\end{eqnarray}
which coincides with equation (3) of the main text upon setting $\alpha(\lambda)=\chi(\lambda)$ and $\chi^{\rm Dr}(\lambda)=\varphi^{\rm Dr}(\lambda)/2$.  The reflection amplitude governing this process is therefore given by 
\begin{eqnarray}
\mathcal{R}(\lambda)=\sin^2[\varphi^{\rm Dr}(\lambda)/2]
\end{eqnarray}
as claimed in the text.  For further details on the relationship between the scattering and diagonal bases we direct the interested reader to the lengthy discussion in the context of the Anderson impurity model contained in~\cite{konik2001transport,konik2002transport}.  

In the limit that $\theta(\lambda)=\Theta(\Lambda-\lambda)$ and we consider a hole placed at $\lambda^h\to\Lambda$ we obtain the magnetoresistance of the model as first obtained in~\cite{andrei1982calculation}.

\section{Entropy Production}\label{sec:Derivations}
In this section we derive the expression for the entropy production caused by the impurity.  The thermodynamic entropy in a stationary state is given by the Yang-Yang entropy~\cite{takahashi1972one},
\begin{eqnarray}
\mathcal{S}&=&\sum_{j}\int{\rm d}\lambda\, \mathcal{S}_j(\lambda)\\
\mathcal{S}_j(\lambda)&=&-\left(\rho_j(\lambda)\log[\vartheta_j(\lambda)]+\rho^h_j(\lambda)\log[1-\vartheta_j(\lambda)]\right)
\end{eqnarray}
Given the form of the GHD equations for the root densities,  it can be shown (see next section for details)  that
\begin{eqnarray}
\partial_t\rho^h_j(\lambda,x,t)+\partial_xv_j(\lambda,x,t)\rho^h_j(\lambda,x,t)=\delta(x)\mathcal{I}_j^h(\lambda,t)
\end{eqnarray}
where $\mathcal{I}^h_j(\lambda,t)$ is the collision integral for the hole distributions.   Using this we then have that 
\begin{eqnarray}
\partial_t \mathcal{S}_j(\lambda,x,t)+\partial_x v_j(\lambda,x,t)\mathcal{S}_j(\lambda,x,t)=-\delta(x)\left(\mathcal{I}_j(\lambda,t)\log[\vartheta_j(\lambda,x,t)]+\mathcal{I}^h_j(\lambda,t)\log[1-\vartheta_j(\lambda,x,t)]\right)
\end{eqnarray}
Integrating this over $\lambda,x$ and summing over the quasiparticle species we have 
\begin{eqnarray}\label{eq:entropy}
\partial_t \mathcal{S}(t)=-\sum_j\int {\rm d}\lambda \left(\mathcal{I}_j(\lambda,t)\log[\vartheta_j(\lambda,0,t)]+\mathcal{I}^h_j(\lambda,t)\log[1-\vartheta_j(\lambda,0,t)]\right).
\end{eqnarray}
The interpretation of this expression is straightforward: in the  first term on the right hand side we see the combination of the rate of change for the particles of species $j$ multiplied by its contribution to the entropy. The second term is the same for the quasiparticle holes.  We can reduce this further by using the form of $\mathcal{I}^h_j$ which we now discuss. 

\subsection{GHD equation for the occupation function and hole distributions}
It is often convenient within GHD to deal with the occupation functions $\vartheta_j(\lambda,x,t)$ rather than the root densities.  Fortunately a GHD equation can be obtained for these quantities also.  Following the steps outlined in~\cite{bertini2016transport, CastroDoyonYoshimura} we arrive at a similar equation in the presence of impurities.  

To carry this out we use the vector notation from the appendix of~\cite{bertini2016transport}, namely we denote by vectors functions carrying a single quasiparticle index
\begin{eqnarray}
[\vec{f}]_j(\lambda)=f_j(\lambda)
\end{eqnarray}
while those carrying two indices are represented by an operator $\hat{T}$ such that
\begin{eqnarray}
[\hat{T}\vec{f}]_j(\lambda)=\sum_k\int \mathrm{d}\mu T_{jk}(\lambda,\mu)f_k(\mu)
\end{eqnarray}
The inverse with respect to this operation is denoted ${\hat{T}}^{-1}$.  It is also convenient to introduce an operator for a single index quantity such that for $f_j$
\begin{eqnarray}
[\hat{f}]_{jk}(\lambda,\mu)=\delta(\lambda-\mu)\delta_{jk}f_k(\lambda).
\end{eqnarray} 
With these we can write the Bethe equations as 
\begin{eqnarray}
\vec{\rho^t}&=&\frac{\vec{p'}}{2\pi}-\hat{T}\vec{\rho}\\\vec{\rho}&=&\hat{\vartheta}\vec{\rho^t}.
\end{eqnarray}
Similarly, denoting by $\epsilon_j=v_j\rho_j$ ,$\epsilon^t_j=v_j\rho^t_j$ we can also write
\begin{eqnarray}
\vec{\epsilon^t}&=&\frac{\vec{\varepsilon'}}{2\pi}-\hat{T}\vec{\epsilon}\\
\vec{\epsilon}&=&\hat{\vartheta}\vec{\epsilon^t}.
\end{eqnarray}
Using these it was shown that GHD equations in the absence of the impurity can be written as 
\begin{eqnarray}
[\hat{\vartheta}^{-1}+\hat{T}]^{-1}\hat{\vartheta}^{-1}\left[\partial_t\hat{\vartheta}+\hat{v}\partial_x\hat{\vartheta}\right]\vec{\rho^t}=0,
\end{eqnarray}
leading to
\begin{eqnarray}
\partial_t\hat{\vartheta}+\hat{v}\partial_x\hat{\vartheta}=0.
\end{eqnarray}  
Following the same steps with the impurity term included we find 
\begin{eqnarray}
[\hat{\vartheta}^{-1}+\hat{T}]^{-1}\hat{\vartheta}^{-1}\left[\partial_t\hat{\vartheta}+\hat{v}\partial_x\hat{\vartheta}-\delta(x)\hat{\mathcal{I}}^\vartheta\right]\vec{\rho^t}=0
\end{eqnarray}
where by comparing to equations (1) of the main text we have
\begin{eqnarray}
[\hat{\vartheta}^{-1}+\hat{T}]^{-1}\hat{\vartheta}^{-1}\hat{\mathcal{I}}^\vartheta \, \vec{\rho^t}=\vec{\mathcal{I}}.
\end{eqnarray}
Thus we have 
\begin{eqnarray}
\partial_t\hat{\vartheta}+\hat{v}\partial_x\hat{\vartheta}=\delta(x)\hat{\mathcal{I}}^\vartheta. 
\end{eqnarray}
Using this one can also derive an expression for $\mathcal{I}^t_j$ and likewise for $\mathcal{I}^h_j$.  To do this we note that since
\begin{eqnarray}
\vec{\rho^t}=\hat{\vartheta}^{-1}[\hat{\vartheta}^{-1}+\hat{T}]^{-1}\frac{\vec{p'}}{2\pi}
\end{eqnarray}
we have that 
\begin{eqnarray}
\partial_t\vec{\rho^t}&=&-\hat{\vartheta}^{-1}\partial_t \hat{\vartheta}\vec{\rho^t}+\vartheta^{-1}[\hat{\vartheta}^{-1}+\hat{T}]^{-1}\hat{\vartheta}^{-1}\partial_t \hat{\vartheta}\vec{\rho^t}\\
&=&(\hat{\vartheta}^{-1}[\hat{\vartheta}^{-1}+\hat{T}]^{-1}-1)\hat{\vartheta}^{-1}\partial_t\hat{\vartheta} \vec{\rho^t}.
\end{eqnarray}
From this which we deduce that
\begin{eqnarray}
\vec{\mathcal{I}^t}=(\hat{\vartheta}^{-1}[\hat{\vartheta}^{-1}+\hat{T}]^{-1}-1)\hat{\vartheta}^{-1}\hat{\mathcal{I}^\vartheta}\vec{\rho^t}.
\end{eqnarray}
Then since $\vec{\rho^t}=\vec{\rho}+\vec{\rho^h}$ we have that 
\begin{eqnarray}
\vec{\mathcal{I}^h}&=&((\hat{\vartheta}^{-1}-1)[\hat{\vartheta}^{-1}+\hat{T}]^{-1}-1)\hat{\vartheta}^{-1}\hat{\mathcal{I}^\vartheta}\vec{\rho^t}\\
&=&-(1+\hat{T})\vec{\mathcal{I}}.
\end{eqnarray}
The first term here is expected in a noninteracting model where the rate of scattering of particles is the negative of that of the holes.  The second term is a correction to this in interacting models. 

Using these expressions in~\eqref{eq:entropy} we find that 
\begin{eqnarray}
\partial_t \mathcal{S}(t)&=&\sum_j\int {\rm d}\lambda \,s_j(\lambda)\mathcal{I}_j(\lambda,t),\\
s_j(\lambda,t)&=&\log{\frac{1-\vartheta_j(\lambda,0,t)}{\vartheta_j(\lambda,0,t)}}+\sum_k\int{\rm d}\mu \,T_{jk}(\lambda-\mu)\log{(1-\vartheta_k(\mu,0,t))}
\end{eqnarray}
which is the expression presented in the main text.

\section{Quench dynamics of the Kane-Fisher model}
\label{sec:KFequations}
In this section we present some relevant details on the quench dynamics of the Kane-Fisher model.   It is useful to start by considering the system in the absence of the impurity. In this case the Bethe Ansatz equations are 
\begin{eqnarray}
\rho^t_{\pm}(\lambda)=\frac{e^\lambda}{2\pi}-\int{\rm{d}}\mu \, \left[T_{\pm +}(\lambda-\mu) \rho_+(\mu)+T_{\pm -}(\lambda-\mu)\right] \rho_-(\mu).
\end{eqnarray}
where $T_{\pm\pm}=T_{\pm\mp}\equiv T(\lambda)$  ,
\begin{eqnarray}
T(\lambda)=\int \frac{{\rm d}\omega }{2\pi}e^{-i\omega \lambda} \frac{\sinh(\frac{\pi}{2}(\gamma-1)\omega)}{2\sinh{(\frac{\pi}{2}\gamma\omega)}\cosh{(\frac{\pi}{2}\omega)}}.
\end{eqnarray}
This  can be rewritten
using $\rho^t_+(\lambda)=\rho^t_-(\lambda)\equiv \rho^t(\lambda)$ to give
\begin{eqnarray}\label{Suppeq:KFBAE}
\rho^t(\lambda)=\frac{e^\lambda}{2\pi}-\int{\rm{d}}\mu \, T(\lambda-\mu) (\vartheta_+(\mu)+\vartheta_-(\mu))\rho^t(\mu).
\end{eqnarray}
Upon choosing a specific $\vartheta_{\pm}(\lambda)$ one can solve these equations to determine, $\rho_\pm(\lambda)$. Typically this is done by numerically integrating the equations however in some cases, one of which we consider, it can be treated analytically.  

\begin{figure}
    \centering
\includegraphics[width=.5\linewidth]{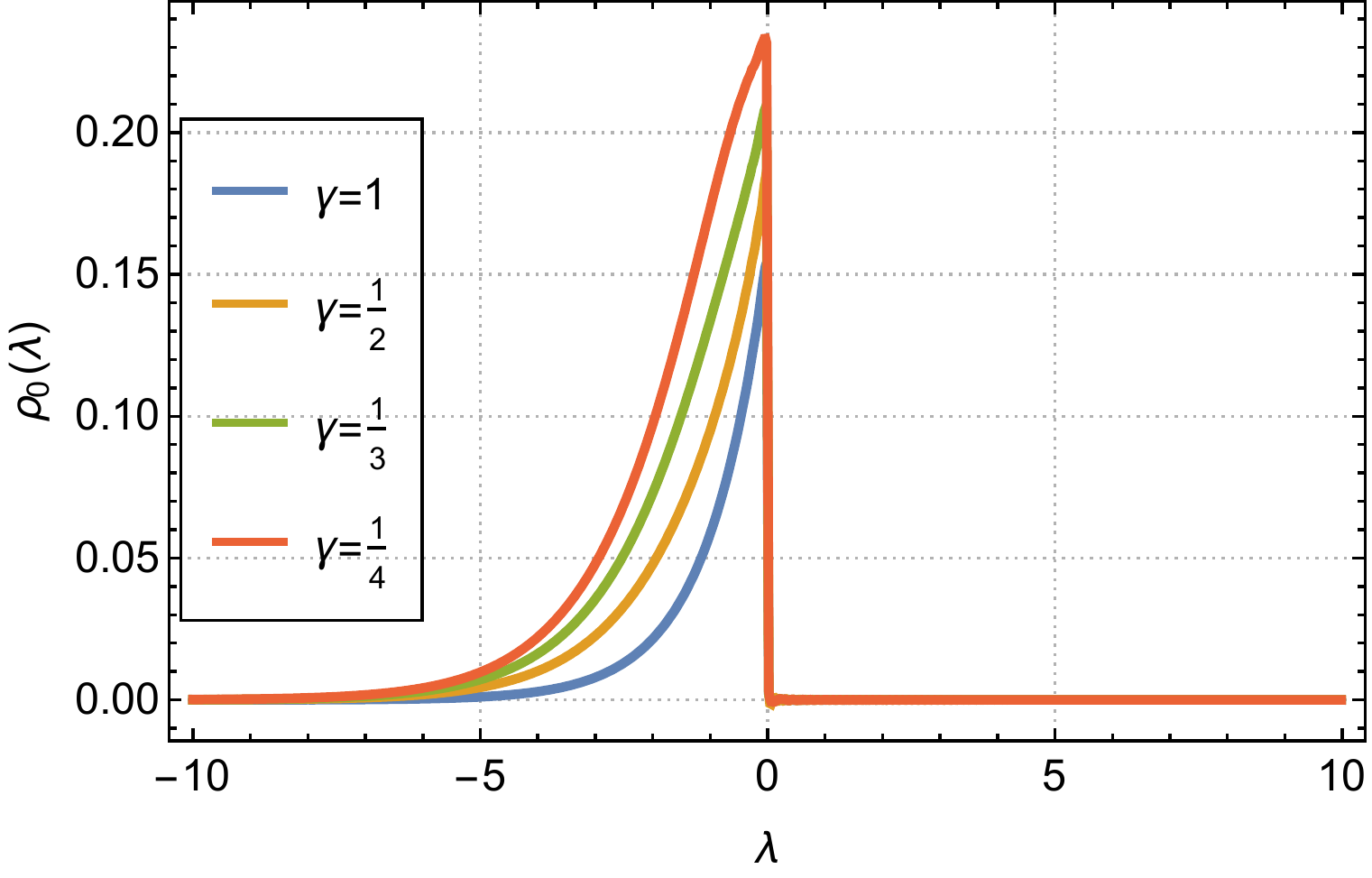}
    \caption{The root density  $\rho_0$ evaluated by numerical integration of~\eqref{eq:rhointial} for $\gamma=1,1/2,1/3,1/4$ with $\Lambda=0$.  }
    \label{fig:rho}
\end{figure}

Our initial state consists of both types of quasiparticles being filled up to a certain rapidity, $\Lambda$ to the left of the impurity. Its occupation functions are $\vartheta_+(\lambda,x,0)=\vartheta_+(\lambda,x,0)=\Theta(\Lambda-\lambda)\Theta(-x)$.  Inserting these into~\eqref{Suppeq:KFBAE} we find that the initial rapidity distributions $\rho_{\pm}(\lambda,x,0)=\Theta(-x)\rho_0(\lambda)$ where $\rho_0(\lambda)$ satisfies
\begin{eqnarray}
\rho_{0}(\lambda)&=&\frac{e^\lambda}{2\pi}-2\int^\Lambda_{-\infty}{\rm{d}}\mu \, T(\lambda-\mu)\rho(\mu),~~\lambda\leq \Lambda,\\
\rho^h_{0}(\lambda)&=&\frac{e^\lambda}{2\pi}-2\int^\Lambda_{-\infty}{\rm{d}}\mu \, T(\lambda-\mu)\rho(\mu),~~\lambda > \Lambda.
\end{eqnarray}
This set of equations can be solved using the Wiener-Hopf integral method, see e.g.~\cite{Takahashi}.  
The result  is 
\begin{eqnarray}\label{eq:rhointial}
\rho_{0}(\lambda)&=&\frac{e^\Lambda}{2\pi}\int_{-\infty}^\infty\frac{{\rm d}\omega}{2 \pi}\frac{G(-i)G(\omega)}{1+i\omega}e^{-i\omega(\lambda-\Lambda)}
\end{eqnarray}
where we have introduced
\begin{eqnarray}
G(\omega)=\sqrt{\frac{1}{\gamma}}\frac{\Gamma\left(1+i\frac{\gamma\omega}{2}\right)\Gamma\left(\frac{1+i\omega}{2}\right)}{\Gamma\left(1+i\frac{\omega}{2}\right)\Gamma\left(\frac{1+i\gamma\omega}{2}\right)}. 
\end{eqnarray}
Some examples for different values of $\gamma$ are plotted in~Fig.\ref{fig:rho}.

When the impurity is included the bare quasi particles are scattered according to the impurity $S$- matrix of the form of equation (3) with~\cite{fendley1994exact} 
\begin{eqnarray}\label{eq:barephase}
\chi(\lambda)=\pi/2-\arctan{e^{
\frac{\lambda-\lambda_U}{\gamma}}}
\end{eqnarray}
and 
\begin{eqnarray}
\alpha(\lambda)=i\int \frac{{\rm d}\omega }{4\omega}e^{-i\omega (\lambda-\lambda_U)} \frac{\tanh(\frac{\pi}{2}\omega)}{\sinh{(\frac{\pi}{2}\gamma\omega)}}+\text{const.}~.
\end{eqnarray}
where the added constant is unimportant. In an arbitrary state these  will be dressed according to 
\begin{eqnarray}\label{eq:dressedchi}
[\chi'(\lambda)]^{\rm dr}=-\frac{1}{2\gamma}\text{sech}\left(\frac{\lambda-\lambda_U}{\gamma}\right)-\int {\rm d}\mu \,T(\lambda-\mu)(\vartheta_+(\mu)+\vartheta_-(\mu))[\chi'(\mu)]^{\rm dr}
\end{eqnarray}
and similarly for $\alpha'(\lambda)$
\begin{eqnarray}\label{eq:dressedchi}
[\alpha'(\lambda)]^{\rm dr}=\alpha'(\lambda)-\int {\rm d}\mu \,T(\lambda-\mu)(\vartheta_+(\mu)+\vartheta_-(\mu))[\alpha'(\mu)]^{\rm dr}.
\end{eqnarray}
 For certain states these can be solved using the Weiner-Hopf method also.

\subsection{GHD Solution }
We look now at the GHD equations for our chosen quench.  As the bulk system is Lorentz invariant and the quasiparticles are massless, the quasiparticle velocity is not dressed by interactions meaning $v_\pm(\lambda)=[\epsilon'_\pm(\lambda)]^{\rm dr}/[p'_\pm(\lambda)]^{\rm dr}=\pm 1$.  Using this the solution at finite time is found to be
\begin{eqnarray}\nonumber
\rho_-(\lambda, x, t)&=&\Theta(-x)\left[\Theta(-x-t)+\Theta(t+x)\mathcal{R}(\lambda,t)\right]
\rho_0(\lambda)\\\label{Rho}
\rho_+(\lambda,x,t)&=&\left[\Theta(-x)+\Theta(x)\Theta(t-x)\mathcal{T}(\lambda,t)\right]\rho_0(\lambda)\end{eqnarray}
where $\mathcal{R}(\lambda,t)=|\sin[\chi^{\rm Dr}(\lambda,t)]|^2$ the dressed reflection amplitude $\mathcal{T}(\lambda,t)=1-\mathcal{R}(\lambda,t)$ the transmission amplitude.  To determine these however we need to find the dressed phase shift $\chi^{\rm Dr}(\lambda,t)$ which gets dressed according to~\eqref{eq:dressedchi} using the state at the origin
\begin{eqnarray}
[\chi'(\lambda,t)]^{\rm dr}=-\frac{1}{2\gamma}\sech{\left(\frac{\lambda-\lambda_U}{\gamma}\right)}
-\int{\rm{d}}\mu \,T(\lambda-\mu)[\vartheta_+(\mu,0,t)+\vartheta_-(\mu,0,t)][\chi'(\mu,t)]^{\rm dr}
\end{eqnarray}
From this we obtain $\chi^{\rm Dr}(\lambda,t)=\int^\lambda{\rm d}\mu[\chi'(\mu,t)]^{\rm dr}$.  
Inserting the form of the finite time solution and using the fact that $\rho^t_{+}(\lambda,x,t)=\rho^t_{+}(\lambda,x,t)$ we find that for $t>0$
\begin{eqnarray}\label{eq:dressedchiprime}
[\chi'(\lambda,t)]^{\rm dr}=-\frac{1}{2\gamma}\sech{\left(\frac{\lambda-\lambda_U}{\gamma}\right)}
-\int_{-\infty}^\Lambda {\rm{d}}\mu \,T(\lambda-\mu)[\chi'(\mu,t)]^{\rm dr}.
\end{eqnarray}
Immediately one sees that this is time independent which serves as a consistency check of the formalism. Indeed GHD is based upon the principle of local equilibrium and so we require that the state at the origin which determines the dressing of the phase shift be stationary.  We can also observe that the dressing of the phase shift is not the same as in the state $\rho_0(\lambda)$ i.e.  the state to the left of the impurity before the quench.  Immediately after the quench one might expect a particle incident from the left to experience a phase shift dressed by $\rho_0$ while one incident from the right to have an undressed phase shift due to that absence of particles.  This is not apparent at the Euler scale and instead the dressed phase shift in the stationary state about the origin, also called the non-equilibrium steady state (NESS),  can be viewed as being  dressed by the average of the initial states to the left and right or the impurity.  

\begin{figure}
    \centering(a)
\includegraphics[trim=0 200 0 200, clip, width=.45\linewidth]{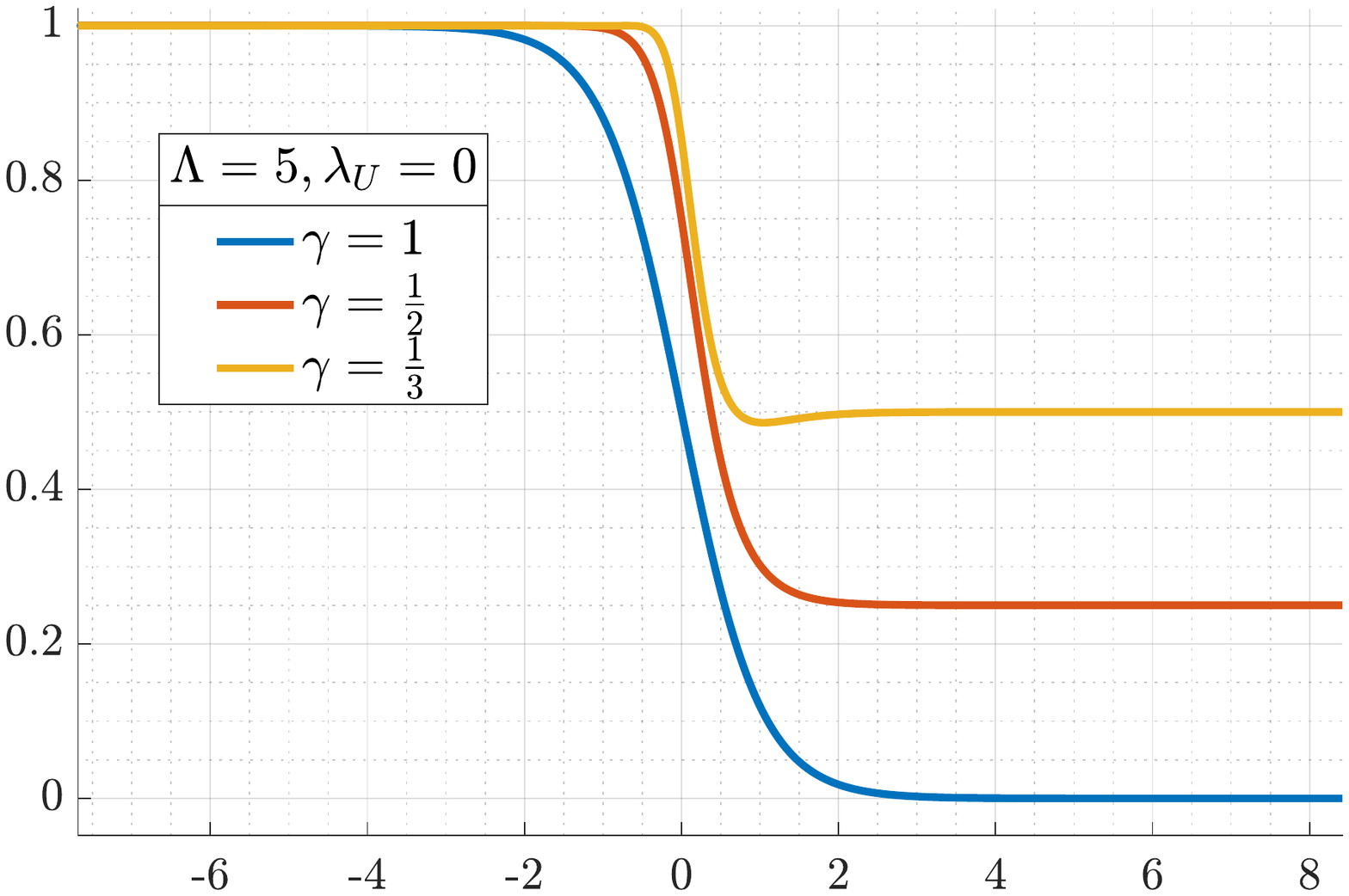}~~(b)
\includegraphics[trim=0 200 0 200, clip, width=.45\linewidth]{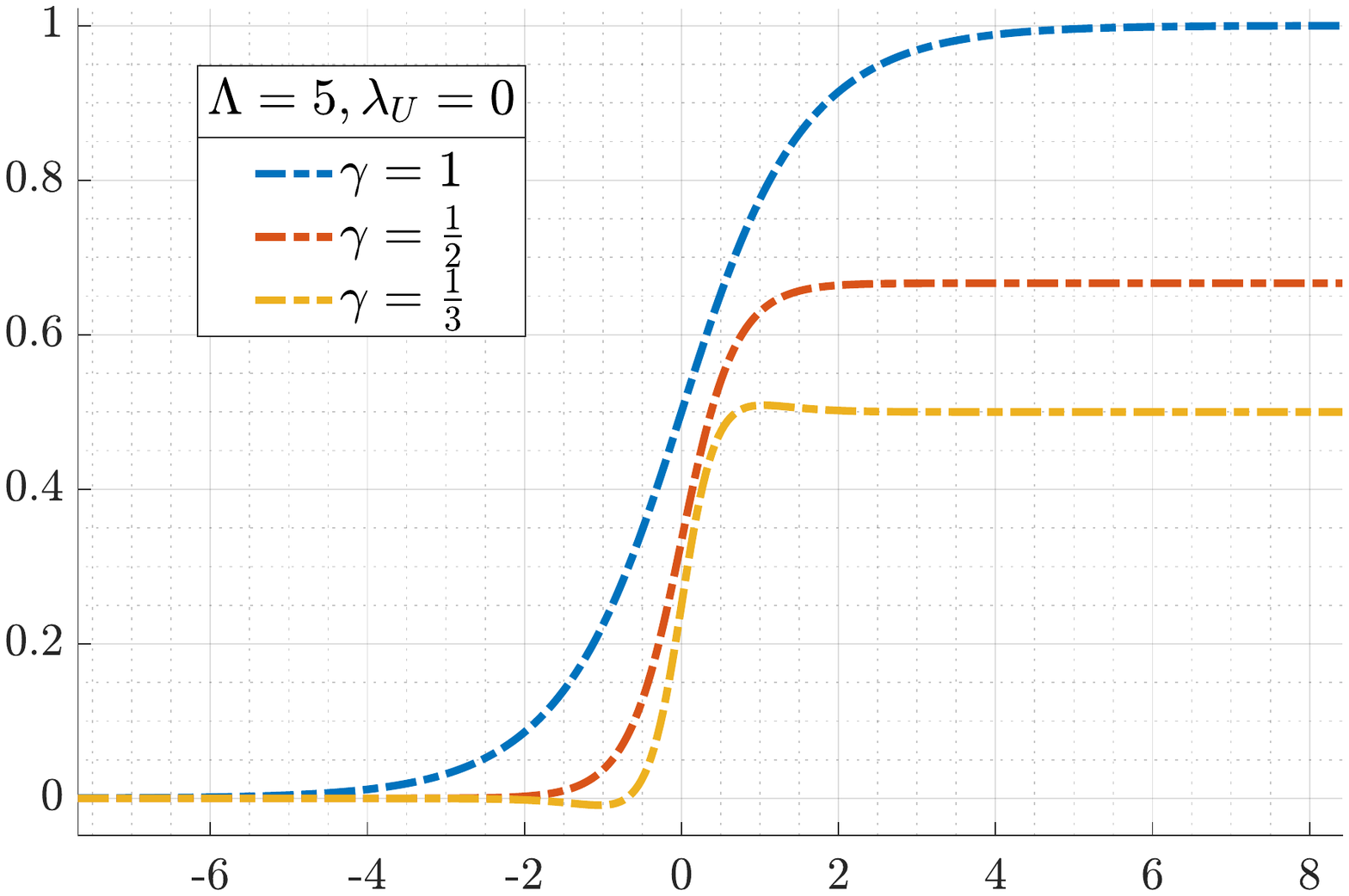}

(c)
\includegraphics[trim=0 200 0 200, clip, width=.45\linewidth]{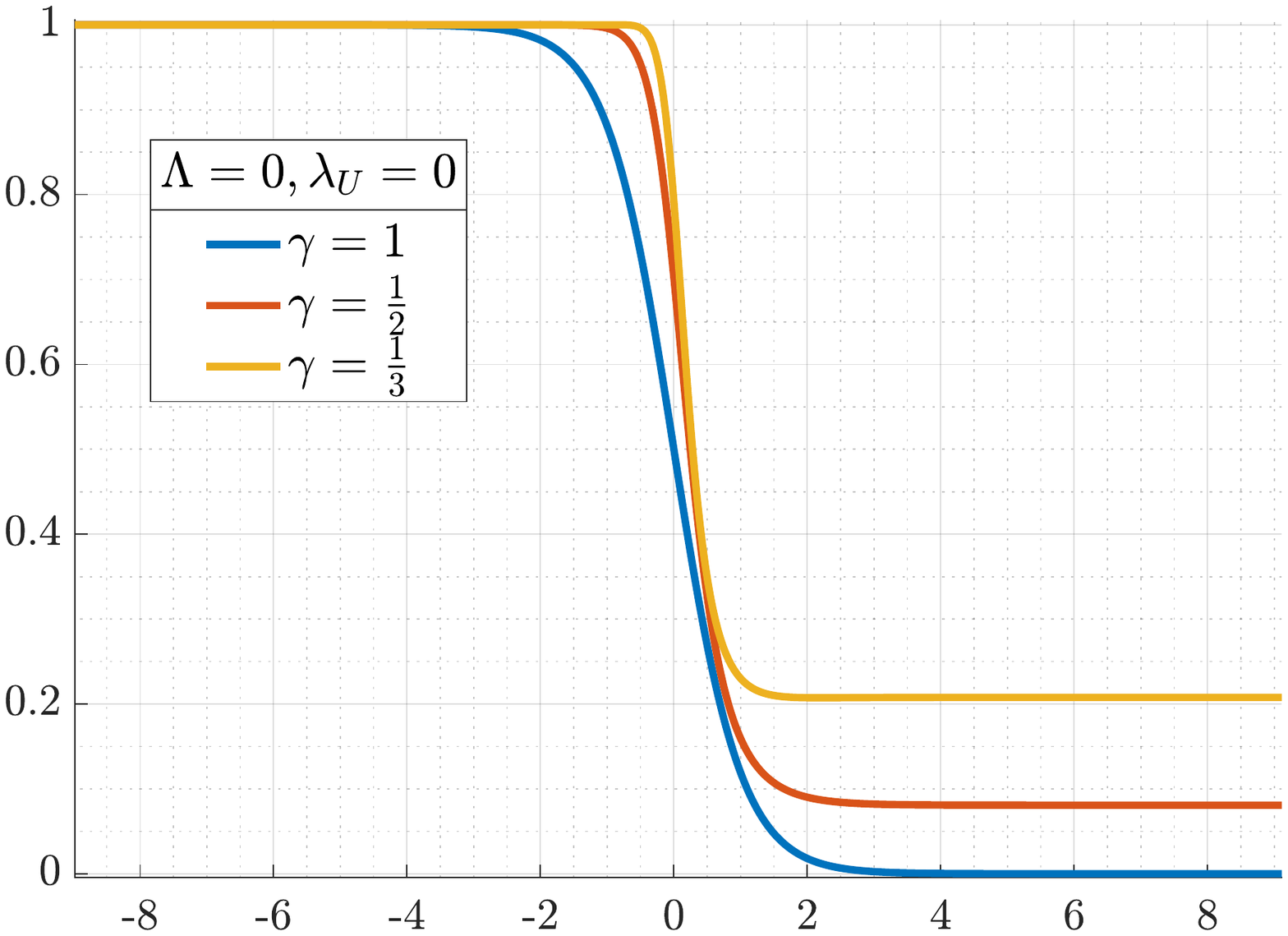}~~(d)
\includegraphics[trim=0 200 0 200, clip, width=.45\linewidth]{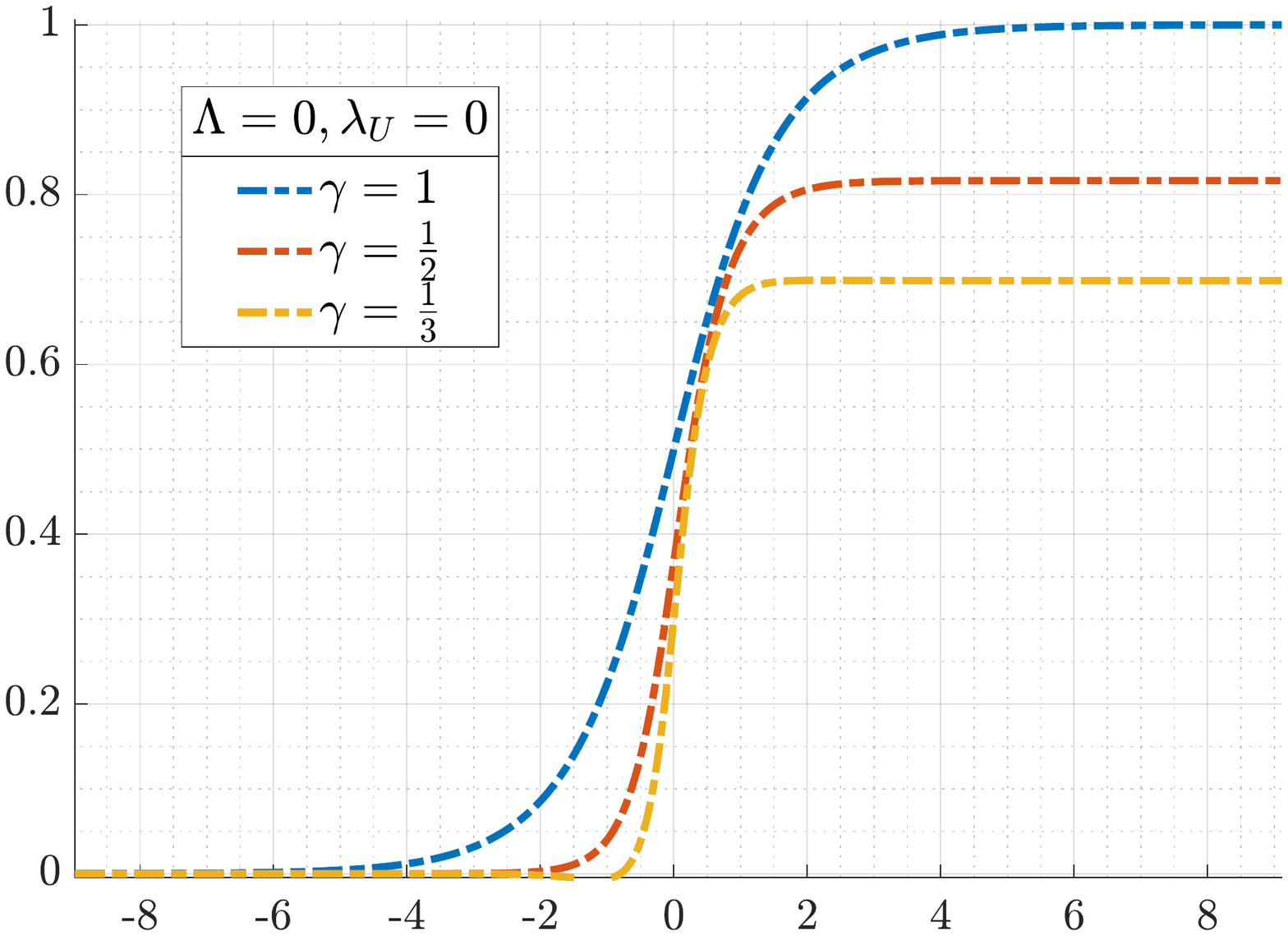}

(e)
\includegraphics[trim=0 200 0 200, clip, width=.45\linewidth]{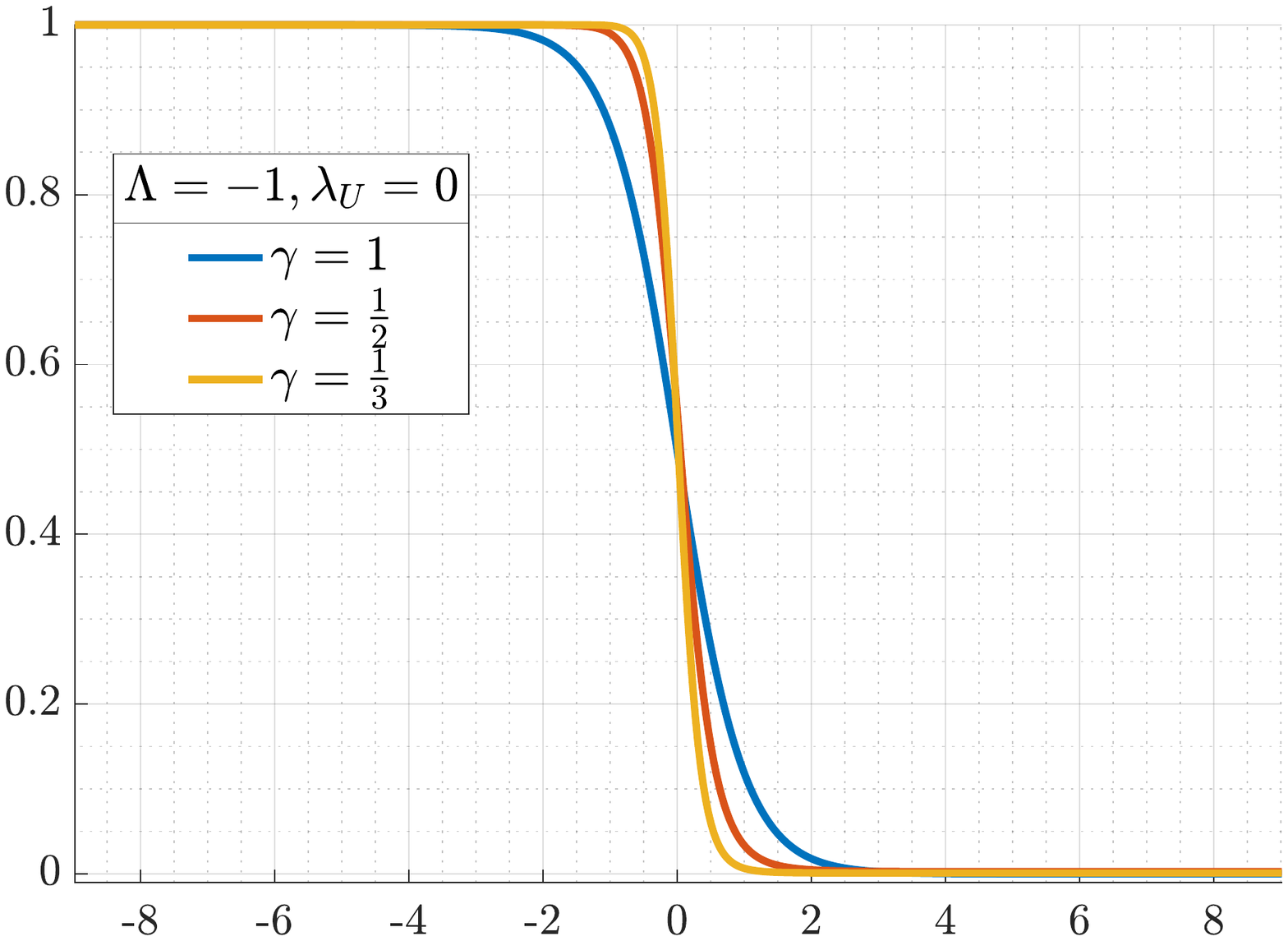}~~(f)
\includegraphics[trim=0 200 0 200, clip, width=.45\linewidth]{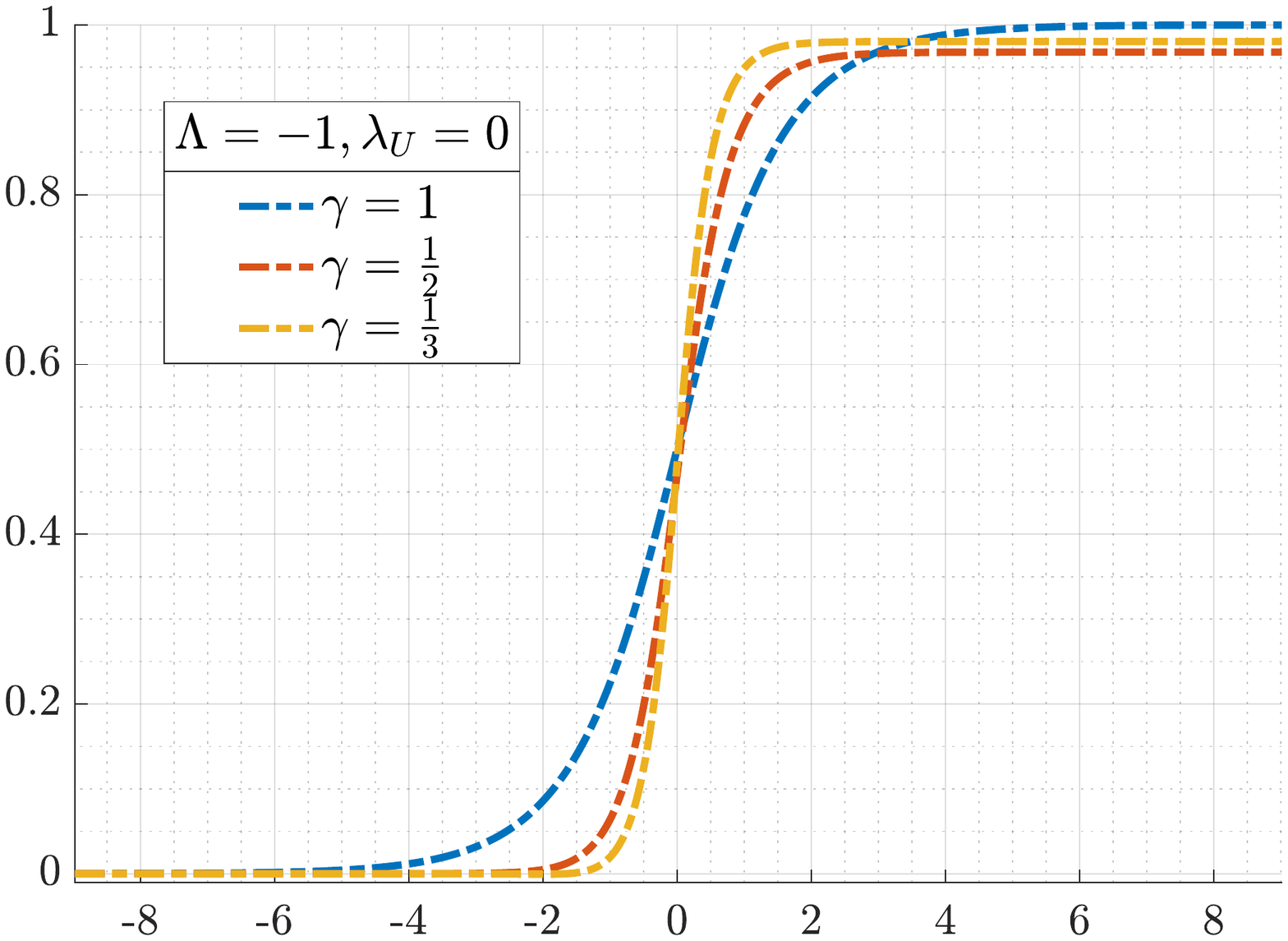}
    \caption{The reflections amplitude, $\mathcal{R}(\lambda)=\sin^2[\chi^{\rm Dr}(\lambda)]$  as a function of $\lambda$ for $\gamma=1,1/2,1/3$ with (a) $\Lambda=5$, $\lambda_U=0$ (c) $\Lambda=0$, $\lambda_U=0$ and (e)  $\Lambda=-1$, $\lambda_U=0$ obtained from~\eqref{eq:dressedchi}. The rescaled and shifted dressed phase shift $1-2\chi^{\rm Dr}(\lambda)/\pi$ for the same values of the parameters are shown in (b), (d) and (f) respectively.  In (b) we see the asymptotic behaviour towards the quasiparticle charge,  $2\gamma/(1+\gamma)$ for $\lambda>0$.  In (d) and (f) the plateaus approach $1$ as the effect of the dressing is reduced.  }
    \label{fig:chiFourier}
\end{figure}

We can solve~\eqref{eq:dressedchiprime} again using the Wiener-Hopf method however it is instructive to examine some  limiting cases first.  Since the driving term is peaked about $\lambda=\lambda_U$ while the scattering kernel is peaked about $\lambda\approx 0$ we see that the phase shift remains undressed when $\lambda_U\gg\Lambda$.  Essentially the scale set by the impurity is much larger than the one determining the quench and so the system is close to equilibrium from this point of view.   Additionally when $\lambda\gg\Lambda$ the dressing is also absent  meaning $[\chi']^{\rm dr}$ has the same $\lambda\pm\infty$ behaviour as $\chi'$.  In the opposite regime when $\Lambda\gg\lambda_U,\lambda$ we can extend the limits of integration to $\Lambda\to\infty$ and solve the system by Fourier transform.  In that case
\begin{eqnarray}\label{eq:chifourier}
\chi^{\rm Dr}(\lambda)\approx\int_{-\infty}^\infty \frac{{\rm d}\omega}{2 i \omega}\frac{\tanh{\left(\frac{\pi}{2}\gamma\omega\right)}\cosh{(\frac{\pi}{2}\omega)}}{\sinh{\left(\frac{\pi}{2}(1+\gamma)\omega\right)}}e^{-i\omega(\lambda-\lambda_U)}+\frac{\pi}{2}
\end{eqnarray}
With the constant fixed by reproducing the proper $\lambda\to-\infty$ behaviour.  For $\gamma=1$ this reproduces the bare result~\eqref{eq:barephase} while we can see the effect of interactions by examining the region $\Lambda\gg\lambda_U$. The integral in this region is dominated by the pole at $\omega=0$ and we find that $\chi^{\rm Dr}$ has a plateau at a value determined by the effective quasiparticle charge in the NESS  for this range of $\lambda$~$2\gamma/(1+\gamma)$,
\begin{equation}
\chi^{\rm Dr}(\lambda)\approx\frac{\pi}{2} \frac{1-\gamma}{1+\gamma}~~\text{for}~~\lambda_U<\lambda\ll\Lambda. 
\end{equation}
This behaviour has consequences for the reflection amplitude which is seen in Fig. ~\ref{fig:chiFourier} where we plot ~\eqref{eq:chifourier} for several values of $\gamma$. 
Combining this with the exponentially decaying  behaviour of  $\rho_0(\lambda)$ allows one to estimate the current through the impurity for $\lambda_U\ll\Lambda$ to be
\begin{eqnarray}
\int{\rm d}\lambda \mathcal{T}(\lambda)\rho_0(\lambda)\approx \cos^2\left(\frac{\pi}{2}\frac{1-\gamma}{1+\gamma}\right)\int{\rm d}\lambda\, \rho_0(\lambda). 
\end{eqnarray}
The full solution for arbitrary $\lambda_U,\Lambda,\lambda$ is given by
\begin{eqnarray}
\chi^{\rm Dr}(\lambda)=\int\frac{{\rm d}\omega{\rm d}z}{8\pi\omega}e^{-i(\lambda-\Lambda)\omega+iz (\lambda_U-\Lambda)}\text{sech}\left(\frac{\pi}{2}\gamma z\right)F(-z)\left[\frac{F(\omega)}{\omega-z+i0^+}-\frac{F(-\omega)^{-1}}{\omega-z-i0^+}\right]
\end{eqnarray}
where now~\cite{fendley1995exact2},
\begin{eqnarray}
F(\omega)=\sqrt{2\pi(1+1/\gamma)}e^{i\omega\Delta}\frac{\Gamma\left(i\frac{(1+\gamma)\omega}{2 }\right)}{\Gamma\left(i\frac{\gamma\omega}{2 }\right)\Gamma\left(\frac{1}{2}+i\frac{\omega}{2 }\right)},~~~~\Delta=-\frac{1}{2}\log(\gamma)-(1+\gamma)\log(1+1/\gamma).
\end{eqnarray}
Again this can be checked to reproduce~\eqref{eq:barephase} for $\gamma=1$.  A series expansion can be obtained for the dressed phase shift by evaluating the integrals by residues, in particular the pole at $\omega=z$  in the first term reproduces the previous expression~\eqref{eq:chifourier} while the second evaluated using this pole gives the bare result with the additional terms providing corrections in the intermediate regions and interpolating between the two.  We plot several examples of $\chi^{\rm Dr}$ and the associated reflections amplitude in~\ref{fig:chiFourier}.

\subsection{Friedel sum rule}
When localized impurities are present in a metal they induce oscillations of the charge density surrounding the impurity known as Friedel oscillations and which result in an excess/defecit of charge being present near the impurity~\cite{mahan2000many}.  Using the Friedel sum rule this can related directly to the sum of the impurity phase shifts at the Fermi surface i.e $\delta N_\text{imp}=\sum_j\varphi_j(E_F)/2\pi$.  This relation remains valid in the NESS so we can use the GHD solution which is valid only at the Euler scale of long wavelengths and times to infer local properties of the system about the impurity.  In particular for the case at hand using the fact that the bare phase shifts are $\alpha\pm\chi$ we have that 
\begin{eqnarray}
\delta N_\text{imp}&=&F(0)\int\frac{{\rm d}\omega}{2\pi i}e^{i\omega (\lambda_U-\Lambda)}\frac{F(-\omega)}{\omega+i 0^+}  \frac{\tanh(\frac{\pi}{2}\omega)}{\sinh{(\frac{\pi}{2}\gamma\omega)}}
\end{eqnarray}
  We can close the contour in the upper or lower half planes depending on the sign of $\lambda_U-\Lambda$, i.e whether the impurity scale is greater or less than the non-equilibrium scale set by the quench.   For $\lambda_U>\Lambda$ we have 
\begin{equation}
\delta N_\text{imp}\simeq-\frac{2}{1+\gamma} +\sum_{n=1}^\infty a_n z^{2n+1}+b_n z^{\frac{2n }{\gamma}}
\end{equation}
for some coefficients $a_n, b_n$ and with $z=e^{\lambda_U-\Lambda}$.  We note here that there is constant term which arises from the pole at $-i 0^+$ indicating a fractional charge deficit.  In the opposite limit, $\lambda_U<\Lambda$  we find
\begin{eqnarray}
\delta N_\text{imp}\simeq \sum_{n=1}^\infty  c_n z^{-2n-1}+d_n z^{-\frac{2n }{\gamma}}+f_n z^{- \frac{2n}{1+\gamma}}
\end{eqnarray}
again with $c_n,d_n,f_n$ being some coefficients coming from the residues.  Here there is as $\delta N_\text{imp}$ should vanish when $\lambda_U\to-\infty$.   In equilibrium expansions of this kind allow one to determine the scaling dimension of relevant or irrelevant operators about a fixed point by comparing the powers of $z$ to perturbative methods.  Since we are out of equilibrium such an interpretation is not appropriate,  nevertheless it is interesting to note that the appearance of different exponents when the impurity scale is smaller than the quench scale.

\end{document}